\documentclass{article}

\newcommand {\bl}{\begin{list}{}{\leftmargin 2em}}
\newcommand {\ite}{\item{}\hspace{-2em}}
\newcommand {\el}{\end{list}}
\newcommand {\beq}{\begin{equation}}
\newcommand {\eeq}{\end{equation}}

\newcommand{\bm}[1]{\mbox{\boldmath $#1$}}

\newcounter{saveeqn}
\newcommand{\alpheqn}{\setcounter{saveeqn}{\value{equation}}%
\stepcounter{saveeqn}\setcounter{equation}{0}%
\renewcommand{\theequation}{\mbox{\arabic{saveeqn}\alph{equation}}}}
\newcommand{\reseteqn}{\setcounter{equation}{\value{saveeqn}}%
\renewcommand{\theequation}{\arabic{equation}}}

\newcounter{EQ1}
\newcounter{EQ1A}
\newcounter{EQ2}
\newcounter{EQ3}
\newcounter{EQ4}
\newcounter{EQ5}
\newcounter{EQ6}
\newcounter{EQ6A}
\newcounter{EQ7}
\newcounter{EQ8}
\newcounter{EQ9}
\newcounter{EQ9A}
\newcounter{EQ9B}
\newcounter{EQ10}

\newcounter{EQ1App}

\usepackage[dvips]{graphicx, color}
\usepackage{exscale, latexsym, makeidx, newlfont}
\usepackage{amsfonts, amsbsy}
\usepackage[mathscr]{euscript}
\begin{document}

\hspace*{-6.5mm} Published on: Communications in Nonlinear Science and Numerical
Simulation, 2018, 59, 515-543 \\
\hspace*{21.5mm} https://doi.org/10.1016/j.cnsns.2017.11.021

\begin{center}
\section*{Measurement-based perturbation theory and differential equation parameter
estimation with applications to satellite gravimetry }
\end{center}

\vspace{5mm}

\hspace*{-5mm}Peiliang Xu \\
 Disaster Prevention Research Institute, Kyoto University, Uji, Kyoto 611-0011, Japan
 \\ pxu@rcep.dpri.kyoto-u.ac.jp

\vspace{5mm}

\begin{center}
\centerline{ \parbox{140mm}{ 
{\bf Abstract}: {\sl The numerical integration method has been routinely used by major
institutions worldwide, for example, NASA Goddard Space Flight Center and German Research
Center for Geosciences (GFZ), to produce global gravitational models from satellite
tracking measurements of CHAMP and/or GRACE types. Such Earth's gravitational products
have found widest possible multidisciplinary applications in Earth Sciences. The method
is essentially implemented by solving the differential equations of the partial
derivatives of the orbit of a satellite with respect to the unknown harmonic coefficients
under the conditions of zero initial values. From the mathematical and statistical point
of view, satellite gravimetry from satellite tracking is essentially the problem of
estimating unknown parameters in the Newton's nonlinear differential equations from
satellite tracking measurements. We prove that zero initial values for the partial
derivatives are incorrect mathematically and not permitted physically. The numerical
integration method, as currently implemented and used in mathematics and statistics,
chemistry and physics, and satellite gravimetry, is groundless, mathematically and
physically. Given the Newton's nonlinear governing differential equations of satellite
motion with unknown equation parameters and unknown initial conditions, we develop three
methods to derive new local solutions around a nominal reference orbit, which are linked
to measurements to estimate the unknown corrections to approximate values of the unknown
parameters and the unknown initial conditions. Bearing in mind that satellite orbits can
now be tracked almost continuously at unprecedented accuracy, we propose the
measurement-based perturbation theory and derive global uniformly convergent solutions to
the Newton's nonlinear governing differential equations of satellite motion for the next
generation of global gravitational models. Since the solutions are global uniformly
convergent, theoretically speaking, they are able to extract smallest possible
gravitational signals from modern and future satellite tracking measurements, leading to
the production of global high-precision, high-resolution gravitational models. By
directly turning the nonlinear differential equations of satellite motion into the
nonlinear integral equations, and recognizing the fact that satellite orbits are measured
with random errors, we further reformulate the links between satellite tracking
measurements and the global uniformly convergent solutions to the Newton's governing
differential equations as a condition adjustment model with unknown parameters, or
equivalently, the weighted least squares estimation of unknown differential equation
parameters with equality constraints, for the reconstruction of global high-precision,
high-resolution gravitational models from modern (and future) satellite tracking
measurements.
 \vspace{2mm} \\ {\bf Key words:} differential equation parameter estimation,
 Earth's gravity field, satellite gravimetry, measurement-based
perturbation, condition adjustment with parameters, nonlinear differential equations,
nonlinear Volterra's integral equations} } }

\end{center}

\vspace{4mm}

\section{Introduction}
The history of geodesy has changed with the launch of the first artificial satellite
Sputnik-1 on 4 October 1957 by the former Soviet Union. Satellites have brought a
revolutionary change of the way we measure the Earth, both geometrically and physically.
Sixty years of satellite gravimetry has since witnessed a profound advance in both
satellite gravity theory and practical production of Earth's gravitational models, in
particular, in the past decade or so, thanks to the launches of three dedicated satellite
gravity missions: the Challenging Mini-satellite Payload (CHAMP) launched in 2000, the
Gravity Recovery and Climate Experiment (GRACE) in 2002, and the Gravity field and
steady-state Ocean Circulation Explorer (GOCE) in 2009. The dedicated satellite gravity
missions have provided frontier challenges and science tools to explore and understand
solid and fluid geophysical processes and dynamics of the Earth (see e.g., NRC 1997; Wahr
et al. 1998; Dickey 2000; Reigber et al. 2003; Tapley et al. 2004a; Rummel et al. 2011a).
From the geodetic point of view, the most expected celebration and continuation of these
sixty years of great achievement of satellite gravimetry may culminate in the launch of
the GRACE Follow-on mission scheduled in 2017. For more information on GRACE Follow-on,
the reader is referred to Flechtner et al. (2014), Christophe et al. (2015) and the
website https://gracefo.jpl.nasa.gov/.

To compute the Earth's gravitational field from satellite tracking measurements, a number
of mathematical methods have been proposed to establish the links between measurements
and the force parameters of the Earth's gravitational field. The major classes of methods
include: (i) linear perturbation methods;  (ii) the dynamical numerical integration
 method; (iii) the orbit-energy-based method; (iv) the two-point (orbital)
boundary value problem theory, which was first solved by Schneider (1968, 1984) and
further developed by Ilk et al. (2005, 2008) (see also Mayer-G\"{u}rr et al. 2005); (v)
the orbit-inverted acceleration approach. The idea of computing accelerations from
GPS-derived coordinates was first proposed by Jekeli and Garcia (1997) for airborne
gravimetry and then applied to reconstruct the Earth's gravitational field from CHAMP
mission by Reubelt et al. (2003)(see also Ditmar and van der Sluijs 2004; Bezd\v{e}k et
al. 2014); and (vi) satellite gradiometry (see e.g., Rummel 1986; Rummel et al. 2011b).
Recently, Xu (2008, 2012) proposed a measurement-based perturbation method, which is
globally convergent uniformly.

In this paper, we will be mainly concerned with the first two types of methods to
reconstruct the Earth's gravitational field from satellite tracking measurements, because
Earth's gravitational models were first derived by using linear perturbation methods, and
because the numerical integration method has now been routinely used to produce global
 Earth's gravitational models. Although the orbit-energy-based method is
mathematically rigorous, as first proposed by Bjerhammer (1967, 1969) and likely also
independently by Wolff (1969) and further modified by Jekeli (1999) in order to account
for technological advance of space observation, it is still not able to fully utilize the
unprecedented accuracy of all modern space measurements (see also Hotine and Morrison
1969). Neither the two-point boundary value problem nor the orbit-inverted acceleration
approach have been used by any major institutions such as NASA Goddard Space Flight
Center and GFZ to produce  global Earth's gravitational models, though they are indeed
used to compute gravitational models from GRACE and CHAMP measurements (see e.g., Reubelt
et al. 2003; Ditmar and van der Sluijs 2004; Mayer-G\"{u}rr et al. 2005; Bezd\v{e}k et
al. 2014). The two-point boundary value problem always has been based on short arcs and,
as a result, can never fully utilize unprecedented accuracy and continuity of missions of
GRACE/CHAMP types to their potential limit of measurement technology. Nevertheless, the
method is only of quasi-linear accuracy but is well known to be difficult to implement,
in particular, for sparse tracking data. Orbit-inverted accelerations can be unreliable
and inaccurate, since the operator of differentiation is ill-posed in nature. As a direct
consequence, position-inverted acceleration signals can be distorted and their noise can
be substantially amplified, the extent of which depends on the noise level of satellite
positions, the time interval used to derive accelerations and regularization. Whenever
possible, one should avoid solving this intermediate inverse ill-posed problem and use
its distorted signals of acceleration to further invert for global gravitational models.
In the case of satellite gradiometry, since we directly measure the second gradients of
the gravitational potential, the observational equations of the gravitational tensors are
mathematically straightforward (Rummel 1986; Rummel et al. 2011b).

Linear perturbation methods are to find an approximate solution to nonlinear Lagrange's
planetary equations and mathematically rigorous, which have been well developed and
documented (see e.g., Brouwer 1944, 1959; Kozai 1959; Brouwer and Clemence 1961; Kaula
1961, 1966; Hagihara 1972; Taff 1985). Soon after the launches of first artificial
satellites in 1950s and 60s, with camera and Doppler tracking measurements, linear
perturbation solutions and simplified variants for small and/or zero divisors were used
to compute the flattening and/or the eccentricity of the Earth (see e.g., Buchar 1958;
Merson and King-Hele 1958; King-Hele and Merson 1959; O'Keefe et al. 1959; Izsak 1961),
lumped and/or resonance-derived harmonic coefficients (see e.g., Cook 1961, 1963, 1967;
Anderle 1965a; Yionoulis 1965; Kloko\v{c}n\'{i}k and Posp\'{i}\v{s}ilov\'{a} 1981;
King-Hele and Walker 1982; Kloko\v{c}n\'{i}k et al. 2013), and gravitational models with
low degrees and orders (see e.g., Kaula 1961a, 1966; Kozai 1961; Izsak 1963; Guier 1963;
Guier and Newton 1965; Cook 1963, 1967; Hagihara 1971). Combined solutions of satellite
tracking measurements with terrestrial gravity data can be found, for example, in Kaula
(1961b), Gaposchkin and Lambeck (1970, 1971) and Gaposchkin (1974). For more information
on the early work on the determination of the Earth's gravitational field from satellite
tracking measurements, the reader is referred to two excellent reviews by Kaula (1963)
and Kozai (1966). Although linear perturbation methods nowadays are mainly used for
mission analysis, they could be revitalized to produce global satellite gravitational
models by implementing the idea of measurement-based perturbation developed by Xu (2008,
2012).

Estimating unknown differential equation parameters has been essential in many areas of
science and engineering. A key mathematical component of such an estimation procedure has
been to solve the derived differential equations of the partial derivatives with respect
to the unknown differential equation parameters under the assumption of zero initial
values for the partial derivatives, as originally published by Gronwall (1919) almost 100
years ago (see also Goddington and Levinson 1955; Howland and Vaillancourt 1961). This
estimation technique has found widest possible applications, for example, in mathematics,
statistics, chemistry, physics and satellite gravimetry, which, however, has now been
best known as the dynamical numerical integration method in geodesy. Actually, in the
community of geoscience, this method seemed to be first hinted at by Anderle (1965b),
likely independent of what had been published by then, since no references were made to
the mathematical literature of Gronwall (1919), Goddington and Levinson (1955) and
Howland and Vaillancourt (1961). The method was then (re-)published in a mathematical
paper by Riley et al. (1967) (see also Ballani 1988; Montenbruck and Gill 2000; Beutler
et al. 2010), again, without any reference to the above-mentioned mathematical
literature.

The dynamical numerical integration method has gained wide spread acceptance without
challenge with the publications of Anderle (1965b) and Riley et al. (1967), and has since
been routinely used by almost all major institutions worldwide to produce global
gravitational models from satellite tracking measurements, likely partly attributed to
the fact that NASA Goddard Space Flight Center used and implemented this numerical
integration idea by Anderle (1965b) as the mathematical foundation to compute Earth's
gravitational models (see e.g., Lerch et al. 1974; Long et al. 1989). Among the most
important gravitational models before the dedicated satellite gravity missions CHAMP,
GRACE and GOCE are the GEM series of gravitational models from the Goddard Space Flight
Center (see e.g., Marsh et al. 1988, 1990) and those from the joint German-French team
(see e.g., Schwintzer et al. 1997; Biancale et al. 2000). For reviews on progress in
satellite gravimetry before the launches of the dedicated satellite gravity missions, the
reader is referred to Lambeck and Coleman (1983) (see also Lambeck and Coleman 1986) for
a brief progress report from 1958 to 1982 and to Nerem et al. (1995) for a summarized
report, retrospective and prospective on gravity observation.

Although the numerical integration method has been widely accepted by almost all major
institutions worldwide as a standard method for making global  gravitational models from
satellite tracking measurements, this is an unbelievable scientific fallacy, since the
method is based on groundless claim and proved to be incorrect, mathematically and
physically (Xu 2009a, 2015a, 2015b). Unfortunately, the method has been used for almost
60 years by all important institutions worldwide to compute global  gravitational models
from satellite tracking measurements for use in geodesy, solid geophysics, ocean
dynamics, hydrology, interaction of ocean and surface water with atmosphere, and far more
beyond, bearing in mind that no technical documentation is available to provide a
complete and mathematically rigorous support for the method, to my best knowledge, after
an extensive search of literature.

In the 1965 U.S. Naval Weapons Laboratory technical report, Anderle (1965b) wrote two
sentences in the section of Procedure to mean this method by saying ``{\em Numerical
integration ... was used to compute the orbit of the satellite. The partial derivatives
of satellite position with respect to orbit and gravity parameters were also obtained by
numerical integration of the perturbation equations}''. No mathematical formulation was
given to provide further technical support and explanation of the idea. In connection
with the idea by Anderle (1965b), a complete publication about the mathematics of the
method was given by Riley et al. (1967) from Hughes Aircraft Company and Aerospace
Corporation, which was soon well known and accepted among geodesists worldwide. Although
Riley et al. (1967) correctly derived the differential equations of the orbit and
velocity of a satellite with respect to differential equation parameters in their
mathematics paper, they simply claimed at the beginning of the second page of the paper
``{\em The initial values ... in general will be zero if $\beta_k$ is a differential
equation parameter}''. As a matter of fact, the differential equations for the partial
derivatives with respect to the equation unknowns and the zero initial values for the
partial derivatives dated back much earlier to Gronwall (1919) and Ritt (1919) (see also
Goddington and Levinson 1955; Howland and Vaillancourt 1961). Mathematically, this
zero-initial-value statement does not derive from the original differential equations but
is nothing more than a claim. Obviously, a key element to decide the particular solution
of differential equations is claimed without any mathematical/physical justification. In
the technical report on Goddard Earth models (5 and 6), Lerch et al. (1974) reported that
they used the numerical integration method from the idea of Anderle (1965b). More
precisely, in the appendix of orbit theory for the software system GEODYN, Lerch et al.
(1974) wrote ``{\em The partial derivatives ... are obtained by direct numerical
integration of the variational equations}'' on page A1-10 and ``{\em Initially, ... the
rest of the matrix (corresponding to the partial derivatives -- notes added by the
author) is zero}'' on page A1-22. To support the production of global gravitational
models from satellite tracking measurements, Goddard Space Flight Center, together with
Computer Sciences Corporation, prepared a lengthy technical report of about 700 pages
(Long et al. 1989). In Section 6.1.4 on page 6-11, although Long et al. (1989) correctly
realized that their differential equations (6-49) required initial conditions, and even
though the initial values are the key to solve the differential equations, they did not
touch the issue of how to determine the initial values but chose to show how to
numerically solve the differential equations as if the initial values had been given. The
numerical integration method was followed by the joint German/French team as well (see
e.g., Reigber 1989). Recently, Xu (2009a, 2015a, 2015b) mathematically proved rigorously
that assigning zero values to the initial partial derivatives violates the physics of
motion of celestial bodies. More will come in Section~\ref{SectNumMeth}.

Profound technological advances in space observation have been achieved in comparison
with those in 1950s and 1960s when linear perturbations were significantly developed and
the numerical integration method was hinted at and published. Two most important features
of these advances are: (i) low Earth orbiting (LEO) satellites can now be tracked and
directly measured, precisely and almost continuously, by using Global Navigation
Satellite Systems (GNSS). In other words, we have precise and continuous orbits of LEO
satellites with an arc of arbitrary length; and (ii) tracking measurements are of
unprecedented  accuracy. With GNSS, the orbital precision of LEO satellites can now
routinely reach the level of 1 cm (see e.g., \v{S}vehla \& Rothacher 2005) and even the
level of millimeters over a short period of time, as demonstrated by experiments on the
ground (Xu et al. 2013). The accuracy of inter-satellite tracking is now at the level of
a few $\mu m$ in rangings and $0.1 \mu m/s$ in range rates (see e.g., Kim 2000). Much
higher accuracy can be expected when the new generations of laser instruments are
operational (see e.g., Bender {\em et al.} 2003; Seeber 2003; Pierce et al. 2008; Sheard
et al. 2012; Turyshev et al. 2014).

Both linear perturbation methods and the numerical integration method (even if it were
correct) are not able to utilize a long orbital arc of continuous tracking and
unprecedented  accuracy of modern space observation technology. Linear perturbation
methods are only valid in a small neighborhood of mean orbital elements and almost break
down in the case of small divisors such as resonances, critical inclinations and circular
orbits. In the case of the numerical integration method, let us treat it as if it were
correct for now, common practice is always to divide a long arc into many small pieces,
say in hours or one day, because the modelling error will increase with time. Up to a
certain epoch, the modelling error will dominate such that a gravitational solution would
not be physically meaningful any more. To control the increase of modelling errors, one
would have to divide a long arc into many short arcs. A direct consequence of this common
practice is that we will not be able to extract small gravitational signals from
satellite tracking measurements, since, small gravitational signals would take time to
show up their effects on the orbit. Thus, to summarize, we would conclude that both
linear perturbation methods and the numerical integration method (as if it were correct)
are too approximate to benefit from and do not match profound technological advances in
modern and future space observation. Actually, Lambeck and Coleman (1986) pointed out
that further improvements both in satellite gravity theory and data evaluation methods
are required before the next generation of satellite gravity missions is launched.

The purposes of this paper are threefold: (i) to briefly review the methods of linear
perturbation  and to prove that the numerical integration method is groundless,
mathematically and physically, though the numerical integration method was essentially
first published in the mathematical literature by Gronwall (1919) (see also Goddington
and Levinson 1955; Howland and Vaillancourt 1961; Riley et al. 1967) and then widely used
in chemistry and physics  (Dickinson et al. 1976; Hwang et al. 1978; Linga et al. 2006),
statistics (Ramsay et al. 2007; Wang and Enright 2013), and likely independently
developed and applied in satellite gravimetry (Anderle 1965b; Riley et al. 1967; Lerch et
al. 1974; Long et al. 1989; Ballani 1988; Reigber 1989; Montenbruck and Gill 2000;
Beutler et al. 2010); (ii) given differential equations with unknown parameters and
unknown initial conditions, we will construct the linearized solutions to the
differential equations in terms of the unknown corrections to approximate values of the
unknown differential equation parameters and the unknown initial conditions by extending
Euler and modified Euler numerical integration methods. These linearized solutions are of
local nature, since they are derived with a nominal reference trajectory. They are
mathematically rigorous and require no assumption of zero initial values for the partial
derivatives with respect to the unknown differential equation parameters, as otherwise
incorrectly documented in the literature of mathematics, statistics, chemistry, physics
and satellite gravimetry. These local solutions should help better understand the
advantages, disadvantages and limitations/problems of linear perturbation methods and the
foundational erroneousness of the numerical integration method; and (iii) to construct
mathematically improved and global uniformly convergent solutions to the governing
differential equations of LEO satellite motion such that they can take full current and
future technological advances of space observation to extract smallest possible
gravitational signals from satellite tracking measurements. As a result, we expect to
produce global high-precision high-resolution gravitational models, which can also be
called the next generation of global gravitational models. From the mathematical point of
view, the accuracy and resolution of the next generation of global gravitational models
can be sufficiently high up to the limit that modern space observation can provide.

The paper is organized as follows. Section 2 will briefly review linear perturbation
methods, with emphasis for the determination of gravitational models. Since the numerical
integration method, though first published by Gronwall (1919) (see also Goddington and
Levinson 1955; Howland and Vaillancourt 1961; Riley et al. 1967), has been widely used by
almost all major institutions worldwide to compute global  gravitational models from
satellite tracking measurements (of CHAMP and GRACE missions), we will first outline the
method and then follow Xu (2009a) to prove that assigning zero initial values to the
partial derivatives of satellite position and velocity with respect to the gravitational
unknown parameters, namely, the harmonic coefficients, is mathematically erroneous and
physically not permitted. We will develop Euler and modified Euler numerical integration
methods to solve differential equations with unknown parameters and unknown initial
conditions in Section~3. As a result, we can represent the linearized local solutions to
the original differential equations in terms of the unknown corrections to approximate
values of the unknown parameters and the unknown initial conditions, which can then be
used to reconstructed from measurements. The content of this section should help
understand correct implementation of numerical integration techniques for gravitational
modelling. In Section~4, by assuming that LEO satellite orbits are precisely measured
with GNSS, we will present a measurement-based perturbation theory, as originally
developed by Xu (2008), which guarantees mathematically global uniform convergence of the
solutions to the Newton's differential equations of satellite motion for satellite orbits
of arbitrary length. Finally, in Section~5, we will propose the method of
measurement-based condition adjustment with unknown parameters to reconstruct global
gravitational models from satellite tracking measurements.

\section{Linear perturbation and the standard-implemented numerical integration method}

\subsection{Linear perturbation methods}
The motion of artificial satellites is governed by Newton's law of gravitation. Almost
all earlier works on satellite gravimetry and celestial mechanics are based on Lagrange's
planetary equations: \alpheqn \beq \label{EqKeplerA} \frac{da}{dt} =
\frac{2}{na}\frac{\partial T}{\partial M}\eeq \beq \label{EqKeplerE} \frac{de}{dt} =
\frac{1-e^2}{na^2e}\frac{\partial T}{\partial M} -
\frac{(1-e^2)^{1/2}}{na^2e}\frac{\partial T}{\partial \omega} \eeq \beq
\label{EqKepleromega} \frac{d\omega}{dt} = -
\frac{\cos\,i}{na^2(1-e^2)^{1/2}\sin\,i}\frac{\partial T}{\partial i} +
\frac{(1-e^2)^{1/2}}{na^2e} \frac{\partial T}{\partial e} \eeq \beq \label{EqKepleri}
\frac{di}{dt} = \frac{\cos\,i}{na^2(1-e^2)^{1/2}\sin\,i}\frac{\partial T}{\partial
\omega} - \frac{1}{na^2(1-e^2)^{1/2}\sin\,i} \frac{\partial T}{\partial \Omega} \eeq \beq
\label{EqKeplerOmega} \frac{d\Omega}{dt} =
\frac{1}{na^2(1-e^2)^{1/2}\sin\,i}\frac{\partial T}{\partial i} \eeq \beq
\label{EqKeplerM} \frac{dM}{dt} = n - \frac{1-e^2}{na^2e} \frac{\partial T}{\partial e} -
\frac{2}{na}\frac{\partial T}{\partial a}
\eeq\reseteqn\setcounter{EQ1}{\value{equation}}(see e.g., Brouwer and Clemence 1961;
Kaula 1966; Hagihara 1972; Taff 1985), where $\mathbf{K}=[a, e, \omega, i, \Omega, M]$
are the six Keplerian orbital elements, which stand for the semi-major axis of the
orbital ellipse, the eccentricity, the argument of the perigee, the inclination of the
orbital plane, the longitude of the ascending node and the mean anomaly, respectively;
$n$ is the mean motion, and $T$ is the disturbing potential (of force or equation
parameters $\mathbf{p}$), which is usually a small quantity. In celestial mechanics of
the solar system, $T$ can come from disturbing planets of extremely small masses when
compared with the solar mass (see e.g., Hagihara 1972); in the case of artificial Earth's
satellites, $T$ can be mainly due to the disturbing potential of the Earth, celestial
bodies of very large distances such as the Sun and the Moon, and/or other disturbing
forces such as the solid earth and ocean tides, the radiation pressure of the Sun and the
air drag of the atmosphere (see e.g., Kaula 1966; Jekeli 1999). Since the general
relativistic dragging, namely, Lense-Thirring effect, has been shown to be of significant
impact on satellite orbits (see e.g., Iorio 2012; Iorio et al. 2011, 2013; Renzetti 2012,
2013), it should be fully taken into account in the future computation of high precision,
high resolution gravitational models from tracking measurements. One may equivalently
rewrite Lagrange's planetary equations (\arabic{EQ1}) in six other orbital elements (see
e.g., Brouwer and Clemence 1961; Taff 1985; Seeber 2003).

Without loss of generality, we denote the general solution to Lagrange's planetary
equations (\arabic{EQ1}) by $\mathbf{K}(t,\mathbf{p},\mathbf{c}_{k})$, where
$\mathbf{c}_{k}$ stands for six arbitrary integration constants. Different integration
constants specify the motions of different satellites. In principle, if the general
solution $\mathbf{K}(t,\mathbf{p},\mathbf{c}_{k})$ would be analytically available, given
the parameters $\mathbf{p}$ and the six integration constants $\mathbf{c}_{k0}$ (or
alternatively an initial point $\mathbf{K}_{t0}$ or any six independent values on
$\mathbf{K}(t,\mathbf{p},\mathbf{c}_{k})$), one can then obtain the particular solution
$\mathbf{K}(t,\mathbf{p},\mathbf{c}_{k0})$ and use it to compute and predict the orbit of
the celestial body at any time $t$. On the other hand, given the general solution
$\mathbf{K}(t,\mathbf{p},\mathbf{c}_{k})$ and a sufficient number of measurements on
$\mathbf{K}(t,\mathbf{p},\mathbf{c}_{k})$, one can then estimate the (unknown) force
parameters $\mathbf{p}$ from the measurements. Unfortunately, the general analytical
solution $\mathbf{K}(t,\mathbf{p},\mathbf{c}_{k})$ can only be obtained for the idealized
two-body problem in which a particle of negligible mass is attracted by another point
mass (see e.g., Brouwer and Clemence 1961; Kaula 1966; Taff 1985; Prussing and Conway
1993). In general, no analytical solution to (\arabic{EQ1}) can be possible, for two
reasons: (i) Lagrange's planetary equations (\arabic{EQ1}) are nonlinear and extremely
difficult to solve analytically. Thus, perturbation theory has been playing a fundamental
role in celestial mechanics and satellite gravimetry (see e.g., Brouwer 1946, 1959; Kozai
1959; Kaula 1966; Hagihara 1972; Cary 1981; Taff 1985), which attempts to construct an
approximate solution to Lagrange's planetary equations (\arabic{EQ1}) through the
procedure of successive approximation. Actually, there are two types of perturbation
methods. One is to construct an approximate solution through the mathematical standard
approach of small parameter perturbation (see e.g., Brouwer 1946, 1959; Hagihara 1972;
Cary 1981; Taff 1985; Nayfeh 2004). The other method is to treat the variables on the
right hand side of Lagrange's planetary equations (\arabic{EQ1}) as constants, except for
the mean anomaly $M$, and then integrate the differential equations (\arabic{EQ1}) to
construct an approximate solution. This latter approach has been widely applied in
satellite geodesy (see e.g., Kozai 1959; Kaula 1961a, 1966); and (ii) the disturbing
potential function $T$ itself may not be exact and/or sufficiently precise. Actually,
perturbation theory, together with astronomical measurements, historically played a
decisive role in correctly identifying an unknown disturbing celestial body to explain
the deviations of the theoretical predictions from measurements by Adams
(https://en.wikipedia.org/wiki/John\_Couch\_Adams), Le Verrier
(https://en.wikipedia.org/wiki/Urbain\_Le\_Verrier) and Lowell (1915) (see also
 https://en.wikipedia.org /wiki/Planets\_beyond\_Neptune), successfully leading to the
great discovery of both Neptune and Pluto in 1846 and 1930, respectively (see e.g.,
Grosser 1964; Hagihara 1972; Lequeux 2013), though the data of Neptune given by Le
Verrier and Adams are in large errors (see e.g., Hubbell and Smith 1992).

As an inverse problem of celestial mechanics, satellite gravimetry is to reconstruct the
unknown force parameters $\mathbf{p}$ in the nonlinear Lagrange's planetary equations
(\arabic{EQ1}) with unknown initial conditions from a sufficient number of measurements
on $\mathbf{K}(t,\mathbf{p},\mathbf{c}_{k})$. In this case, we assume that the disturbing
potential function $T$ itself is precisely given but can contain a number of unknown
parameters $\mathbf{p}$, though part of $T$ may be directly measured and corrected. We
also implicitly assume that there exist no other unknown sources that can contribute to
$T$ in a non-negligible way. In satellite geodesy, the disturbing potential $T$ is mainly
attributed to the rotating Earth, the attraction by the Sun, the Moon and large planets,
the solid earth and ocean tides, the radiation pressure of the Sun, the air drag of the
atmosphere and other conservative and non-conservative forces (see e.g., Kaula 1966;
Reigber 1989; Jekeli 1999; Seeber 2003). From a mathematical point of view, if parts of
$T$ can be directly observable, such effects can be treated as known or given; otherwise,
they are modelled with unknown parameters. In physical geodesy, we usually write the
disturbing potential $T$ of the Earth in the non-inertial earth-fixed reference frame as
follows: \beq \label{TPotential} T = \frac{GM}{r}\sum \limits^{N_{\scriptsize
\textrm{max}}} \limits_{l=2} \sum \limits_{m=0} \limits^l \left( \frac{R}{r}\right)^l[
C_{lm} \cos(m\lambda) + S_{lm} \sin(m\lambda) ] P_{lm}(\cos\theta), \eeq (see e.g.,
Groves 1961; Kaula 1966; Heiskanen and Moritz 1967), where $N_{\scriptsize \textrm{max}}$
is a maximum number of degrees and orders, $R$ is the mean radius of the Earth, $C_{lm}$
and $S_{lm}$ are the unknown normalized, dimensionless harmonic coefficients which will
be collected into the unknown vector $\mathbf{p}$ and to be estimated from satellite
tracking measurements, $\lambda$ and $\theta$ are the longitude and colatitude of the
satellite, respectively, and $P_{lm}(t)$ is the normalized Legendre function. However, in
satellite gravimetry and celestial mechanics, the differential equations of motion of a
satellite are almost always given in the inertial reference frame (see e.g., Kaula 1966;
Taff 1985; Seeber 2003). In this case, we will need the transformation of coordinates
from the earth-fixed reference frame into the inertial reference frame through rotations
(see e.g., Seeber 2003). More specifically, if we use spherical coordinate systems, then
we need the following transformation: \alpheqn \beq \label{lambda2alpha} \lambda = \alpha
+ \delta \alpha - \omega_e t, \eeq \beq \label{theta2zeta} \theta = \zeta + \delta \zeta,
\eeq\reseteqn\setcounter{EQ1A}{\value{equation}}(see e.g., Jekeli 1999), where $\delta
\alpha$ and $\delta \zeta$ are the corrections to $\alpha$ and $\zeta$, which depend on
precession, nutation, the Earth's rotation and polar motion and can be computed from
theory and measurements (see e.g., Seeber 2003), $\alpha$ and $\zeta$ are the right
ascension and co-declination in the inertial reference frame of epoch J2000.0, $\omega_e$
is the rate of the Earth's rotation. The disturbing potential $T$ in the inertial
reference frame is clearly a function of time.

To rigorously determine mathematically the unknown force parameters $C_{lm}$ and $S_{lm}$
from satellite tracking measurements, we have to first exactly solve Lagrange's planetary
equations (\arabic{EQ1}), with the disturbing potential $T$ given by (\ref{TPotential}),
link the exact solution to the satellite tracking measurements and finally estimate
$C_{lm}$ and $S_{lm}$. Unfortunately, the nonlinear differential equations (\arabic{EQ1})
are too complicated to exactly solve analytically. Thus, perturbation theory is always
used to construct an approximate solution to (\arabic{EQ1}) (see e.g., Brouwer 1959;
Kozai 1959; Groves 1960; Kaula 1961, 1966; Hagihara 1972; Taff 1985). The most complete
perturbation theory was fully developed for the determination of $C_{lm}$ and $S_{lm}$
from satellite tracking measurements by Kaula (1961, 1966), following the approach of
Kozai (1959) and given the representation of $T$ expressed in terms of the six Keplerian
orbital elements by Groves (1960). It is nowadays well known as {\em Kaula linear
perturbation theory}. More specifically, since the exact analytical solution is generally
hard or almost impossible to obtain, Kaula (1961, 1966) derived the linear perturbation
solution to Lagrange's planetary equations (\arabic{EQ1}) by treating all the orbital
elements as constants and/or by replacing them with the mean orbital elements, except for
the rapidly time-varying element of the mean anomaly $M$, and then integrating all the
terms on the right hand side of (\arabic{EQ1}).

The major advantage of Kaula's linear perturbation solution is its suitability to analyze
the physical properties of the solution. Since $T$ is expressed in terms of six Keplerian
orbital elements, the physical features of $T$ can be further classified into secular,
long-periodic and short-periodic (see e.g., Kozai 1959; Groves 1960; Kaula 1961, 1966).
More precisely, the terms as a function of the mean anomaly $M$ will change periodically
(and rapidly) and are {\em short-periodic}; the terms as a function of $\omega$ but not
$M$ are {\em long-periodic}; and the terms irrelevant of $\omega$ nor $M$ are {\em
secular}. In other words, secular terms changes slowly but approximately linearly with
time. In addition, one can also identify, in the linear perturbation solution, the
physically interesting phenomenon of orbital mean-motion resonance when the rotation rate
of the Earth and the mean motion of a satellite are commensurable (see e.g., Cook 1961;
Yionoulis 1965; Kaula 1966; Hagihara 1972; Blitzer and Anderson 1981; Taff 1985;
Kloko\v{c}n\'{i}k et al. 2013). Nevertheless, Kaula's linear perturbation solution is
only valid locally around the neighbourhood of the mean orbital elements and will diverge
with the increase of time. Thus, Kaula's linear perturbation solution will not be able to
fully utilize a long orbital arc and high precision of modern space observation to
estimate $\mathbf{p}$.

\subsection{The standard-implemented numerical integration method}\label{SectNumMeth}
In the Cartesian coordinate system, the motion of an artificial satellite can also be
mathematically written alternatively by the following nonlinear vector differential
equations: \beq \label{NewtonLAW} \ddot{\mathbf{x}} = \mathbf{a}_E(t, \mathbf{x},
\dot{\mathbf{x}}, \mathbf{p}) + \mathbf{a}_M(t, \mathbf{x}, \dot{\mathbf{x}},
\mathbf{p}_M), \eeq (see e.g., Brouwer \& Clemence 1961; Kaula 1966; Taff 1985; Seeber
2003), where $\mathbf{x}$ is the position vector of the satellite in the inertial
reference frame, $\mathbf{a}_E(t, \mathbf{x}, \dot{\mathbf{x}}, \mathbf{p})$ is the
Earth's gravitational attraction exerted on the satellite, and $\mathbf{a}_M(t,
\mathbf{x}, \dot{\mathbf{x}}, \mathbf{p}_M)$ stands for all other forces which may
include solid earth and ocean tides, atmospherical drag, solar radiation pressure and
third-body effects (see e.g., Taff 1985; Reigber 1989; Jekeli 1999; Seeber 2003), with
$\mathbf{p}_M$ standing for the unknown parameters (if any) of these force models. For
the next generation of high precision, high resolution global  gravitational models from
satellite tracking, the force model $\mathbf{a}_M(t, \mathbf{x}, \dot{\mathbf{x}},
\mathbf{p}_M)$ should include the general relativistic dragging, namely, the
Lense-Thirring effect, which has been shown to have a significant effect on satellite
orbits, satellite-to-satellite ranges and range-rates for a sufficiently lengthy arc (see
e.g., Iorio 2012; Iorio et al. 2011, 2013; Renzetti 2012, 2013) but has not been
considered up to the present in the production of global  gravitational models.

In the remainder of this paper, we will use $\ddot{\mathbf{x}}$ and $\dot{\mathbf{x}}$ to
stand for the first and second derivatives of $\mathbf{x}$ with time, respectively. (We
use the notation $d\mathbf{x}/dt$ to stand for time derivative in (\arabic{EQ1}), since
the dot of $i$ there does not look good) Since the acceleration $\mathbf{a}_E(t,
\mathbf{x}, \dot{\mathbf{x}}, \mathbf{p})$ is independent of $\dot{\mathbf{x}}$, it can
be rewritten as follows: \beq \label{Two_dotX} \mathbf{a}_E(t, \mathbf{x}, \mathbf{p}) =
- \frac{GM}{r^3}\mathbf{x} + \frac{\partial T}{\partial \mathbf{x}}, \eeq where $GM$ is
the product of the Earth's mass $M$ and the gravitational constant $G$, $r=\|\mathbf{x}
\|$, and $T$ is the disturbing potential of the Earth's gravitational field with the
parameters of harmonic coefficients. Because (\ref{NewtonLAW}) is formulated in an
inertial reference system (see e.g., Taff 1985; Jekeli 1999), the earth-fixed $\lambda$
and $\theta$ in the disturbing potential $T$ of (\ref{TPotential}) must first be
transformed through (\arabic{EQ1A}) into the inertial reference frame (see e.g., Jekeli
1999). Since the second term $\mathbf{a}_M(t, \mathbf{x}, \dot{\mathbf{x}},
\mathbf{p}_M)$ of (\ref{NewtonLAW}) adds no new mathematical difficulty to solve the
nonlinear differential equations (\ref{NewtonLAW}), we will limit ourselves to the
Earth's gravitational field in the remainder of this paper. Thus, the nonlinear
differential equations (\ref{NewtonLAW}) can be simplified as follows: \beq
\label{EarthNewtonLAW} \ddot{\mathbf{x}} = \mathbf{a}_E(t, \mathbf{x}, \mathbf{p}), \eeq
where $\mathbf{p}$ is an unknown vector of equation parameters. Initial conditions to
(\ref{EarthNewtonLAW}) are unknown as well.

We should note, however, that if $\mathbf{a}_M(t, \mathbf{x}, \dot{\mathbf{x}},
\mathbf{p}_M)$ exists but is neither estimated together with $\mathbf{p}$ nor corrected
with a sufficiently precise model, the effect of $\mathbf{a}_M(t, \mathbf{x},
\dot{\mathbf{x}}, \mathbf{p}_M)$ will be absorbed into the estimate of $\mathbf{p}$. Such
an effect is theoretically systematic and of signal nature. It cannot be filtered out,
since, in principle, a filter can only reduce the level of noise but would mess up or
smear the true signal of concern, if the signals are not constant inside the window of
the filter. In other words, filtering, as currently used in satellite gravimetry, will
definitely distort the true signal of interest, because gravity signals are clearly not
constant inside the window of filtering. To test the Lense-Thirring dragging from
satellite measurements, for example, one should probably have to either estimate the
effect together with the gravitational model or to develop statistical hypothesis testing
methods for weak continuous functions; the latter is of theoretical interest by itself
and deserves a separate full research.

Given a number of (geometrical) tracking measurements to the satellite such as positions
of the satellite, ranges, range rates and directions to the satellite, denoted by $y_1,
y_2, ...,y_n$ or in the vector form $\mathbf{y}$, the problem of satellite gravimetry is
to use the tracking measurements $\mathbf{y}$ to determine the gravitational parameters
$\mathbf{p}$. Mathematically, this is essentially the problem of estimating the unknown
parameters $\mathbf{p}$ of the differential equations (\ref{EarthNewtonLAW}) with unknown
initial conditions from satellite tracking measurements. If there are a number of LEO
satellites, say $s$ satellites, the motion of each satellite being governed by the same
differential equations of type (\ref{EarthNewtonLAW}) with the same gravitational
parameters $\mathbf{p}$ but with different initial conditions or different integration
constants. If we collect satellite tracking measurements $\mathbf{y}_i$ on the {\em i}th
satellite, then we will have to combine all these measurements
$\mathbf{y}_1,\mathbf{y}_2,...,\mathbf{y}_s$ together to solve for the parameters
$\mathbf{p}$. In the following development of the method, without loss of generality, we
will confine ourselves to one satellite.

To determine $\mathbf{p}$ from $\mathbf{y}$, one of the most important steps is to
represent each $y_i$ in terms of $\mathbf{p}$. Since a geometrical satellite tracking
measurement $y_i$ is generally a function of the position and velocity of the satellite
at the {\em i}th epoch, in principle, we have to first solve the differential equations
(\ref{EarthNewtonLAW}). Let $\mathbf{x}(t,\mathbf{p}, \mathbf{c})$ denote the general
solution to the nonlinear differential equations (\ref{EarthNewtonLAW}), with
$\mathbf{c}$ standing for six arbitrary integration constants. These six integration
constants $\mathbf{c}$ are mathematically independent of the equation parameters
$\mathbf{p}$. In other words, the general solution $\mathbf{x}(t,\mathbf{p}, \mathbf{c})$
mathematically represents an infinite number of solutions to the differential equations
(\ref{EarthNewtonLAW}), which can physically describe the motions of different
satellites. As far as initial values are properly given, one can then use them to fix six
arbitrary integration constants $\mathbf{c}$ and obtain the specific solution to uniquely
describe the motion of the satellite. Mathematically, the vector $\mathbf{c}$ for this
particular solution can now be expressed as the functions of $\mathbf{p}$ and the initial
conditions. In satellite geodesy, initial values are often three initial position
coordinates $\mathbf{x}_0$ and three initial velocity components $\mathbf{v}_0$ of the
satellite at the initial epoch $t_0$. As a result, the orbital position solution of
motion of the satellite can be implicitly written as $\mathbf{x}(t,\mathbf{p},
\mathbf{c}(\mathbf{p}, \mathbf{x}_0, \mathbf{v}_0))$. We emphasize that six arbitrary
integration constants $\mathbf{c}$ can also be alternatively determined from any six
independent values on the solution, instead of the initial position and velocity
conditions, since any specific value on the particular solution contains the information
on $\mathbf{c}$. Except for the idealized two-body problem with the point mass model, it
is almost impossible to obtain an analytical solution to the nonlinear differential
equations (\ref{EarthNewtonLAW}) with the initial conditions $\mathbf{x}_0$ and
$\mathbf{v}_0$. Thus, we can only use the implicit orbit $\mathbf{x}(t_i,\mathbf{p},
\mathbf{c}(\mathbf{p}, \mathbf{x}_0, \mathbf{v}_0))$ to develop observational equations
for geometrical tracking measurements of any kind. Taking a (velocity-independent)
geometrical tracking measurement $y_i$ as an example, we can symbolically write its
observational equation as follows: \beq \label{YiOBSEquationOriginal} y_i =
f(\mathbf{x}(t_i,\mathbf{p}, \mathbf{c}(\mathbf{p}, \mathbf{x}_0, \mathbf{v}_0))) +
\epsilon_i, \eeq where $f(\cdot)$ stands for a nonlinear functional and $\epsilon_i$ is
the random error of the measurement $y_i$.

Under the framework of the numerical integration method, as described in Lerch et al.
(1974), Long et al. (1989) and Reigber (1989) for satellite gravimetry, we will have to
first linearize the observational equation (\ref{YiOBSEquationOriginal}). Given a set of
approximate values $\mathbf{p}^0$, $\mathbf{x}_0^0$ and $\mathbf{v}_0^0$ for
$\mathbf{p}$, $\mathbf{x}_0$ and $\mathbf{v}_0$, respectively, one can then numerically
integrate the differential equations (\ref{EarthNewtonLAW}) and obtain the approximate
position of the satellite at time epoch $t_i$, which is denoted by $\mathbf{x}^0_i$.
Thus, the observational equation (\ref{YiOBSEquationOriginal}) can be {\em formally}
linearized as follows: \alpheqn \beq \label{YiOBSEquation} \delta y_i = \mathbf{a}_{ix}
\Delta \mathbf{x}_0 + \mathbf{a}_{iv} \Delta \mathbf{v}_0 + \mathbf{a}_{ip} \Delta
\mathbf{p} + \epsilon_i \eeq at the approximate values of $\mathbf{p}^0$,
$\mathbf{x}_0^0$ and $\mathbf{v}_0^0$, where
$$ \delta y_i = y_i - f(\mathbf{x}^0_i), $$
 $$ \Delta \mathbf{x}_0 = \mathbf{x}_0 - \mathbf{x}_0^0, $$
$$ \Delta \mathbf{v}_0 = \mathbf{v}_0 - \mathbf{v}_0^0, $$ and
$$ \Delta \mathbf{p} = \mathbf{p} - \mathbf{p}^0. $$ The three row vectors
$\mathbf{a}_{ix}$,  $\mathbf{a}_{iv}$ and $\mathbf{a}_{ip}$ are all computed at the
approximate values $\mathbf{p}^0$, $\mathbf{x}_0^0$ and $\mathbf{v}_0^0$, and defined,
respectively,  as follows: \begin{eqnarray} \label{PartialX} \mathbf{a}_{ix} & = &
\frac{\partial f(\mathbf{x}(t_i,\mathbf{p}, \mathbf{c}(\mathbf{p}, \mathbf{x}_0,
\mathbf{v}_0)))}{\partial \mathbf{x}_0^T} \nonumber
\\ & = & \frac{\partial f(\mathbf{x}(t_i,\mathbf{p}, \mathbf{c}(\mathbf{p}, \mathbf{x}_0,
\mathbf{v}_0)))}{\partial \mathbf{x}^T} \frac{\partial \mathbf{x}}{\partial
\mathbf{x}_0^T}, \end{eqnarray} \begin{eqnarray} \label{PartialV} \mathbf{a}_{iv} & = &
\frac{\partial f(\mathbf{x}(t_i,\mathbf{p}, \mathbf{c}(\mathbf{p}, \mathbf{x}_0,
\mathbf{v}_0)))}{\partial \mathbf{v}_0^T} \nonumber \\
& = & \frac{\partial f(\mathbf{x}(t_i,\mathbf{p}, \mathbf{c}(\mathbf{p}, \mathbf{x}_0,
\mathbf{v}_0)))}{\partial \mathbf{x}^T} \frac{\partial \mathbf{x}}{\partial
\mathbf{v}_0^T}, \end{eqnarray} \begin{eqnarray} \label{PartialP} \mathbf{a}_{ip} & = &
\frac{\partial f(\mathbf{x}(t_i,\mathbf{p}, \mathbf{c}(\mathbf{p}, \mathbf{x}_0,
\mathbf{v}_0)))}{\partial \mathbf{p}^T} \nonumber \\
 & = & \frac{\partial f(\mathbf{x}(t_i,\mathbf{p}, \mathbf{c}(\mathbf{p}, \mathbf{x}_0,
\mathbf{v}_0)))}{\partial \mathbf{x}^T} \frac{\partial \mathbf{x}}{\partial
\mathbf{p}^T}.
\end{eqnarray}\reseteqn\setcounter{EQ2}{\value{equation}}

As a key step to estimate $\mathbf{p}$ from $\mathbf{y}$, we have to compute the vectors
$\mathbf{a}_{ix}$,  $\mathbf{a}_{iv}$ and $\mathbf{a}_{ip}$. The common matrix of the
partial derivatives $\partial f(\cdot)/\partial \mathbf{x}^T$ in (\ref{PartialX}) to
(\ref{PartialP}) can be readily obtained, as widely available (see e.g., Kaula 1961,
1966; Lerch et al. 1974; Long et al. 1989; Tapley et al. 2004b). If we would treat
$\mathbf{p}$ as if it were given for now, the problem is turned into a standard problem
of statistical orbit determination with the differential equations
(\ref{EarthNewtonLAW}), the initial conditions $\mathbf{x}_0$ and $\mathbf{v}_0$, and the
measurements $\mathbf{y}$. In this case, computing the matrices of the partial
derivatives $\partial \mathbf{x}/\partial \mathbf{x}^T_0$ and $\partial
\mathbf{x}/\partial \mathbf{v}^T_0$ is theoretically equivalent to finding the state
transition matrix for the state of position and velocity from the initial epoch $t_0$ to
the current epoch $t_i$. This problem has been completely solved and well documented in,
for example, Long et al. (1989), Tapley (1989), Tapley et al. (2004b) and  Gunter et al.
(2006).

Now, to complete the final construction of the observational equation
(\ref{YiOBSEquation}), the key issue is to compute the partial derivatives of
$\mathbf{x}(t_i,\mathbf{p}, \mathbf{c}(\mathbf{p}, \mathbf{x}_0, \mathbf{v}_0))$ with
respect to $\mathbf{p}$ in (\ref{PartialP}), namely, $\partial \mathbf{x}/\partial
\mathbf{p}^T =\partial \mathbf{x}(t_i,\mathbf{p}, \mathbf{c}(\mathbf{p}, \mathbf{x}_0,
\mathbf{v}_0))/\partial \mathbf{p}^T$ for conciseness of notations. Since we do not have
an analytical solution $\mathbf{x}(t_i,\mathbf{p}, \mathbf{c}(\mathbf{p}, \mathbf{x}_0,
\mathbf{v}_0))$, it is not possible to directly compute its partial derivatives with
respect to the parameters $\mathbf{p}$. Instead, one has attempted to obtain these
partial derivatives through solving their differential equations.

To start with, let us collect the satellite position $\mathbf{x}$ and velocity
$\mathbf{v}$ at time $t$ in the vector $\mathbf{z}$ and denote the partial derivatives of
$\mathbf{z}$ with respect to $\mathbf{p}$ by $\mathbf{S}(t, \mathbf{p})$, namely,
\alpheqn \beq \label{VectorZ} \mathbf{z} = ( \mathbf{x}^T, \mathbf{v}^T)^T, \eeq \beq
\label{MatrixS} \mathbf{S}(t,\mathbf{p}) = \frac{\partial \mathbf{z}}{\partial
\mathbf{p}^T}.\eeq\reseteqn\setcounter{EQ3}{\value{equation}}The partial derivatives
$\partial \mathbf{x}/\partial \mathbf{p}^T$ is obviously part of a more general matrix
$\mathbf{S}(t,\mathbf{p})$ of partial derivatives. It has been rigorously shown
mathematically that $\mathbf{S}(t,\mathbf{p})$ can be directly derived from the original
differential equations (\ref{EarthNewtonLAW}) (see, e.g., Riley et al. 1967; Ballani
1988; Montenbruck \& Gill 2000) and is governed by the following system of differential
equations:  \beq \label{PartialCS} \frac{\partial \mathbf{S}(t,\mathbf{p})}{\partial t} =
\left(
\begin{array}{cc} \mathbf{0} & \mathbf{I}
\\  \partial \mathbf{a}_E(t, \mathbf{x}, \mathbf{p})/\partial
\mathbf{x}^T &  \mathbf{0} \end{array} \right) \mathbf{S}(t) + \left(
\begin{array}{c} \mathbf{0} \\ \partial \mathbf{a}_E(t, \mathbf{x},
\mathbf{p})/\partial \mathbf{p}^T \end{array} \right). \eeq Obviously, the derived
equations (\ref{PartialCS}) of the partial derivatives do not mathematically add any new
information on the original problem of satellite gravimetry. Actually, for a general
differential equation with unknown parameters, the differential equations of type
(\ref{PartialCS}) were already given by Gronwall (1919) and Ritt (1919) almost 100 years
ago (see also Goddington and Levinson 1955).

Although we have the differential equations (\ref{PartialCS}) for the partial derivatives
$\mathbf{S}(t,\mathbf{p})$, they can still be useless, unless {\em the} initial
conditions of $\mathbf{S}(t,\mathbf{p})$ at the time epoch $t_0$ are available. We use
the italic font for ``the'' before ``initial conditions'' to emphasize that the initial
conditions of $\mathbf{S}(t,\mathbf{p})$ cannot be arbitrarily given but must comply with
the original problem. Unfortunately, the original problem of satellite gravimetry,
namely, the governing differential equations (\ref{EarthNewtonLAW}) and the satellite
tracking measurements $\mathbf{y}$, do not provide any direct hint/clue on what values
$\mathbf{S}(t_0,\mathbf{p})$ can take on.

With no way out, the claims on $\mathbf{S}(t_0,\mathbf{p})$ were made and accepted in
satellite geodesy, as cited verbatim from some of the publications in the introduction.
Riley et al. (1967) claimed that $\mathbf{S}(t_0,\mathbf{p})$ is generally zero. Lerch et
al. (1974) treated $\mathbf{S}(t_0,\mathbf{p})$ as zero for the software system GEODYN
(see also Beutler et al. 2010). Long et al. (1989) mentioned the importance of the
initial values $\mathbf{S}(t_0,\mathbf{p})$ for solving the differential equations
(\ref{PartialCS}) but without saying what they should be. Reigber (1989) referred the
reader to the earlier version of the report by Long et al. (1989). Others simply avoid
mentioning the initial conditions for $\mathbf{S}(t,\mathbf{p})$ (see e.g., Rowlands et
al. 2002; Gunter et al. 2006). In the next subsection, we will use a counter example in
Xu (2009a) to prove that setting the  initial values $\mathbf{S}(t_0,\mathbf{p})$ to zero
is mathematically incorrect and physically not permitted. It is unbelievable to see that
the numerical integration method ends up on mathematically vain ground without rhyme or
reason, though it has been widely used by almost all major institutions worldwide to
produce global  gravitational models from satellite tracking measurements, with widest
possible applications in Earth Sciences.

\subsection{No zero initial values for $\mathbf{S}(t_0,\mathbf{p})$ permitted
mathematically and physically} In this part of the paper, we will prove that no zero
initial values for $\mathbf{S}(t_0,\mathbf{p})$ can be permitted from both the
mathematical and physical points of view. We will use a counter example reported in Xu
(2009a, 2015) for this purpose, though one can readily construct many other counter
examples. Then we will use strictly logical reasonings to explain why
$\mathbf{S}(t_0,\mathbf{p})$ cannot be zero. For some of the arguments, the reader is
referred to Xu (2009a) for details.

As a general rule in mathematics, we need nothing more than a counter example to disprove
that something is incorrect. Let us start with the second counter example in Xu (2009a),
which is rewritten as follows: \beq \label{CounterEX} \ddot{y} + p_1^2 y - p_2\cos(p_1t)
= 0, \eeq where $p_1$ and $p_2$ are two equation (unknown) parameters. By directly
solving the differential equation (\ref{CounterEX}), we obtain the general solution: \beq
\label{GSolutionY} y(t) = \frac{p_2}{2p_1^2}\{\cos(p_1t)+p_1t\sin(p_1t)\} + c_1
\sin(p_1t) + \frac{c_2}{p_1}\cos(p_1t), \eeq where $c_1$ and $c_2$ are two arbitrary
integration constants. Mathematically, integration constants $c_1$ and $c_2$ are
independent of $p_1$ and $p_2$. As far as $c_1$ and $c_2$ are given specific values,
which can be implicitly defined, for example, through assuming two values of $y(t)$ at
two different time epochs, we will then obtain the particular solution of
(\ref{GSolutionY}). With the true general solution (\ref{GSolutionY}), we can easily
compute and obtain the true values of the derivatives of $y(t)$ with respect to $p_1$ and
$p_2$, which are simply given as follows: \alpheqn
\begin{eqnarray} \label{Partial_E2p1} \frac{\partial y(t)}{\partial p_1} & = &
- \frac{p_2}{p_1^3}\{ \cos(p_1t) + p_1t\sin(p_1t) \} +
\frac{p_2}{2p_1}t^2\cos(p_1t) \nonumber \\
&   &  + c_1t\cos(p_1t) - \frac{c_2}{p_1^2}\cos(p_1t) - \frac{c_2t}{p_1}\sin(p_1t),
\end{eqnarray} \beq \label{Partial_E2p2} \frac{\partial y(t)}{\partial p_2} =
\frac{1}{2p_1^2} \{ \cos(p_1t) + p_1t\sin(p_1t) \}. \eeq
\reseteqn\setcounter{EQ4}{\value{equation}}

For an arbitrary $t_0$, the derivatives $\partial y(t)/\partial p_1$ of
(\ref{Partial_E2p1}) and $\partial y(t)/\partial p_2$ of (\ref{Partial_E2p2}) clearly
cannot be zero. Actually, if the derivatives (\ref{Partial_E2p1}) and
(\ref{Partial_E2p2}) would be equal to zero at the time epoch $t_0$, we would readily
have two equations for two unknowns $p_1$ and $p_2$ and would be able to solve them
without any value of $y(t)$. For example, if the initial values for these derivatives
would be allowed to be equal to zero, then the second derivative (\ref{Partial_E2p2})
would be turned into the following equation: \beq \label{EquationP1} \cos(p_1t_0) +
p_1t_0\sin(p_1t_0) = 0, \eeq for any non-zero $p_1$. By solving this equation, we could
obtain the value(s) of $p_1$ (if solutions exist). Obviously, this is logically
ridiculous, since this would indicate that we would be able to determine the unknown
parameter $p_1$ in the differential equation (\ref{CounterEX}) without any information on
$y(t)$. In the case of $t_0=0$, (\ref{EquationP1}) becomes $1 = 0$ --- an even more
ridiculous expression. The source of errors is clearly with the assumption of setting the
values of partial derivatives (\ref{Partial_E2p1}) and (\ref{Partial_E2p2}) to zero at
the initial epoch $t_0$.

For the satellite gravimetry problem (\ref{EarthNewtonLAW}) with the geometrical
satellite tracking measurements $\mathbf{y}$, we will simply make some logical reasonings
and explanations. The mathematical proof and physical explanations of no zero initial
values for $\mathbf{S}(t_0,\mathbf{p})$ can be found in Xu (2009a).

{\em Remark 1:} For the general orbit solution $\mathbf{x}(t,\mathbf{p}, \mathbf{c})$ of
(\ref{EarthNewtonLAW}) without given initial conditions, the problem of satellite
gravimetry is to determine both $\mathbf{p}$ and $\mathbf{c}$ from the geometrical
satellite tracking measurements $\mathbf{y}$. The orbital position at the time epoch
$t_i$ can be mathematically written symbolically as $\mathbf{x}(t_i,\mathbf{p},
\mathbf{c})$. $\mathbf{x}(t_i,\mathbf{p}, \mathbf{c})$ is only a point of the general
solution $\mathbf{x}(t,\mathbf{p}, \mathbf{c})$ of the satellite at the time epoch $t_i$.
Logically, no orbital position at one time epoch is superior to any other orbital
positions of the same orbit but at different time epochs. Since $t_0$ is arbitrary, any
orbital position can equally serve as an initial condition. If
$\mathbf{S}(t_0,\mathbf{p})$ could be set to zero, then all the other
$\mathbf{S}(t_i,\mathbf{p}) \, (t_i\neq t_0)$ could be treated in the same manner as zero
from the mathematical point of view, implying physically that $\mathbf{x}(t,\mathbf{p},
\mathbf{c})$ would not be a function of $\mathbf{p}$. Obviously, this conclusion violates
our starting differential equations (\ref{EarthNewtonLAW}) with $\mathbf{p}$. Thus,
$\mathbf{S}(t_0,\mathbf{p})$ cannot be equal to zero.

{\em Remark 2:} According to Xu (2015a, 2015b), for the particular orbit solution
$\mathbf{x}(t,\mathbf{p}, \mathbf{c}_{x0})$ of (\ref{EarthNewtonLAW}), with the
integration constants $\mathbf{c}_{x0}$ fixed/given, for example, from two orbital
positions at two different epochs, the partial derivatives $\mathbf{S}(t,\mathbf{p})$ at
any time epoch $t$ must be unique. If $\mathbf{S}(t_0,\mathbf{p})=\mathbf{0}$, then we
can compute $\mathbf{S}(t,\mathbf{p})$ by solving the differential equations
(\ref{PartialCS}). To emphasize the starting time $t_0$, we denote the values of
$\mathbf{S}(t,\mathbf{p})$ by $\mathbf{S}(t,\mathbf{p}, t_0)$. Now let us assume a
different starting time epoch, say $t_{01}\neq t_0$. Since the initial values for
$\mathbf{S}(t,\mathbf{p})$ are assumed to be zero, we should have
$\mathbf{S}(t_{01},\mathbf{p})=\mathbf{0}$ and obtain its corresponding values of the
partial derivatives $\mathbf{S}(t,\mathbf{p}, t_{01})$. Following the same logic, let us
assume another starting time epoch $t_{0i}$, which can be arbitrarily different from
either $t_{01}$ or $t_0$. By the claim of Riley et al. (1967) (see also Lerch et al.
1974; Long et al. 1989), we have $\mathbf{S}(t_{0i},\mathbf{p})=\mathbf{0}$, with which
we can further obtain $\mathbf{S}(t,\mathbf{p}, t_{0i})$ by solving the differential
equations (\ref{PartialCS}). For three arbitrarily different time epochs $t_0$, $t_{01}$
and $t_{0i}$, their corresponding partial derivatives at the same time epoch t, namely,
$\mathbf{S}(t,\mathbf{p}, t_0)$, $\mathbf{S}(t,\mathbf{p}, t_{01})$ and
$\mathbf{S}(t,\mathbf{p}, t_{0i})$, will not be equal to each other. This obviously
contradicts the fact that $\mathbf{S}(t,\mathbf{p})$ is unique for this particular orbit.
The source of errors again certainly comes from the incorrect assumption of zero initial
values for $\mathbf{S}(t,\mathbf{p})$.

{\em Remark 3:} If $\mathbf{S}(t,\mathbf{p})$ could be set to zero at the initial epoch
$t_0$, by following the same logical reasonings as in (\arabic{EQ4}) and/or
(\ref{EquationP1}), we would be able to solve for $\mathbf{p}$ from the system of
equations $\mathbf{S}(t_0,\mathbf{p})=\mathbf{0}$, since the number of equations
$\mathbf{S}(t_0,\mathbf{p})=\mathbf{0}$ is exactly equal to that of the unknown
parameters $\mathbf{p}$, implying that we could determine the unknown harmonic
coefficients $\mathbf{p}$ without any satellite tracking measurements; this is again an
unacceptable result. Actually, on the other hand, if
$\mathbf{S}(t_0,\mathbf{p})=\mathbf{0}$, then we could solve the differential equations
(\ref{PartialCS}) and obtain {\em the} orbital solution, denoted by
$\mathbf{x}_{t0}(t,\mathbf{p}, \mathbf{c}(\mathbf{p}, \mathbf{x}_0, \mathbf{v}_0))$,
which implies that we do not need any initial conditions $\mathbf{x}_0$ and
$\mathbf{v}_0$ to find the particular orbital solution. Since $t_0$ is arbitrary, we
could obtain an infinite number of different solutions for the same satellite gravimetry
problem. All these are certainly incorrect mathematically, again with the source of
errors in the assumption of $\mathbf{S}(t_0,\mathbf{p})=\mathbf{0}$.

The above counter example, together with all the mathematical, physical and logical
reasonings, has all clearly nullified the claim by Riley {\em et al.} (1967) that the
initial values of the partial derivatives with respect to equation parameters are
generally zero. Actually, this claim is also used as a starting point for the Goddard
Space Flight Center software system GEODYN (see e.g., Lerch et al. 1974; Long et al.
1989) and in Europe (see e.g., Reigber 1989) for the production of global gravitational
models from satellite tracking measurements. Bearing in mind that a variety of global
 gravitational model products has been widely used in and far more beyond geodesy
such as solid geophysics, hydrology, continental water variation and so on, we believe
that software systems must be first updated onto a solid mathematical foundation right
now before continuing to produce and circulate such gravitational products from satellite
tracking measurements.
Finally, we state a theorem in Xu (2009a) to conclude this subsection as follows: \vspace{1mm} \\
{\bf Theorem 1:} {\em Given the governing vector differential equations}
(\ref{EarthNewtonLAW}) {\em with the unknown harmonic coefficients} $\mathbf{p}$ {\em in}
(\ref{TPotential}), {\em then setting the initial values of the partial derivatives of
the orbit and velocity} {\em with respect to the unknown harmonic coefficients}
$\mathbf{p}$ {\em to zero at any specified initial time epoch} $t_0$ {\em is not
permitted, mathematically and physically}.

Before closing this section, I should point out that zero initial partial derivatives
with respect to the parameters of differential equations has been routinely used beyond
geodesy. Completely independent of the development in satellite geodesy, for simplicity
but without loss of generality, given an ordinary differential equation $\dot{y} =
f(t,y,p)$, Gronwall (1919) correctly derived the differential equation of $y$ with
respect to the unknown equation parameter $p$, as in the case of (\ref{PartialCS}) (see
also Ritt 1919), but incorrectly claimed that its initial value is equal to zero without
providing any reasons or arguments. The work of Gronwall (1919) was then further spread
through the book on ordinary differential equations by Goddington and Levinson (1955).
Actually, the solution to the given ordinary differential equation can be symbolically
written as follows:
 \beq \label{GoddingtonL1955} y(t,p) = y(t_0,p) + \int_{t_0}^tf(t,y,p) dt, \eeq from which we
 can only obtain the following identity:
 \beq \label{GoddingtonL1955Derivative} \left. \frac{dy(t,p)}{dp}\right|_{t=t_0} =
 \frac{dy(t_0,p)}{dp}, \eeq but certainly not the zero initial derivative,
as we have proved in this paper. Since all the geodetic literature on satellite
gravimetry has not cited or mentioned any of these mathematical publications, it seems
that the claim of zero initial partial derivatives with respect to the equation
parameters has been taken for granted everywhere for almost 100 years, though
incorrectly, as we have proved in this paper and Xu (2009). The incorrect claim of zero
initial derivatives now still continues to spread, as can be seen, for example, in
Howland and Vaillancourt (1961), Dickinson et al. (1976), Hwang et al. (1978), Linga et
al. (2006), Ramsay et al. (2007) and Wang and Enright (2013).

Instead of solving a differential equation and using the solution to estimate the
equation parameters, researchers have also chosen to use splines and/or basis function
expansion to approximately represent the solution to the differential equation and then
to estimate the equation parameters (see e.g., Ramsay et al. 2007; Liang and Wu 2008).
One may either first estimate the coefficients of the fitting basis functions and then
further use the fitted solution to estimate the differential equation parameters or
choose to simultaneously estimate both the unknown basis function coefficients and the
differential equation parameters. Nevertheless, the disadvantages of the basis function
approach with a finite number of unknown coefficients could be threefold: (i) it
generally does not satisfy the original differential equation; (ii) we need to estimate
many more extra unknown coefficients of basis functions. If we simultaneously estimate
both the unknown basis function coefficients and the differential equation parameters
from measurements, this new estimation can generally be nonlinear. Even worse, the total
number of the unknown coefficients of basis functions and the differential equation
parameters may become larger than the number of measurements such that the new estimation
problem becomes rank-deficient, though the number of measurements can be far more than
sufficient to estimate the (original) differential equation parameters; and (iii) it will
create modelling errors as a consequence of (i), whose extent would depend on the
difference between the approximate solution as a finite series of basis functions and the
(true) solution to the original differential equation. If the modelling errors are larger
than the noise level of measurements, it would become impossible to extract maximum
information on the equation parameters from measurements at the level of random
measurement errors.

\section{Linearization and numerical integration techniques for estimation of unknown
differential equation parameters from measurements}\label{AppromAnalNumINT} If
$\mathbf{S}(t_0,\mathbf{p})\neq\mathbf{0}$ and is unknown, then the differential
equations (\ref{PartialCS}) are not useful. As a consequence, we are not able to compute
$\mathbf{a}_{ip}$ of (\ref{PartialP}) to complete the construction of the observational
equation (\ref{YiOBSEquation}). The question now is how we can properly implement
numerical integration techniques to determine the Earth's gravitational field from
satellite tracking measurements. In principle, given the satellite tracking measurements
$\mathbf{y}$ with a corresponding weighting matrix $\mathbf{W}$, we can write the least
squares objective function as follows: \beq \label{1stLSProb} \textrm{min:} \,\,
[\mathbf{y}-\mathbf{f}(\mathbf{x}(t_{yi},\mathbf{p},\mathbf{c}))]^T\mathbf{W}
[\mathbf{y}-\mathbf{f}(\mathbf{x}(t_{yi},\mathbf{p},\mathbf{c}))] \eeq subject to the
equality constraint defined by the differential equations (\ref{EarthNewtonLAW}), where
$\mathbf{f}(\cdot)$ are the theoretical values of the measurements $\mathbf{y}$, and each
$\mathbf{x}(t_{yi},\mathbf{p},\mathbf{c})$ satisfies (\ref{EarthNewtonLAW}) and stands
for the theoretical orbital position of the satellite at the time epoch $t_{yi}$ when the
tracking measurement $y_i$ is collected. If initial conditions are available, the
constants $\mathbf{c}$ can be alternatively expressed in terms of these initial
conditions.

Since the equality constraints are given in the form of differential equations, we cannot
use conventional optimization methods to solve the minimization problem
(\ref{1stLSProb}), subject to (\ref{EarthNewtonLAW}). We have to use numerical techniques
to discretize the differential equations (\ref{EarthNewtonLAW}) such that we can
represent $\mathbf{x}(t_{yi},\mathbf{p},\mathbf{c}))$ in terms of the unknown
differential equation parameters $\mathbf{p}$. As in the case of (\ref{YiOBSEquation}),
given some approximate values $\mathbf{p}^0$, $\mathbf{x}^0_0$ and $\mathbf{v}^0_0$ of
$\mathbf{p}$, $\mathbf{x}_0$ and $\mathbf{v}_0$, respectively, we can obtain the
numerical solution $\mathbf{x}^0(t,\mathbf{p}^0, \mathbf{c}(\mathbf{p}^0, \mathbf{x}_0^0,
\mathbf{v}_0^0))$ (or simply $\mathbf{x}^0(t)$ for conciseness of notations) by
numerically solving the following nonlinear differential equations: \beq
\label{EarthNewtonLAWX0} \ddot{\mathbf{x}}^0 = \mathbf{a}_E(t, \mathbf{x}^0,
\mathbf{p}^0) \eeq under the given initial values of $\mathbf{x}^0_0$ and
$\mathbf{v}^0_0$. Accordingly, the solution of $\mathbf{v}$ is denoted by
$\mathbf{v}^0(t)(=\dot{\mathbf{x}}^0(t))$.

Since the differential equations (\ref{EarthNewtonLAW}) are nonlinear, we may attempt to
find their approximate solutions in terms of $\mathbf{p}$ by either directly linearizing
(\ref{EarthNewtonLAW}) or using numerical integration methods. For convenience, we
rewrite the second order differential equations (\ref{EarthNewtonLAW}) as an equivalent
system of first order differential equations: \begin{eqnarray}
\label{1stOrderEarthNewtonLAW}
\dot{\mathbf{z}}(t)  =  \left[ \begin{array}{c} \dot{\mathbf{x}}(t) \\
\dot{\mathbf{v}}(t)
\end{array} \right]
  =  \left[ \begin{array}{c} \mathbf{v}(t) \\ \mathbf{a}_E(t, \mathbf{x}, \mathbf{p}) \end{array}
\right]. \end{eqnarray} As in the case of (\ref{EarthNewtonLAW}), both the equation
parameters $\mathbf{p}$ and initial conditions to (\ref{1stOrderEarthNewtonLAW}) are
unknown.

\subsection{The linearized local solution}
Subtracting $\mathbf{z}^0(t)$ from (\ref{1stOrderEarthNewtonLAW}), we have
\begin{eqnarray} \label{1stOrderEarthNewtonLAWV1} \dot{\mathbf{z}}(t) - \dot\mathbf{z}^0(t) &
=& \left[ \begin{array}{c} \dot{\mathbf{x}}(t) - \dot{\mathbf{x}}^0(t) \\
\dot{\mathbf{v}}(t) - \dot{\mathbf{v}}^0(t) \end{array} \right] \nonumber \\
 & = & \left[ \begin{array}{c} \mathbf{v}(t) - \mathbf{v}^0(t) \\
 \mathbf{a}_E(t, \mathbf{x}, \mathbf{p}) -
\mathbf{a}_E(t, \mathbf{x}^0, \mathbf{p}^0) \end{array} \right]. \end{eqnarray} Denoting
$$ \Delta\dot{\mathbf{z}}(t) = \dot{\mathbf{z}}(t) - \dot\mathbf{z}^0(t), $$
$$ \Delta\mathbf{x}(t) = \mathbf{x}(t) - \mathbf{x}^0(t), $$
 $$ \Delta\mathbf{v}(t) = \mathbf{v}(t) - \mathbf{v}^0(t), $$ and then linearizing the
 right hand side of (\ref{1stOrderEarthNewtonLAWV1}), we have \alpheqn
\begin{eqnarray} \label{1stOrderEarthNewtonLAWV2} \Delta\dot{\mathbf{z}}(t) & = &
\left[ \begin{array}{c} \Delta\mathbf{v}(t)  \\
\mathbf{F}_{ax}(t)\Delta\mathbf{x}(t) + \mathbf{F}_{ap}(t)\Delta\mathbf{p} \end{array}
\right] \nonumber \\
 & = & \left[ \begin{array}{cc} \mathbf{0} & \mathbf{I} \\
     \mathbf{F}_{ax}(t)  &   \mathbf{0} \end{array} \right] \left[
     \begin{array}{c} \Delta\mathbf{x}(t) \\
     \Delta\mathbf{v}(t)  \end{array} \right] + \left[
     \begin{array}{c} \mathbf{0} \\
     \mathbf{F}_{ap}(t)  \end{array} \right] \Delta\mathbf{p} \nonumber \\
 & = & \left[ \begin{array}{cc} \mathbf{0} & \mathbf{I} \\
     \mathbf{F}_{ax}(t)  &   \mathbf{0} \end{array} \right] \Delta\mathbf{z}(t) + \left[
     \begin{array}{c} \mathbf{0} \\
     \mathbf{F}_{ap}(t)  \end{array} \right] \Delta\mathbf{p},
\end{eqnarray} which is a standard linear dynamical
system of differential equations, where \beq \label{matrFax} \mathbf{F}_{ax}(t) = \left.
\frac{\partial \mathbf{a}_E(t, \mathbf{x}, \mathbf{p})}{\partial \mathbf{x}^T}
\right|_{\mathbf{x}=\mathbf{x}^0(t), \,\mathbf{p}=\mathbf{p}^0}, \eeq \beq
\label{matrFap} \mathbf{F}_{ap}(t) = \left. \frac{\partial \mathbf{a}_E(t, \mathbf{x},
\mathbf{p})}{\partial \mathbf{p}^T} \right|_{\mathbf{x}=\mathbf{x}^0(t),
\,\mathbf{p}=\mathbf{p}^0}, \eeq\reseteqn\setcounter{EQ5}{\value{equation}}and
$\mathbf{I}$ is a $(3\times 3)$ identity matrix.

Given the initial conditions $\mathbf{x}_0$ and $\mathbf{v}_0$ for the original problem
of satellite gravimetry, we can have the corresponding initial conditions
$\Delta\mathbf{z}_0$ for $\Delta\mathbf{z}(t)$. Thus, according to Stengel (1986) and
Grewal and Andrews (1993), we can readily write the solution to the linear differential
equations (\ref{1stOrderEarthNewtonLAWV2}) as follows: \beq \label{LinSolZ}
\Delta\mathbf{z}(t) = \bm{\Phi}(t,t_0) \Delta\mathbf{z}_0 + \int_{t_0}^t
\bm{\Phi}(t,\tau) \left[
     \begin{array}{c} \mathbf{0} \\
     \mathbf{F}_{ap}(\tau)  \end{array} \right] d\tau \Delta\mathbf{p}, \eeq or equivalently,
\alpheqn \beq \label{LinSolX} \mathbf{z}(t) = \mathbf{z}^0(t) + \bm{\Phi}(t,t_0)
\Delta\mathbf{z}_0 +
\int_{t_0}^t \bm{\Phi}(t,\tau) \left[      \begin{array}{c} \mathbf{0} \\
\mathbf{F}_{ap}(\tau)  \end{array} \right] d\tau \Delta\mathbf{p}, \eeq where
$\bm{\Phi}(t,t_0)$ is the state transition matrix and is equal to \beq \label{matrTran}
\bm{\Phi}(t,t_0) = \bm{\Phi}(t) \bm{\Phi}^{-1}(t_0), \eeq (see e.g., Grewal and Andrews
1993), and the fundamental matrix $\bm{\Phi}(t)$ is the solution to the following matrix
differential equations: \beq \label{matrTranEQ}
\dot{\bm{\Phi}}(t) = \left[ \begin{array}{cc} \mathbf{0} & \mathbf{I} \\
     \mathbf{F}_{ax}(t)  &   \mathbf{0} \end{array} \right] \bm{\Phi}(t), \eeq under the
     initial matrix conditions \beq \label{matrTranConditions} \bm{\Phi}(t_0) = \mathbf{I}_6,
\eeq\reseteqn\setcounter{EQ6}{\value{equation}}with $\mathbf{I}_6$ is a $(6\times 6)$
identity matrix. For more properties about $\bm{\Phi}(t)$, including its uniqueness and
non-singularity, the reader is referred to Grewal and Andrews (1993).

It is clear that the solution (\ref{LinSolX}) of the satellite orbit and velocity is a
linear vector function of the corrections $\Delta\mathbf{z}_0$ to the approximate initial
values $[\mathbf{x}^0_0,\, \mathbf{v}^0_0]$ and the corrections $\Delta\mathbf{p}$ to the
approximate values $\mathbf{p}^0$. Therefore, we can readily linearize the original
satellite tracking measurement (\ref{YiOBSEquationOriginal}) with respect to
$\Delta\mathbf{z}_0$ and $\Delta\mathbf{p}$.

\subsection{Numerical integration methods to construct local solutions}
When numerical integration methods are required, one always assumes that the functions to
be integrated are given and/or known, and the target is to use such methods to
numerically compute the integration of the functions, as can be found in any standard
textbooks on numerical analysis and numerical integration (see e.g., Stoer and Burlirsch
2002; Teodorescu et al. 2013). These techniques can be directly used to compute an
approximate reference orbit of a satellite, given initial conditions $[\mathbf{x}_0^0, \,
\mathbf{v}_0^0]$ and $\mathbf{p}^0$. However, in satellite gravimetry from satellite
tracking measurements, since both initial conditions $[\mathbf{x}_0, \, \mathbf{v}_0]$
and the differential equation parameters $\mathbf{p}$ are unknown, it is impossible to
exactly compute satellite orbits by directly implementing any well documented numerical
integral methods.

In this part of the paper, unlike standard textbooks on numerical integration to compute
the integral of a given function (without any unknown parameters) (see e.g., Stoer and
Burlirsch 2002; Teodorescu et al. 2013), our basic idea is to construct a solution to
nonlinear differential equations with unknown equation parameters and unknown initial
conditions, with the aid of numerical integration methods. More precisely, in the case of
satellite gravimetry from tracking measurements, we will represent the orbital solution
$\mathbf{x}(t,\mathbf{p}, \mathbf{c}(\mathbf{p}, \mathbf{x}_0, \mathbf{v}_0))$ to the
Newton's differential equations (\ref{EarthNewtonLAW}) in terms of its approximate value,
the unknown corrections $\Delta\mathbf{z}_0$ to the approximate initial values
$[\mathbf{x}_0^0, \,\mathbf{v}_0^0]$ and the unknown corrections $\Delta\mathbf{p}$ of
the harmonic coefficients by numerically solving the nonlinear differential equations
(\ref{EarthNewtonLAW}) under the initial (unknown) conditions $\mathbf{x}_0$ and
$\mathbf{v}_0$. Recall that for each measurement $y_i$ at the time epoch $t_{yi}$, we
obtain the nominal approximate orbit $\mathbf{x}^0(t_{yi},\mathbf{p}^0,
\mathbf{c}(\mathbf{p}^0, \mathbf{x}_0^0, \mathbf{v}_0^0))$ by numerically solving the
nonlinear differential equations (\ref{EarthNewtonLAWX0}) under the initial conditions
$\mathbf{x}_0^0$ and $\mathbf{v}_0^0$ at the initial time epoch $t_0$. The procedure of
numerical integration has to partition the time interval $[t_0,\, t_{yi}]$ into a number
of sub-intervals, usually equidistant such that $$ t_j = t_0 + j h, \,\, j = 1, 2, ...,
m_{yi}
$$ where $h=(t_{yi}-t_0)/m_{yi}$. One can then apply numerical integration methods to
progressively compute all the nominal reference positions $\mathbf{x}^0(t_j,\mathbf{p}^0,
\mathbf{c}(\mathbf{p}^0, \mathbf{x}_0^0, \mathbf{v}_0^0))$.

However, for the satellite gravimetry problem with tracking measurements, we do not have
the true values of the satellite position and velocity at an initial time epoch $t_0$ but
can only assume their approximate values. In addition, the harmonic coefficients
$\mathbf{p}$ are unknown as well. Since the differential equations (\ref{EarthNewtonLAW})
are nonlinear, it is not likely to use analytical methods to directly derive a
convergent, analytical representation of $\mathbf{x}(t,\mathbf{p}, \mathbf{c}(\mathbf{p},
\mathbf{x}_0, \mathbf{v}_0))$ in terms of $\Delta\mathbf{z}_0$ and $\Delta\mathbf{p}$,
unless one is satisfied with a linearized, one-iteration solution. Thus, we will focus on
explicit numerical integration methods to progressively solve the nonlinear differential
equations (\ref{EarthNewtonLAW}) in the remainder of this section.

In what follows, we will use the Euler method and the modified Euler method to
demonstrate the construction of $\mathbf{x}(t_j,\mathbf{p}, \mathbf{c}(\mathbf{p},
\mathbf{x}_0, \mathbf{v}_0))$ in terms of $\Delta\mathbf{z}_0$ and $\Delta\mathbf{p}$.
Other numerical integration methods such as Heun's method, Runge-Kutta methods of any
order and/or the Newton-Cotes method can be treated in the same manner and will be
omitted here. The interested reader can work them out by himself or herself. For
conciseness of notations, we will denote the right hand side of
(\ref{1stOrderEarthNewtonLAW}) by $\mathbf{g}(t, \mathbf{z}(t),\mathbf{p})$ and rewrite
(\ref{1stOrderEarthNewtonLAW}) as follows: \beq \label{EulerOrderEarthNewtonLAW}
\dot{\mathbf{z}}(t) = \mathbf{g}(t, \mathbf{z}(t),\mathbf{p}) \eeq under the (unknown)
initial conditions $\mathbf{z}_0$ (namely, $\mathbf{x}_0$ and $\mathbf{v}_0$).

To start the Euler method, we have  \beq \label{EulerStart} \mathbf{z}(t_1) =
\mathbf{z}(t_0) +  h\mathbf{g}(t_0, \mathbf{z}(t_0),\mathbf{p}), \eeq (see e.g., Stoer
and Burlirsch 2002; Teodorescu et al. 2013). Linearizing the vector functions
$\mathbf{g}(\cdot)$ at $(\mathbf{z}_0^0,\, \mathbf{p}^0)$ and bearing in mind the
approximate orbit $\mathbf{z}^0(t, \mathbf{z}^0(t),\mathbf{p}^0)$, we can rewrite
(\ref{EulerStart}) into: $$ \mathbf{z}^0(t_1) + \Delta\mathbf{z}(t_1) = \mathbf{z}^0_0 +
\Delta\mathbf{z}_0 +  h \mathbf{g}(t_0, \mathbf{z}^0_0,\mathbf{p}^0) +
h\mathbf{G}_{gz0}\Delta\mathbf{z}_0 + h\mathbf{G}_{gp0}\Delta\mathbf{p}, $$ or
equivalently,  \beq \label{EulerStartLin} \Delta\mathbf{z}(t_1) = \delta\mathbf{z}^0_{01}
+ \left[ \mathbf{I}_6  + h\mathbf{G}_{gz0} \right] \Delta\mathbf{z}_0 +
h\mathbf{G}_{gp0}\Delta\mathbf{p}, \eeq where \alpheqn \beq \delta\mathbf{z}^0_{01} =
\mathbf{z}^0_0+ h\mathbf{g}(t_0, \mathbf{z}^0_0,\mathbf{p}^0) - \mathbf{z}^0(t_1), \eeq
 \beq \label{matrGgz0}
   \mathbf{G}_{gz0} = \left. \frac{\partial \mathbf{g}(t, \mathbf{z}(t),
 \mathbf{p})}{\partial \mathbf{z}^T} \right|_{\mathbf{z}(t)=\mathbf{z}^0_0,
\,\mathbf{p}=\mathbf{p}^0},  \eeq  and \beq \label{matrGgp0} \mathbf{G}_{gp0} = \left.
\frac{\partial \mathbf{g}(t, \mathbf{z}(t),
 \mathbf{p}) }{\partial \mathbf{p}^T} \right|_{\mathbf{z}(t)=\mathbf{z}^0_0,
\,\mathbf{p}=\mathbf{p}^0}. \eeq\reseteqn\setcounter{EQ6A}{\value{equation}}

To progress from $t_1$ to $t_2$, the Euler method takes the form of formula:
 \beq \label{EulerStartT1T2Orignal} \mathbf{z}(t_2) =
\mathbf{z}(t_1) +  h\mathbf{g}(t_1, \mathbf{z}(t_1),\mathbf{p}). \eeq Following the same
procedure as described above, and neglecting all the terms of order $h^2$, we can rewrite
the above formula as follows: \beq \label{EulerStartT1T2} \Delta\mathbf{z}(t_2)  =
\delta\mathbf{z}^0_{12} + \left[ \mathbf{I}_6  + h\mathbf{G}_{gz1} \right]
\Delta\mathbf{z}(t_1) + h\mathbf{G}_{gp1}\Delta\mathbf{p}, \eeq where
$$ \delta\mathbf{z}^0_{12} = \mathbf{z}^0(t_1)+ h\mathbf{g}(t_1,
\mathbf{z}^0(t_1),\mathbf{p}^0) - \mathbf{z}^0(t_2). $$ Inserting (\ref{EulerStartLin})
into (\ref{EulerStartT1T2}) and after some rearrangement, we have
\begin{eqnarray} \label{EulerFinalT1T2} \Delta\mathbf{z}(t_2) & = &
\delta\mathbf{z}^0_{12}+ \delta\mathbf{z}^0_{01} +
h\mathbf{G}_{gz1}\delta\mathbf{z}^0_{01} + \left[ \mathbf{I}_6  + h\mathbf{G}_{gz0}+
h\mathbf{G}_{gz1} \right] \Delta\mathbf{z}_0 +
[\mathbf{G}_{gp0}+\mathbf{G}_{gp1}]\Delta\mathbf{p} \nonumber \\
& = & \delta\mathbf{z}^0_{02}+ h\mathbf{G}_{gz1}\delta\mathbf{z}^0_{01} +
\left[\mathbf{I}_6 + h\sum\limits_{j=0}\limits^1
\mathbf{G}_{gzj}\right]\Delta\mathbf{z}_0 + h\sum\limits_{j=0}\limits^1
\mathbf{G}_{gpj}\Delta\mathbf{p}, \end{eqnarray} where $$ \delta\mathbf{z}^0_{02} =
\mathbf{z}^0_0+ h\sum\limits_{j=0}\limits^1\mathbf{g}(t_j,
\mathbf{z}^0(t_j),\mathbf{p}^0) - \mathbf{z}^0(t_2). $$ The matrices $\mathbf{G}_{gz1}$
and $\mathbf{G}_{gp1}$ are computed in the same manner as in (\ref{matrGgz0}) and
(\ref{matrGgp0}) but at the point of $\mathbf{z}^0(t_1)$.

Repeating the same procedure as described in the above, we can finally obtain the
representation of the corrections $\Delta\mathbf{z}(t_{yi})$ as follows: \beq
\label{EulerFinalYi} \Delta\mathbf{z}(t_{yi}) = \delta\mathbf{z}^0_{0t_{yi}} +
h\sum\limits_{j=1}\limits^{m_{yi}-1}\mathbf{G}_{gzj}\delta\mathbf{z}^0_{0j} + \left[
\mathbf{I}_6 + h\sum\limits_{j=0}\limits^{m_{yi}-1} \mathbf{G}_{gzj} \right]
\Delta\mathbf{z}_0 + h\sum\limits_{j=0}\limits^{m_{yi}-1}
\mathbf{G}_{gpj}\Delta\mathbf{p}, \eeq where $$ \delta\mathbf{z}^0_{0t_{yi}} =
\mathbf{z}^0_0+ h\sum\limits_{j=0}\limits^{m_{yi}-1}\mathbf{g}(t_j,
\mathbf{z}^0(t_j),\mathbf{p}^0) - \mathbf{z}^0(t_{yi}). $$ In the similar manner, one can
then work out the corrections for all the satellite tracking measurements $\mathbf{y}$,
continue to linearize (\ref{YiOBSEquationOriginal}) and complete the construction of the
observational equations for $\mathbf{y}$. Probably, we should note that the corrections
$\Delta\mathbf{z}(t_{yi})$ of (\ref{EulerFinalYi}) contain constant calibrated terms,
depending on $\delta\mathbf{z}^0_{0t_{yi}}$ and
$\mathbf{G}_{gzj}\delta\mathbf{z}^0_{0j}$, plus the terms with the unknown orbital
position and velocity corrections $\Delta\mathbf{z}_0$ and the unknown corrections
$\Delta\mathbf{p}$ of the harmonic coefficients.

To further show that the representation of the corrections $\Delta\mathbf{z}(t_{yi})$
will change with different methods of numerical integration, we will now derive such a
representation by using the modified Euler method. Given the initial (unknown) conditions
$[\mathbf{x}_0,\, \mathbf{v}_0]$ and the unknown parameters $\mathbf{p}$, the modified
Euler method formally starts with the following recursive formula: \beq
\label{MEulerStart} \mathbf{z}(t_j) = \mathbf{z}(t_{j-1}) + \frac{h}{2} \left[
\mathbf{g}(t_{j-1}, \mathbf{z}(t_{j-1}),\mathbf{p}) + \mathbf{g}\{t_j,
\mathbf{z}(t_{j-1})+h\mathbf{g}(t_{j-1}, \mathbf{z}(t_{j-1}),\mathbf{p}),\mathbf{p}\}
\right], \eeq for $j=1,2,...,m_{yi}$ (see e.g., Teodorescu et al. 2013), with the nominal
reference orbit $\mathbf{z}^0(t, \mathbf{z}^0(t),\mathbf{p}^0)$.

Following the same technical procedure as in the case of the Euler method, we can finally
obtain the representation of $\Delta\mathbf{z}(t_{yi})$ for the modified Euler method, as
follows:
\begin{eqnarray} \label{ModifiedEulerFinalYi} \Delta\mathbf{z}(t_{yi}) & = &
\delta\mathbf{z}^{0M}_{0t_{yi}} + \frac{h}{2}\sum\limits_{j=1}\limits^{m_{yi}-1} \left[
\mathbf{G}_{gzj} + \mathbf{G}_{gz(j+1)} \right] \delta\mathbf{z}^{0M}_{0j} \nonumber \\
&  & + \frac{h}{2}\sum\limits_{j=0}\limits^{m_{yi}-1}\mathbf{G}_{gz(j+1)}
\delta\mathbf{z}^0_{j(j+1)}  + \left[\mathbf{I}_6 +
\frac{h}{2}\sum\limits_{j=0}\limits^{m_{yi}-1} \left\{
\mathbf{G}_{gzj}+\mathbf{G}_{gz(j+1)} \right\} \right]\Delta\mathbf{z}_0 \nonumber \\
&  & + \frac{h}{2}\sum\limits_{j=0}\limits^{m_{yi}-1} \left[
\mathbf{G}_{gpj}+\mathbf{G}_{gp(j+1)} \right] \Delta\mathbf{p}, \end{eqnarray} where $$
\delta\mathbf{z}^{0M}_{0k} = \mathbf{z}_0^0 + \frac{h}{2}\sum\limits_{j=0}\limits^{k-1}
\left[ \mathbf{g}(t_j, \mathbf{z}^0(t_j),\mathbf{p}^0) + \mathbf{g}\{t_{j+1},
\mathbf{z}^0(t_{j+1}),\mathbf{p}^0\} \right] - \mathbf{z}^0(t_k), $$ and
$$ \delta\mathbf{z}^0_{j(j+1)} = \mathbf{z}^0(t_j)+ h\mathbf{g}(t_j,
\mathbf{z}^0(t_j),\mathbf{p}^0) - \mathbf{z}^0(t_{j+1}). $$ The technical derivation of
 (\ref{ModifiedEulerFinalYi}) is given in the appendix.

It is clear from (\ref{EulerFinalYi}) and (\ref{ModifiedEulerFinalYi}) that different
numerical integration methods will result in different representations of the orbital
position and velocity corrections $\Delta\mathbf{z}(t_{yi})$, even though the formulae
can be coded and the coefficients of both $\Delta\mathbf{z}_0$ and $\Delta\mathbf{p}$ can
be automatically computed. We should note that numerical integration schemes can be
different for precisely computing the nominal orbital solution $\mathbf{z}^0(t,
\mathbf{z}^0(t),\mathbf{p}^0)$ and for representing the corrections
$\Delta\mathbf{z}(t_{yi})$ in terms of $\Delta\mathbf{z}_0$ and $\Delta\mathbf{p}$.
Precise numerical integration methods should be used to compute the nominal reference
orbit of a satellite, given approximate initial conditions $[\mathbf{x}^0_0,\,
\mathbf{v}^0_0]$ and a set of approximate harmonic coefficients $\mathbf{p}^0$.
Implementations and interpretations of numerical integration methods in satellite geodesy
are fundamentally different for computing the nominal orbital solution $\mathbf{z}^0(t,
\mathbf{z}^0(t),\mathbf{p}^0)$ by solving the differential equations
(\ref{EarthNewtonLAWX0}) under the initial conditions $[\mathbf{x}^0_0,\,
\mathbf{v}^0_0]$ and for inverting for the unknown equation parameters $\mathbf{p}$ under
the unknown initial conditions $[\mathbf{x}_0,\, \mathbf{v}_0]$ from satellite tracking
measurements. The former is actually the problem of numerical orbit determination with
given initial values and force parameters, but is only a first step towards the latter.

\section{Measurement-based perturbation theory}\label{GlobalAnalIntEQ}
Perturbation has been commonly carried out, either with a small parameter mathematically
or around mean orbital elements in celestial mechanics (see e.g., Kozai 1959; Hagihara
1971; Taff 1985; Nayfeh 2004), though all the six orbital elements are the functions of
time in reality. Although approximate perturbed solutions are useful to gain some
physical insights into the orbit of a celestial body, they are too approximate to
precisely invert for the unknown equation parameters from modern space observation. Kaula
linear perturbation theory is a local approximate solution to Lagrange's planetary
equations around mean orbital elements, which will be divergent with the increase of time
and cannot utilize full advantages of unprecedented  accuracy and continuity of tracking
measurements from modern space observation technology. On the other hand, the numerical
integration method, though widely used by major institutions worldwide to produce global
 Earth's gravitational models from satellite tracking measurements for highly
multidisciplinary applications, has been proved to be groundless, mathematically and
physically. To fully use profound technological advance in space observation for the next
generation of global  gravitational models, mathematical solutions to the differential
equations of motion of an LEO satellite must be sufficiently precise to extract small
gravitational signals in modern space observation.

In this section, we will derive two perturbation solutions: one is local and the other is
global. Our interest in constructing a local perturbation solution is mainly motivated to
demonstrate how to properly use the nominal reference orbit $\mathbf{z}^0(t,
\mathbf{z}^0(t),\mathbf{p}^0)$ to mathematically solve the governing differential
equations (\ref{EarthNewtonLAW}) of motion of an LEO satellite, in addition to the
approximate analytical and numerical integration solutions in
section~\ref{AppromAnalNumINT}. To take full advantages of modern and future (or next
generation of) space observation technology, the key mathematics has to construct a
global perturbation solution, which should meet the following two requirements: (i) the
solution is either better than what modern and future space observation technology can
provide physically, or at least, sufficiently precise at the noise level of such
technology; and (ii) the solution is global uniformly convergent over arcs of any length.
Since small gravitational signals will accumulate their effect on satellite orbits over
time, this second condition will guarantee that we are able to extract smallest possible
gravitational signals in modern and future space observation to its limit, and as a
result, to produce high-precision, high-resolution global Earth's gravitational models.

As in the case of Xu (2008), we will work out the perturbation solutions in the inertial
reference frame. Nevertheless, to avoid any potential confusion of notations, we will now
switch to the spherical coordinates $(\alpha,\zeta)$ in the inertial reference frame,
instead of continuing to use the notations $(\lambda,\theta)$ of Xu (2008). Since the
disturbing potential $T$ of (\ref{PotentialTCS}) is in the spherical coordinate system,
we will need coordinate transformation between the spherical coordinates $(\alpha, \zeta,
r)$ and the Cartesian coordinates $\mathbf{x}=(x_1, x_2, x_3)^T$. To prepare for the
derivations in the remainder of this section, we symbolically rewrite the disturbing
potential (\ref{TPotential}), with $(\alpha,\zeta)$ of (\arabic{EQ1A}) in the inertial
reference frame, as follows:
 \alpheqn \beq \label{PotentialTCS}  T = \sum
\limits^{N_{\scriptsize \textrm{max}}} \limits_{l=2} \sum \limits_{m=0} \limits^l
T_{lm}^c(\alpha,\zeta, r)  C_{lm} + \sum \limits^{N_{\scriptsize \textrm{max}}}
\limits_{l=2} \sum \limits_{m=0} \limits^l T_{lm}^s(\alpha,\zeta, r)  S_{lm}, \eeq where
\beq T_{lm}^c(\alpha,\zeta, r) = \frac{GM}{r} \left( \frac{R}{r}\right)^l
\cos\{m(\alpha+\delta \alpha-\omega_e t)\}P_{lm}\{\cos(\zeta+\delta \zeta)\}, \eeq \beq
T_{lm}^s(\alpha,\zeta, r) = \frac{GM}{r} \left( \frac{R}{r}\right)^l
\sin\{m(\alpha+\delta \alpha-\omega_e t)\} P_{lm}\{\cos(\zeta+\delta \zeta)\}. \eeq
\reseteqn\setcounter{EQ7}{\value{equation}}The partial derivatives of $T$ with respect to
$\mathbf{x}$ are given as follows: \alpheqn \beq \frac{\partial T}{\partial \mathbf{x}} =
\sum \limits^{N_{\scriptsize \textrm{max}}} \limits_{l=2} \sum \limits_{m=0} \limits^l
\mathbf{p}_{lm}^c(\alpha,\zeta, r)  C_{lm} + \sum \limits^{N_{\scriptsize \textrm{max}}}
\limits_{l=2} \sum \limits_{m=0} \limits^l \mathbf{p}_{lm}^s(\alpha,\zeta, r)  S_{lm},
\eeq where \beq \mathbf{p}_{lm}^c(\alpha,\zeta, r) = \mathbf{R}(\alpha,\zeta, r)\,
\frac{\partial T_{lm}^c(\alpha,\zeta, r)}{\partial (\alpha,\zeta, r)^T}, \eeq \beq
\mathbf{p}_{lm}^s(\alpha,\zeta, r) = \mathbf{R}(\alpha,\zeta, r) \, \frac{\partial
T_{lm}^s(\alpha,\zeta, r)}{\partial (\alpha,\zeta, r)^T}, \eeq \beq
\mathbf{R}(\alpha,\zeta, r) = \frac{\partial (\alpha,\zeta, r)}{\partial \mathbf{x}} =
\left[
\begin{array}{ccc}
-sin\alpha/(r\sin\zeta) & \cos\alpha/(r\sin\zeta) & 0 \\
   \cos\alpha\cos\zeta/r & \sin\alpha\cos\zeta/r & -\sin\zeta/r  \\
   \cos\alpha\sin\zeta  & \sin\alpha\sin\zeta & \cos\zeta
   \end{array} \right]^T, \eeq
\beq \frac{\partial T_{lm}^c(\alpha,\zeta, r)}{\partial (\alpha,\zeta, r)^T} = -
\frac{GM}{r}\left( \frac{R}{r}\right)^l \left[ \begin{array}{c}
 m\, \sin\{m(\alpha+\delta \alpha-\omega_e t)\} P_{lm}\{\cos(\zeta+\delta \zeta)\} \\
 \cos\{m(\alpha+\delta \alpha-\omega_e t)\} \sin(\zeta+\delta \zeta) \dot{P}_{lm}\{\cos(\zeta+\delta \zeta)\} \\
 (l+1) \cos\{m(\alpha+\delta \alpha-\omega_e t)\} P_{lm}\{\cos(\zeta+\delta \zeta)\}/r \end{array} \right], \eeq
\beq \frac{\partial T_{lm}^s(\alpha,\zeta, r)}{\partial (\alpha,\zeta, r)^T} = -
\frac{GM}{r}\left( \frac{R}{r}\right)^l \left[ \begin{array}{c}
 - m\, \cos\{m(\alpha+\delta \alpha-\omega_e t)\} P_{lm}\{\cos(\zeta+\delta \zeta)\} \\
\sin\{m(\alpha+\delta \alpha-\omega_e t)\} \sin(\zeta+\delta \zeta)
 \dot{P}_{lm}\{\cos(\zeta+\delta \zeta)\} \\
 (l+1) \sin\{m(\alpha+\delta \alpha-\omega_e t)\}
 P_{lm}\{\cos(\zeta+\delta \zeta)\}/r \end{array} \right], \eeq
   \reseteqn\setcounter{EQ8}{\value{equation}}and $\dot{P}_{lm}(t)$ stands for the derivatives of
the normalized Legendre function $P_{lm}(t)$ (see e.g., Koop 1993).

\subsection{Local perturbation around a nominal reference orbit}\label{LocalPerturbXV}
In section~\ref{AppromAnalNumINT}, we have used linearization and numerical integration
methods to construct local approximate solutions to (\ref{EarthNewtonLAW}). Given the
approximate solution $\mathbf{z}^0(t, \mathbf{z}^0(t),\mathbf{p}^0)$, we will use the
idea of Xu (2008) to derive new local solutions by turning the nonlinear differential
equations (\ref{EarthNewtonLAW}) into the equivalent nonlinear integral equations. More
precisely, for convenience, we combine the
 nonlinear differential equations (\ref{EarthNewtonLAW}), the Earth's gravitational
acceleration (\ref{Two_dotX}) and the Earth's disturbing potential $T$ of
(\ref{TPotential}), together with the unknown initial conditions $[\mathbf{z}_0,\,
\mathbf{v}_0]$, and rewrite the complete system of nonlinear differential equations as
follows: \beq \label{NewtonLAWFinal} \ddot{\mathbf{x}} = - \frac{GM}{r^3}\mathbf{x} +
\sum \limits^{N_{\scriptsize \textrm{max}}} \limits_{l=2} \sum \limits_{m=0} \limits^l
\mathbf{p}_{lm}^c(\alpha,\zeta, r)  C_{lm} + \sum \limits^{N_{\scriptsize \textrm{max}}}
\limits_{l=2} \sum \limits_{m=0} \limits^l \mathbf{p}_{lm}^s(\alpha,\zeta, r)  S_{lm},
\eeq under the initial conditions $[\mathbf{x}_0,\, \mathbf{v}_0]$.

The solution to (\ref{NewtonLAWFinal}) can be formally written as follows:
\begin{eqnarray} \label{XSolutLocal} \mathbf{x}(t) & = & -
\int_{t_0}^t\int_{t_0}^{\eta} \frac{GM}{r^3(\tau)}\mathbf{x}(\tau)d\tau d\eta + \sum
\limits^{N_{\scriptsize \textrm{max}}} \limits_{l=2} \sum \limits_{m=0} \limits^l C_{lm}
\int_{t_0}^t\int_{t_0}^{\eta}  \mathbf{p}_{lm}^c(\alpha(\tau),\zeta(\tau), r(\tau))
 d\tau d\eta  \nonumber \\
 &  & + \sum \limits^{N_{\scriptsize
\textrm{max}}} \limits_{l=2} \sum \limits_{m=0} \limits^l S_{lm}
\int_{t_0}^t\int_{t_0}^{\eta} \mathbf{p}_{lm}^s(\alpha(\tau),\zeta(\tau), r(\tau)) d\tau
d\eta + \mathbf{v}_0(t-t_0) + \mathbf{x}_0,
\end{eqnarray} where the notations $[\alpha(\tau),\zeta(\tau), r(\tau)]$ are the same as those of
$[\alpha,\zeta, r]$ but to explicitly emphasize that they are all the functions of time.
Obviously, we have turned the nonlinear differential equations (\ref{NewtonLAWFinal})
into the nonlinear Volterra's integral equations (\ref{XSolutLocal}) of the second kind.

Linearizing $\mathbf{x}(\tau)$ around the approximate orbit $\mathbf{x}^0(\tau)$ and
bearing in mind that $\mathbf{x}^0(\tau)$ is essentially computed by integrating the same
equations (\ref{XSolutLocal}) with the initial values
$[\mathbf{x}_0^0,\mathbf{v}_0^0,\mathbf{p}^0]$, we have  \alpheqn
\begin{eqnarray} \label{ExpandSolT} \Delta \mathbf{x}(t) & = &
 - \int_{t_0}^t\!\!\int_{t_0}^{\eta}
\mathbf{A}^x(\mathbf{x}^0(\tau))\Delta \mathbf{x}({\tau})d\tau d\eta   \nonumber \\
 &   & + \sum \limits^{N_{\scriptsize
\textrm{max}}} \limits_{l=2} \sum \limits_{m=0} \limits^l \Delta C_{lm}
\int_{t_0}^t\!\!\int_{t_0}^{\eta}  \mathbf{p}_{lm}^c(\alpha^0(\tau), \zeta^0(\tau),
r^0(\tau)) d\tau d\eta  \nonumber \\ &  & + \sum \limits^{N_{\scriptsize \textrm{max}}}
\limits_{l=2} \sum \limits_{m=0} \limits^l C_{lm}^0 \int_{t_0}^t\!\!\int_{t_0}^{\eta}
 \mathbf{A}_{lm}^c(\alpha^0(\tau), \zeta^0(\tau), r^0(\tau)) \Delta \mathbf{x}({\tau}) d\tau d\eta
\nonumber \\
&   & + \sum \limits^{N_{\scriptsize \textrm{max}}} \limits_{l=2} \sum \limits_{m=0}
\limits^l \Delta S_{lm} \int_{t_0}^t\!\!\int_{t_0}^{\eta}
\mathbf{p}_{lm}^s(\alpha^0(\tau), \zeta^0(\tau), r^0(\tau))  d\tau d\eta \nonumber \\ & &
+ \sum \limits^{N_{\scriptsize \textrm{max}}} \limits_{l=2} \sum \limits_{m=0} \limits^l
S_{lm}^0 \int_{t_0}^t\!\!\int_{t_0}^{\eta}
 \mathbf{A}_{lm}^s(\alpha^0(\tau), \zeta^0(\tau), r^0(\tau)) \Delta \mathbf{x}({\tau}) d\tau d\eta
\nonumber \\ &   & + \Delta \mathbf{v}_0(t-t_0) + \Delta\mathbf{x}_0,
\end{eqnarray} where $C_{lm}^0$ and $S_{lm}^0$ are the approximate values of the harmonic coefficients
used in computing the nominal reference orbit $\mathbf{x}^0(\tau)$, and \beq
\mathbf{x}(\tau) = \mathbf{x}^0(\tau) + \Delta \mathbf{x}(\tau), \eeq
\begin{eqnarray} \mathbf{A}^x(\mathbf{x}(\tau)) & = & \left. GM \frac{\partial}{\partial
\mathbf{x}^T}\left( \frac{\mathbf{x}}{r^3}\right)
\right|_{\mathbf{x}= \mathbf{x}^0(\tau) } \nonumber \\
  & = & GM \left\{ \frac{1}{r_0^3(\tau)} \mathbf{I} -
  \frac{3}{r_0^5(\tau)}\mathbf{x}^0(\tau)\left[\mathbf{x}^0(\tau)\right]^T  \right\}, \end{eqnarray}
  \beq r_0(\tau) = \sqrt{\left[\mathbf{x}^0(\tau)\right]^T\mathbf{x}^0(\tau)}, \eeq
 \begin{eqnarray} \mathbf{A}_{lm}^c(\alpha^0(\tau), \zeta^0(\tau), r^0(\tau))
 & = & \left. \frac{\partial
 \mathbf{p}_{lm}^c(\alpha, \zeta, r)}{\partial \mathbf{x}^T}
 \right|_{\mathbf{x}= \mathbf{x}^0(\tau)} \nonumber \\
  & = & \left. \frac{\partial
 \mathbf{p}_{lm}^c(\alpha, \zeta, r)}{\partial (\alpha, \zeta, r)}
    \right|_{\mathbf{x}= \mathbf{x}^0(\tau)} [\mathbf{R}(\alpha^0(\tau), \zeta^0(\tau), r^0(\tau))]^T,
    \end{eqnarray}
\begin{eqnarray} \mathbf{A}_{lm}^s(\alpha^0(\tau), \zeta^0(\tau), r^0(\tau)) & = & \left. \frac{\partial
 \mathbf{p}_{lm}^s(\alpha, \zeta, r)}{\partial \mathbf{x}^T}
 \right|_{\mathbf{x}= \mathbf{x}^0(\tau)} \nonumber \\
  & = & \left. \frac{\partial
 \mathbf{p}_{lm}^s(\alpha, \zeta, r)}{\partial (\alpha, \zeta, r)}
    \right|_{\mathbf{x}= \mathbf{x}^0(\tau)} [\mathbf{R}(\alpha^0(\tau), \zeta^0(\tau), r^0(\tau))]^T.
\end{eqnarray}\reseteqn\setcounter{EQ9}{\value{equation}}

The linearized Volterra's integral equations (\ref{ExpandSolT}) can be solved
successively (see e.g., Kondo 1991; Hackbusch 1995). To start with, one can set  $\Delta
\mathbf{x}({\tau})$ on the right hand side of (\ref{ExpandSolT}) to zero and obtain the
zeroth approximate (or quasi-linear) solution as follows:
\begin{eqnarray} \label{LocalSolutQUASI} \Delta \mathbf{x}(t) & = &
 \sum \limits^{N_{\scriptsize
\textrm{max}}} \limits_{l=2} \sum \limits_{m=0} \limits^l \Delta C_{lm}
\int_{t_0}^t\!\!\int_{t_0}^{\eta} \mathbf{p}_{lm}^c(\alpha^0(\tau), \zeta^0(\tau), r^0(\tau)) d\tau d\eta  \nonumber \\
&   & + \sum \limits^{N_{\scriptsize \textrm{max}}} \limits_{l=2} \sum \limits_{m=0}
\limits^l \Delta S_{lm} \int_{t_0}^t\!\!\int_{t_0}^{\eta}
\mathbf{p}_{lm}^s(\alpha^0(\tau), \zeta^0(\tau), r^0(\tau))  d\tau d\eta \nonumber \\
 &   & + \Delta \mathbf{v}_0(t-t_0) + \Delta\mathbf{x}_0. \end{eqnarray}

Inserting the quasi-linear solution (\ref{LocalSolutQUASI}) into the right hand side of
the integral equations (\ref{ExpandSolT}) and neglecting all the second order terms of
the harmonic coefficients and the cross-product terms of the harmonic coefficients and
$\Delta \mathbf{x}({\tau})$, we can derive the linear approximation solution as follows:
\begin{eqnarray} \label{LocalSolutLinear} \Delta \mathbf{x}(t) & = &
 - \int_{t_0}^t\!\!\int_{t_0}^{\eta}d\tau d\eta
\mathbf{A}^x(\mathbf{x}^0(\tau)) \Big\{ \sum \limits^{N_{\scriptsize \textrm{max}}}
\limits_{l=2} \sum \limits_{m=0} \limits^l \int_{t_0}^{\tau}\!\!\int_{t_0}^{t_1} \Delta
C_{lm} \mathbf{p}_{lm}^c(\alpha^0(t_2), \zeta^0(t_2), r^0(t_2)) dt_2 dt_1  \nonumber \\
&   & + \sum \limits^{N_{\scriptsize \textrm{max}}} \limits_{l=2} \sum \limits_{m=0}
\limits^l \int_{t_0}^{\tau}\!\!\int_{t_0}^{t_1}\Delta S_{lm}
\mathbf{p}_{lm}^s(\alpha^0(t_2), \zeta^0(t_2), r^0(t_2))  dt_2 dt_1 \nonumber \\
 &   & + \Delta \mathbf{v}_0(\tau-t_0) + \Delta\mathbf{x}_0 \Big\} \nonumber \\
 &   & + \sum \limits^{N_{\scriptsize
\textrm{max}}} \limits_{l=2} \sum \limits_{m=0} \limits^l
\int_{t_0}^t\!\!\int_{t_0}^{\eta} \Delta C_{lm} \mathbf{p}_{lm}^c(\alpha^0(\tau),
\zeta^0(\tau), r^0(\tau)) d\tau d\eta  \nonumber \\
&   & + \sum \limits^{N_{\scriptsize \textrm{max}}} \limits_{l=2} \sum \limits_{m=0}
\limits^l \int_{t_0}^t\!\!\int_{t_0}^{\eta}\Delta S_{lm}
\mathbf{p}_{lm}^s(\alpha^0(\tau), \zeta^0(\tau), r^0(\tau))  d\tau d\eta \nonumber \\ & &
+ \Delta \mathbf{v}_0(t-t_0) + \Delta\mathbf{x}_0,
\end{eqnarray} which, after some re-arrangement, becomes: \alpheqn
\begin{eqnarray}
\label{LocalLinearXFinal} \Delta \mathbf{x}(t)  = \mathbf{D}_v^x \Delta \mathbf{v}_0 +
\mathbf{D}_x^x \Delta\mathbf{x}_0 + \sum \limits^{N_{\scriptsize \textrm{max}}}
\limits_{l=2} \sum \limits_{m=0} \limits^l \mathbf{d}^{cx}_{lm} \Delta C_{lm} + \sum
\limits^{N_{\scriptsize \textrm{max}}} \limits_{l=2} \sum \limits_{m=0} \limits^l
\mathbf{d}^{sx}_{lm}\Delta S_{lm},
\end{eqnarray} where
 \beq \label{CoeffDvx} \mathbf{D}_v^x = \mathbf{I}\,(t-t_0) - \int_{t_0}^t\!\!\int_{t_0}^{\eta}
\mathbf{A}^x(\mathbf{x}^0(\tau)) (\tau-t_0)d\tau d\eta, \eeq \beq \label{CoeffDxx}
\mathbf{D}_x^x = \mathbf{I} - \int_{t_0}^t\!\!\int_{t_0}^{\eta}
\mathbf{A}^x(\mathbf{x}^0(\tau)) d\tau d\eta, \eeq
 \begin{eqnarray} \label{Coeffdcxlm} \mathbf{d}^{cx}_{lm} & = & \int_{t_0}^t\!\!\int_{t_0}^{\eta}
 \mathbf{p}_{lm}^c(\alpha^0(\tau), \zeta^0(\tau), r^0(\tau))
d\tau d\eta \nonumber \\  &   &  - \int_{t_0}^t\!\!\int_{t_0}^{\eta} d\tau d\eta
\mathbf{A}^x(\mathbf{x}^0(\tau)) \left\{ \int_{t_0}^{\tau}\!\!\int_{t_0}^{t_1}
\mathbf{p}_{lm}^c(\alpha^0(t_2), \zeta^0(t_2), r^0(t_2))  dt_1 dt_2 \right\},
\end{eqnarray}
\begin{eqnarray} \label{Coeffdsxlm} \mathbf{d}^{sx}_{lm} & = & \int_{t_0}^t\!\!\int_{t_0}^{\eta}
 \mathbf{p}_{lm}^s(\alpha^0(\tau), \zeta^0(\tau), r^0(\tau))
d\tau d\eta \nonumber \\  &   &  - \int_{t_0}^t\!\!\int_{t_0}^{\eta} d\tau d\eta
\mathbf{A}^x(\mathbf{x}^0(\tau)) \left\{ \int_{t_0}^{\tau}\!\!\int_{t_0}^{t_1}
\mathbf{p}_{lm}^s(\alpha^0(t_2), \zeta^0(t_2), r^0(t_2))  dt_1 dt_2 \right\}.
\end{eqnarray}\reseteqn\setcounter{EQ9A}{\value{equation}}

If one is interested in constructing the second order solution of $\mathbf{x}(t)$, one
will have to expand the nonlinear integral equations (\ref{XSolutLocal}) into the Taylor
series and truncate it up to the second order approximation. Then one can repeat the
above procedure,  insert the linear solution (\ref{LocalLinearXFinal}) into the truncated
second order integral equations, and finally obtain the second order solution of $\Delta
\mathbf{x}(t)$ in terms of the unknown corrections $\Delta\mathbf{x}_0$, $\Delta
\mathbf{v}_0$, $\Delta C_{lm}$ and $\Delta S_{lm}$. Because the solutions derived in the
above are only of local nature, we will not go further for the second order local
solutions.

In the case of velocity of satellite motion, again bearing in mind that $\mathbf{v}^0(t)$
has been equivalently computed by integrating the right hand side of the following
integral equations with $[\mathbf{x}_0^0,\,\mathbf{v}_0^0]$ and $\mathbf{p}^0$, we
linearize the following integral equations of velocity:
\begin{eqnarray} \label{VSolutLocal} \mathbf{v}(t) & = & -
\int_{t_0}^t \frac{GM}{r^3(\tau)}\mathbf{x}(\tau)d\tau  + \sum \limits^{N_{\scriptsize
\textrm{max}}} \limits_{l=2} \sum \limits_{m=0} \limits^l C_{lm}
\int_{t_0}^t \mathbf{p}_{lm}^c(\alpha(\tau), \zeta(\tau), r(\tau)) d\tau \nonumber \\
 &  & + \sum \limits^{N_{\scriptsize
\textrm{max}}} \limits_{l=2} \sum \limits_{m=0} \limits^l S_{lm} \int_{t_0}^t
\mathbf{p}_{lm}^s(\alpha(\tau), \zeta(\tau), r(\tau)) d\tau  + \mathbf{v}_0,
\end{eqnarray} around the nominal reference orbit and velocity $[\mathbf{x}^0(\tau),\,
\mathbf{v}^0(\tau)]$ and obtain:
\begin{eqnarray} \label{ExpandSolTV} \Delta \mathbf{v}(t) & = &
 - \int_{t_0}^t\mathbf{A}^x(\mathbf{x}^0(\tau))\Delta \mathbf{x}({\tau})d\tau \nonumber \\
 &   & + \sum \limits^{N_{\scriptsize
\textrm{max}}} \limits_{l=2} \sum \limits_{m=0} \limits^l \Delta C_{lm} \int_{t_0}^t
\mathbf{p}_{lm}^c(\alpha^0(\tau), \zeta^0(\tau), r^0(\tau)) d\tau   \nonumber \\ &  & +
\sum \limits^{N_{\scriptsize \textrm{max}}} \limits_{l=2} \sum \limits_{m=0} \limits^l
C_{lm}^0 \int_{t_0}^t
 \mathbf{A}_{lm}^c(\alpha^0(\tau), \zeta^0(\tau), r^0(\tau)) \Delta \mathbf{x}({\tau}) d\tau
\nonumber \\
&   & + \sum \limits^{N_{\scriptsize \textrm{max}}} \limits_{l=2} \sum \limits_{m=0}
\limits^l \Delta S_{lm} \int_{t_0}^t \mathbf{p}_{lm}^s(\alpha^0(\tau), \zeta^0(\tau),
r^0(\tau))  d\tau \nonumber \\ &  & + \sum \limits^{N_{\scriptsize \textrm{max}}}
\limits_{l=2} \sum \limits_{m=0} \limits^l S_{lm}^0 \int_{t_0}^t
 \mathbf{A}_{lm}^s(\alpha^0(\tau), \zeta^0(\tau), r^0(\tau)) \Delta \mathbf{x}({\tau}) d\tau
  + \Delta \mathbf{v}_0.
\end{eqnarray}

As in the case of (\ref{LocalSolutQUASI}), by setting $\Delta \mathbf{x}({\tau})$ on the
right hand side of (\ref{ExpandSolTV}) to zero, we obtain the zeroth order approximate
(or quasi-linear) solution of the velocity as follows:
\begin{eqnarray} \label{VLocalSolutQUASI} \Delta \mathbf{v}(t) & = &
 \sum \limits^{N_{\scriptsize
\textrm{max}}} \limits_{l=2} \sum \limits_{m=0} \limits^l \Delta C_{lm} \int_{t_0}^t
\mathbf{p}_{lm}^c(\alpha^0(\tau), \zeta^0(\tau), r^0(\tau)) d\tau   \nonumber \\
&   & + \sum \limits^{N_{\scriptsize \textrm{max}}} \limits_{l=2} \sum \limits_{m=0}
\limits^l \Delta S_{lm} \int_{t_0}^t \mathbf{p}_{lm}^s(\alpha^0(\tau), \zeta^0(\tau),
r^0(\tau))  d\tau + \Delta \mathbf{v}_0. \end{eqnarray} To derive the linear solution of
the velocity, by using the same approach as in deriving the linear solution for the
orbit, we can insert the quasi-linear solution (\ref{LocalSolutQUASI}) into the right
hand side of (\ref{ExpandSolTV}) and obtain the linear solution as follows:
\begin{eqnarray}
\label{LocalLinearVFinal} \Delta \mathbf{v}(t)  = \mathbf{D}_v^v \Delta \mathbf{v}_0 +
\mathbf{D}_x^v \Delta\mathbf{x}_0 + \sum \limits^{N_{\scriptsize \textrm{max}}}
\limits_{l=2} \sum \limits_{m=0} \limits^l \mathbf{d}^{cv}_{lm} \Delta C_{lm} + \sum
\limits^{N_{\scriptsize \textrm{max}}} \limits_{l=2} \sum \limits_{m=0} \limits^l
\mathbf{d}^{sv}_{lm}\Delta S_{lm},
\end{eqnarray} where \alpheqn \beq \label{coeffLocalDvv}
  \mathbf{D}_v^v = \mathbf{I} - \int_{t_0}^t
\mathbf{A}^x(\mathbf{x}^0(\tau)) (\tau-t_0)d\tau , \eeq \beq \label{coeffLocalDxv}
\mathbf{D}_x^v =  - \int_{t_0}^t\! \mathbf{A}^x(\mathbf{x}^0(\tau)) d\tau, \eeq
 \begin{eqnarray} \label{coeffLocaldcvlm} \mathbf{d}^{cv}_{lm} & = & \int_{t_0}^t\!
 \mathbf{p}_{lm}^c(\alpha^0(\tau), \zeta^0(\tau), r^0(\tau))
d\tau \nonumber \\  &   &  - \int_{t_0}^t\! d\tau  \mathbf{A}^x(\mathbf{x}^0(\tau))
\left\{ \int_{t_0}^{\tau}\!\!\int_{t_0}^{t_1} \mathbf{p}_{lm}^c(\alpha^0(t_2),
\zeta^0(t_2), r^0(t_2))  dt_1 dt_2 \right\},
\end{eqnarray}
\begin{eqnarray} \label{coeffLocaldsvlm} \mathbf{d}^{sv}_{lm} & = & \int_{t_0}^t\!
 \mathbf{p}_{lm}^s(\alpha^0(\tau), \zeta^0(\tau), r^0(\tau))
d\tau \nonumber \\  &   &  - \int_{t_0}^t\! d\tau  \mathbf{A}^x(\mathbf{x}^0(\tau))
\left\{ \int_{t_0}^{\tau}\!\!\int_{t_0}^{t_1} \mathbf{p}_{lm}^s(\alpha^0(t_2),
\zeta^0(t_2), r^0(t_2))  dt_1 dt_2 \right\}.
\end{eqnarray}\reseteqn\setcounter{EQ9B}{\value{equation}}

\subsection{Global uniformly convergent measurement-based perturbation}
\begin{figure}
\centering
\includegraphics[totalheight=70mm, width =110mm]{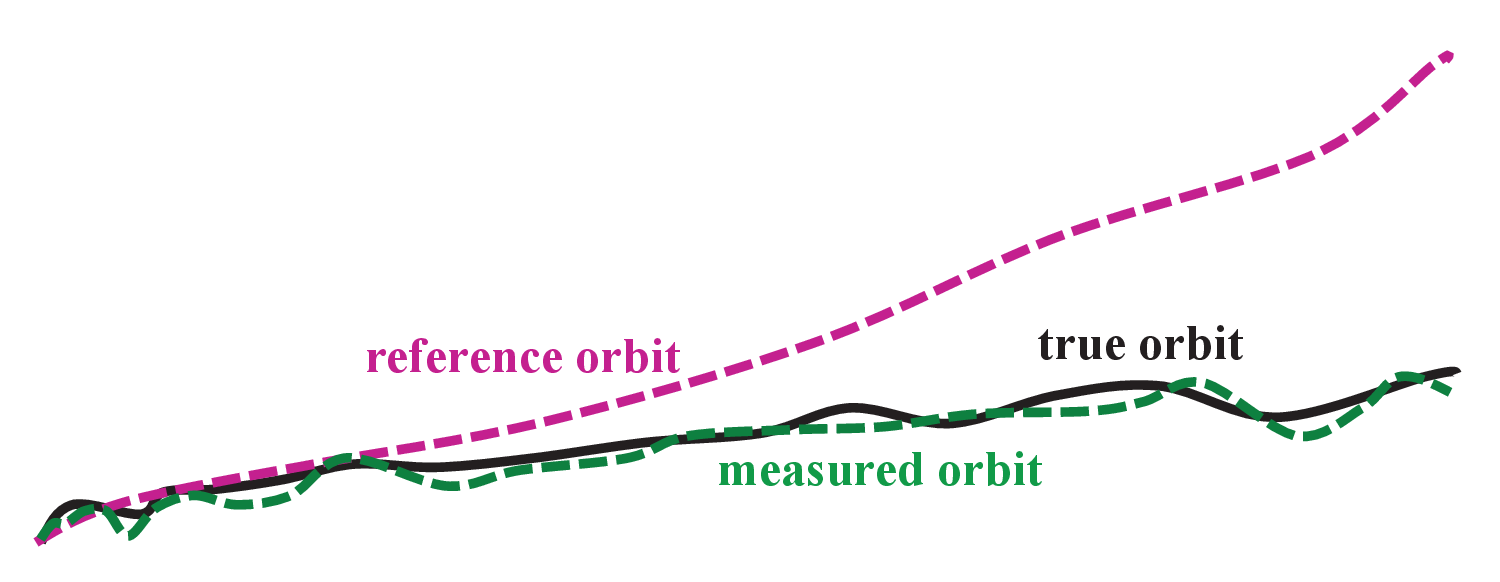}
\caption{Illustration of the relationship among the nominal reference orbit (pink-dashed
line), the precisely measured orbit (green-dashed line) and the true (but unknown) orbit
(black line) of an LEO satellite (modified after Xu 2012, 2015b). } \label{LEOSat3Orbits}
\end{figure}

With the technological advance in GNSS systems and GNSS receiver hardware, orbits of LEO
satellites can now be measured almost continuously (at the sampling rate of 100 Hz or
likely even 200 Hz in the near future) and precisely (at the cm and/or even mm level of
accuracy). Thus, without loss of generality, we will assume a precisely measured orbit
for an LEO gravity satellite, which is denoted analytically as $\{ \mathbf{x}_o(\tau)\,|
\,t_0\leq \tau \leq t\}$ over the whole arc of orbit, with the subscript $o$ standing for
{\em observed}. This measured orbit $\mathbf{x}_o(\tau)$ is only slightly different from
the true orbit $\mathbf{x}(\tau)$ at the level of random errors of measurements. Instead
of using the nominal reference orbit $\{ \mathbf{x}^0(\tau)\,| \,t_0\leq \tau \leq t\}$
to derive local solutions for $\Delta \mathbf{x}(t)$ and $\Delta \mathbf{v}(t)$, we
should certainly use the measured orbit $\mathbf{x}_o(\tau)$ to construct the solutions
to the nonlinear differential equations (\ref{NewtonLAWFinal}), which will then never
diverge with the increase of time. The relationship among the nominal reference orbit,
the precisely measured orbit and the true but unknown orbit of an LEO satellite is
illustrated in Fig.~\ref{LEOSat3Orbits}. In other words, since the measured orbit is very
precise, all the solutions to be derived here are only different from the true orbit and
velocity at the level of measurement noise and are guaranteed to converge globally
uniformly, no matter how long an orbital arc can be.

For the nominal reference orbit $\mathbf{x}^0(\tau)$, the relative error
$|\mathbf{x}(\tau)-\mathbf{x}^0(\tau)|/r$ will be unbounded with the increase of time.
Thus, perturbation solutions with the approximate $\mathbf{x}^0(\tau)$ are only valid
locally and will diverge with the increase of time. Since we have precisely measured
orbits $\mathbf{x}_o(\tau)$, $|\mathbf{x}(\tau)-\mathbf{x}_o(\tau)|/r$ will remain small.
Bearing in mind that the orbit of an LEO satellite can be geometrically measured at the
cm and/or even mm level of accuracy with GNSS, for simplicity, say 1 cm, and by assuming
that an LEO satellite is of altitude of 230 km, with 6371 km as the mean radius of the
Earth, then the relative error $|\mathbf{x}(\tau)-\mathbf{x}_o(\tau)|/r$ would be roughly
as small as $1.515\times 10^{-9}$, irrelevant to the length of an orbital arc. In this
case, the second and higher order terms can be negligible in the expansion of the
nonlinear integral equations (\ref{XSolutLocal}) around $\mathbf{x}_o(\tau)$.

To start with, we denote \beq \label{DeltaXMeasured} \mathbf{x}(\tau) =
\mathbf{x}_o(\tau) + \Delta \mathbf{x}(\tau). \eeq Unlike $\mathbf{x}^0(\tau)$,
$\mathbf{x}_o(\tau)$ is directly measured and does not satisfy the governing differential
equations (\ref{NewtonLAWFinal}); thus, we cannot simply replace $\mathbf{x}^0(\tau)$
with $\mathbf{x}_o(\tau)$ in section~\ref{LocalPerturbXV} to obtain the corresponding
quasi-linear and linear solutions for $\mathbf{x}(t)$ and $\mathbf{v}(t)$. Instead, we
linearize the nonlinear integral equations (\ref{XSolutLocal}) around
$\mathbf{x}_o(\tau)$ and obtain
 \alpheqn
\begin{eqnarray} \label{LinearizedSolT} \mathbf{x}_o(t) + \Delta \mathbf{x}(t) & = &
- \mathbf{x}_E^0(t) - \int_{t_0}^t\!\!\int_{t_0}^{\eta} \mathbf{A}^x(\mathbf{x}_o)\Delta
\mathbf{x}(\tau)d\tau d\eta   \nonumber \\
 &   & + \sum \limits^{N_{\scriptsize
\textrm{max}}} \limits_{l=2} \sum \limits_{m=0} \limits^l
\int_{t_0}^t\!\!\int_{t_0}^{\eta} C_{lm} \mathbf{p}_{lm}^c(\alpha_o(\tau), \zeta_o(\tau),
r_o(\tau))
d\tau d\eta  \nonumber \\
&  & + \sum \limits^{N_{\scriptsize \textrm{max}}} \limits_{l=2} \sum \limits_{m=0}
\limits^l \int_{t_0}^t\!\!\int_{t_0}^{\eta} C_{lm}
 \mathbf{A}_{lm}^c(\alpha_o(\tau), \zeta_o(\tau), r_o(\tau)) \Delta \mathbf{x}(\tau) d\tau d\eta
\nonumber \\
&   & + \sum \limits^{N_{\scriptsize \textrm{max}}} \limits_{l=2} \sum \limits_{m=0}
\limits^l \int_{t_0}^t\!\!\int_{t_0}^{\eta}S_{lm} \mathbf{p}_{lm}^s(\alpha_o(\tau),
\zeta_o(\tau), r_o(\tau))  d\tau d\eta \nonumber \\ &  & + \sum \limits^{N_{\scriptsize
\textrm{max}}} \limits_{l=2} \sum \limits_{m=0} \limits^l
\int_{t_0}^t\!\!\int_{t_0}^{\eta} S_{lm}
 \mathbf{A}_{lm}^s(\alpha_o(\tau), \zeta_o(\tau), r_o(\tau)) \Delta \mathbf{x}(\tau) d\tau d\eta
\nonumber \\
&   & + \mathbf{v}_0^0(t-t_0) + \Delta\mathbf{v}_0(t-t_0)+ \mathbf{x}_0^0 +
\Delta\mathbf{x}_0,
\end{eqnarray} where \beq \mathbf{x}_E^0(t) =
\int_{t_0}^t\!\!\int_{t_0}^{\eta}\frac{GM}{r^3_o(\tau)}\mathbf{x}_o(\tau) d\tau d\eta.
\eeq\reseteqn\setcounter{EQ10}{\value{equation}}Other notations have been defined as in
section~\ref{LocalPerturbXV} but have to be computed by replacing the nominal orbit $\{
\mathbf{x}^0(\tau)\,| \,t_0\leq \tau \leq t\}$ with the measured orbit $\{
\mathbf{x}_o(\tau)\,| \,t_0\leq \tau \leq t\}$.

As in the case of (\ref{LocalSolutQUASI}), by setting $\Delta \mathbf{x}({\tau})$ on the
right hand side of (\ref{LinearizedSolT}) to zero, we obtain the zeroth order
approximation or quasi-linear solution: \begin{eqnarray} \label{GlobalSolutQUASI} \Delta
\mathbf{x}(t) & = & \delta
\mathbf{x}_0(t) + \Delta\mathbf{v}_0(t-t_0) + \Delta\mathbf{x}_0    \nonumber \\
 &   & + \sum \limits^{N_{\scriptsize
\textrm{max}}} \limits_{l=2} \sum \limits_{m=0} \limits^l C_{lm}
\int_{t_0}^t\!\!\int_{t_0}^{\eta} \mathbf{p}_{lm}^c(\alpha_o(\tau), \zeta_o(\tau),
r_o(\tau))  d\tau d\eta  \nonumber \\
&   & + \sum \limits^{N_{\scriptsize \textrm{max}}} \limits_{l=2} \sum \limits_{m=0}
\limits^l S_{lm} \int_{t_0}^t\!\!\int_{t_0}^{\eta}\mathbf{p}_{lm}^s(\alpha_o(\tau),
\zeta_o(\tau), r_o(\tau))  d\tau d\eta,
\end{eqnarray} where  $$ \delta \mathbf{x}_0(t) = - \mathbf{x}_o(t) - \mathbf{x}_E^0(t) +
 \mathbf{v}_0^0(t-t_0) + \mathbf{x}_0^0. $$ Mathematically, the quasi-linear
solution (\ref{GlobalSolutQUASI}) is essentially equivalent to treating the measured
orbit $\mathbf{x}_o(\tau)$ as the true (and given) orbit and substituting the unknown
true orbit $\mathbf{x}(\tau)$ on the right hand side of (\ref{XSolutLocal}) with this
measured orbit $\mathbf{x}_o(\tau)$. Since all the integrals on the right hand side of
(\ref{XSolutLocal}) can be directly computed numerically with $\mathbf{x}_o(\tau)$, the
solution $\mathbf{x}(t)$ on the left hand side of (\ref{XSolutLocal}) can naturally be
represented in terms of the unknown harmonic coefficients $C_{lm}$ and $S_{lm}$. In other
words, the solution (\ref{GlobalSolutQUASI}) can be alternatively expressed as follows:
\begin{eqnarray} \label{GlobalSolutXQuasi} \mathbf{x}(t) & = & -
\int_{t_0}^t\int_{t_0}^{\eta} \frac{GM}{r^3_o(\tau)}\mathbf{x}_o(\tau)d\tau d\eta + \sum
\limits^{N_{\scriptsize \textrm{max}}} \limits_{l=2} \sum \limits_{m=0} \limits^l C_{lm}
\int_{t_0}^t\int_{t_0}^{\eta}  \mathbf{p}_{lm}^c(\alpha_o(\tau), \zeta_o(\tau),
r_o(\tau)) d\tau d\eta  \nonumber \\
 &  & + \sum \limits^{N_{\scriptsize \textrm{max}}} \limits_{l=2}
 \sum \limits_{m=0} \limits^l S_{lm} \int_{t_0}^t\int_{t_0}^{\eta}
\mathbf{p}_{lm}^s(\alpha_o(\tau), \zeta_o(\tau), r_o(\tau)) d\tau d\eta  +
\mathbf{v}_0(t-t_0) + \mathbf{x}_0.
\end{eqnarray}

In a similar manner, by inserting (\ref{GlobalSolutQUASI}) into (\ref{LinearizedSolT}),
we can then construct the linear perturbation solution. If we neglect the small terms
$C_{lm}\Delta \mathbf{x}(\tau)$ and $S_{lm}\Delta \mathbf{x}(\tau)$ on the right hand
side of (\ref{LinearizedSolT}), we can obtain the linear perturbation solution of $\Delta
\mathbf{x}(t)$ as follows: \begin{eqnarray} \label{GlobalLinearXFinal} \Delta
\mathbf{x}(t) & = & \mathbf{l}_0^x(t) + \mathbf{D}_v^x
\Delta\mathbf{v}_0 + \mathbf{D}_x^x \Delta\mathbf{x}_0 \nonumber \\
&   &  + \sum \limits^{N_{\scriptsize \textrm{max}}} \limits_{l=2} \sum \limits_{m=0}
\limits^l \mathbf{d}^{cx}_{lm} C_{lm} + \sum \limits^{N_{\scriptsize \textrm{max}}}
\limits_{l=2} \sum \limits_{m=0} \limits^l \mathbf{d}^{sx}_{lm} S_{lm},
\end{eqnarray} where $$ \mathbf{l}_0^x(t) = \delta
\mathbf{x}_0(t)  - \int_{t_0}^t\!\!\int_{t_0}^{\eta} \mathbf{A}^x(\mathbf{x}_o(\tau))
\delta \mathbf{x}_0(\tau) d\tau d\eta, $$ $\mathbf{D}_v^x$, $\mathbf{D}_x^x$,
$\mathbf{d}^{cx}_{lm}$ and $\mathbf{d}^{sx}_{lm}$ have been defined as in
(\ref{CoeffDvx}) to (\ref{Coeffdcxlm}), respectively, but with $\mathbf{x}^0(\tau)$ there
replaced by $\mathbf{x}_o(\tau)$ for use in (\ref{GlobalLinearXFinal}).

Instead of completely neglecting the terms $C_{lm}\Delta \mathbf{x}(\tau)$ and
$S_{lm}\Delta \mathbf{x}(\tau)$ altogether, since $C_{20}$ is larger than other harmonic
coefficients by an order of about $1,000$, one may like to consider the term
$C_{20}\Delta \mathbf{x}(\tau)$ to construct another linear perturbation solution. In
this case, this new linear perturbation solution with the term $C_{20}\Delta
\mathbf{x}(\tau)$ will become:
\begin{eqnarray} \label{GlobalLinearXFinalVersion2} \Delta
\mathbf{x}(t) & = & \mathbf{l}_0^x(t) + \mathbf{D}_v^x
\Delta\mathbf{v}_0 + \mathbf{D}_x^x \Delta\mathbf{x}_0 \nonumber \\
&   &  + \sum \limits^{N_{\scriptsize \textrm{max}}} \limits_{l=2} \sum \limits_{m=0}
\limits^l \mathbf{d}^{cx}_{lm} C_{lm} + \sum \limits^{N_{\scriptsize \textrm{max}}}
\limits_{l=2} \sum \limits_{m=0} \limits^l \mathbf{d}^{sx}_{lm} S_{lm} \nonumber \\
 &  & + C_{20} \int_{t_0}^t\!\!\int_{t_0}^{\eta}
 \mathbf{A}_{lm}^c(\alpha_o(\tau), \zeta_o(\tau), r_o(\tau)) \Big\{
\delta \mathbf{x}_0(\tau) + \Delta\mathbf{v}_0(\tau-t_0) + \Delta\mathbf{x}_0    \nonumber \\
 &   & + \sum \limits^{N_{\scriptsize
\textrm{max}}} \limits_{l=2} \sum \limits_{m=0} \limits^l
\int_{t_0}^{\tau}\!\!\int_{t_0}^{t_1} C_{lm} \mathbf{p}_{lm}^c(\alpha_o(t_2),
\zeta_o(t_2), r_o(t_2))  dt_2 dt_1  \nonumber \\
&   & + \sum \limits^{N_{\scriptsize \textrm{max}}} \limits_{l=2} \sum \limits_{m=0}
\limits^l \int_{t_0}^{\tau}\!\!\int_{t_0}^{t_1} S_{lm} \mathbf{p}_{lm}^s(\alpha_o(t_2),
\zeta_o(t_2), r_o(t_2))  dt_2 dt_1 \Big\} d\tau d\eta \nonumber \\
& = & \mathbf{l}_0^x(t) + \mathbf{D}_v^x
\Delta\mathbf{v}_0 + \mathbf{D}_x^x \Delta\mathbf{x}_0 \nonumber \\
&   &  + \sum \limits^{N_{\scriptsize \textrm{max}}} \limits_{l=2} \sum \limits_{m=0}
\limits^l \mathbf{d}^{cx}_{lm} C_{lm} + \sum \limits^{N_{\scriptsize \textrm{max}}}
\limits_{l=2} \sum \limits_{m=0} \limits^l \mathbf{d}^{sx}_{lm} S_{lm} \nonumber \\
&  & + C_{20} \int_{t_0}^t\!\!\int_{t_0}^{\eta} \mathbf{A}_{lm}^c(\alpha_o(\tau),
\zeta_o(\tau), r_o(\tau))\delta \mathbf{x}_0(\tau) d\tau d\eta \nonumber \\
 &   & + C_{20}\Delta\mathbf{v}_0 \int_{t_0}^t\!\!\int_{t_0}^{\eta}
\mathbf{A}_{lm}^c(\alpha_o(\tau), \zeta_o(\tau),
r_o(\tau))(\tau-t_0) d\tau d\eta \nonumber \\
 &  &  + C_{20}\Delta\mathbf{x}_0
\int_{t_0}^t\!\!\int_{t_0}^{\eta} \mathbf{A}_{lm}^c(\alpha_o(\tau), \zeta_o(\tau),
r_o(\tau)) d\tau d\eta \nonumber \\
 &  & + C_{20} \int_{t_0}^t\!\!\int_{t_0}^{\eta}
 \mathbf{A}_{lm}^c(\alpha_o(\tau),
\zeta_o(\tau), r_o(\tau)) \Big\{
 \sum \limits^{N_{\scriptsize
\textrm{max}}} \limits_{l=2} \sum \limits_{m=0} \limits^l
\int_{t_0}^{\tau}\!\!\int_{t_0}^{t_1} C_{lm} \mathbf{p}_{lm}^c(\alpha_o(t_2),
\zeta_o(t_2), r_o(t_2))  dt_2 dt_1  \nonumber \\
&   & + \sum \limits^{N_{\scriptsize \textrm{max}}} \limits_{l=2} \sum \limits_{m=0}
\limits^l \int_{t_0}^{\tau}\!\!\int_{t_0}^{t_1} S_{lm} \mathbf{p}_{lm}^s(\alpha_o(t_2),
\zeta_o(t_2), r_o(t_2)) dt_2 dt_1 \Big\} d\tau d\eta.
\end{eqnarray}
We should note that although (\ref{GlobalLinearXFinalVersion2}) is a linear perturbation
solution to the nonlinear Volterra's integral equations (\ref{XSolutLocal}) of the second
kind, it is clearly nonlinear with respect to the unknown parameters, namely,
$\Delta\mathbf{v}_0$, $\Delta\mathbf{x}_0$, and the harmonic coefficients $C_{lm}$ and
$S_{lm}$.

With the measured orbit $\mathbf{x}_o(\tau)$ in hands, we can also construct the global
uniformly convergent quasi-linear and linear solutions to the velocity of satellite
motion, which can be obtained by using the same approach as in deriving the solutions
(\ref{GlobalSolutQUASI}), (\ref{GlobalLinearXFinal}) and
(\ref{GlobalLinearXFinalVersion2}). More precisely, by substituting the unknown (true)
orbit $\mathbf{x}(\tau)$ (and equivalently, $[\alpha(\tau),\zeta(\tau), r(\tau)]$) on the
right hand side of (\ref{VSolutLocal}) with the measured orbit $\mathbf{x}_o(\tau)$, we
can readily construct the quasi-linear perturbation solution of the velocity of satellite
motion as follows:
\begin{eqnarray} \label{VGlobalSolutQuasi} \mathbf{v}(t) & = & -
\int_{t_0}^t \frac{GM}{r^3_o(\tau)}\mathbf{x}_o(\tau)d\tau  + \sum
\limits^{N_{\scriptsize \textrm{max}}} \limits_{l=2} \sum \limits_{m=0} \limits^l
 C_{lm} \int_{t_0}^t \mathbf{p}_{lm}^c(\alpha_o(\tau),
\zeta_o(\tau), r_o(\tau)) d\tau \nonumber \\
 &  & + \sum \limits^{N_{\scriptsize
\textrm{max}}} \limits_{l=2} \sum \limits_{m=0} \limits^l S_{lm} \int_{t_0}^t
\mathbf{p}_{lm}^s(\alpha_o(\tau), \zeta_o(\tau), r_o(\tau)) d\tau  + \mathbf{v}_0^0 +
\Delta \mathbf{v}_0. \end{eqnarray} The quasi-linear solution (\ref{VGlobalSolutQuasi})
is obviously global uniformly convergent, since the measured $\mathbf{x}_o(\tau)$
($t_0\leq \tau \leq t$) is a (precisely measured) realization of the unknown, true orbit
$\mathbf{x}(\tau)$ ($t_0\leq \tau \leq t$), no matter how lengthy the arc of orbit is.

In a similar manner, we can linearize the integral equations (\ref{VSolutLocal}) around
the measured orbit $\mathbf{x}_o(\tau)$ ($t_0\leq \tau \leq t$), substitute the
incremental $\Delta \mathbf{x}(\tau)$ with the quasi-linear solution
(\ref{GlobalSolutQUASI}),  neglect the terms of $C_{lm}\Delta \mathbf{x}(\tau)$ and
$S_{lm}\Delta \mathbf{x}(\tau)$, and finally obtain the linear perturbation solution of
the velocity as follows:
\begin{eqnarray}
\label{VGlobalSolutLinear} \mathbf{v}(t)  = \mathbf{l}^v_0(t) + \mathbf{D}_v^v \Delta
\mathbf{v}_0 + \mathbf{D}_x^v \Delta\mathbf{x}_0 + \sum \limits^{N_{\scriptsize
\textrm{max}}} \limits_{l=2} \sum \limits_{m=0} \limits^l \mathbf{d}^{cv}_{lm} C_{lm} +
\sum \limits^{N_{\scriptsize \textrm{max}}} \limits_{l=2} \sum \limits_{m=0} \limits^l
\mathbf{d}^{sv}_{lm} S_{lm},
\end{eqnarray} where
 $$ \mathbf{l}^v_0(t) = - \int_{t_0}^t \frac{GM}{r^3_o(\tau)}\mathbf{x}_o(\tau)d\tau -
\int_{t_0}^t \mathbf{A}^x(\mathbf{x}_o(\tau)) \delta \mathbf{x}_0(\tau) d\tau +
\mathbf{v}_0^0, $$ the coefficient vectors $\mathbf{D}_v^v$,
 $\mathbf{D}_x^v$, $\mathbf{d}^{cv}_{lm}$ and $\mathbf{d}^{sv}_{lm}$ have been defined as
 in (\ref{coeffLocalDvv}) to (\ref{coeffLocaldsvlm}), respectively, but computed with the
measured orbit $\mathbf{x}_o(\tau)$ instead of the approximate nominal orbit
$\mathbf{x}^0(\tau)$. If one would be interested in constructing the linear solution of
the velocity with the term $C_{20}\Delta \mathbf{x}(\tau)$, one can follow the same
approach as in the derivation of (\ref{GlobalLinearXFinalVersion2}), which is omitted
here, nevertheless.

In the previous two sections, i.e. sections~\ref{AppromAnalNumINT} and
\ref{GlobalAnalIntEQ}, we have solved for the orbital position and velocity solutions to
the Newton's governing differential equations (\ref{NewtonLAWFinal}) of satellite motion
and represented them in terms of the unknown equation parameters and the unknown initial
conditions, given either the nominal reference orbit $\mathbf{x}^0(\tau)$ ($t_0\leq \tau
\leq t$) or the measured orbit $\mathbf{x}_o(\tau)$ ($t_0\leq \tau \leq t$). The derived
orbital and velocity solutions can then be used to establish the links between satellite
tracking measurements and the unknown gravitational parameters and the corrections to the
approximate initial values. For more details on observational equations of space
measurements, including satellite tracking measurements and satellite-to-satellite
tracking measurements, the reader is referred to Kaula (1961a, 1966), Lerch et al.
(1974), Long et al. (1989) and Xu (2008). By directly applying the least squares
principle to the (nonlinear and/or linearized) observational equations, one can obtain
the optimal estimate of the Earth's gravitational field. We may like to point out that in
the case of quasi-linear perturbation with measured orbits, no iteration will be needed
in the least squares estimation of the unknown gravitational parameters, since the
perturbation solutions are linear with respect to the gravitational parameters and global
uniformly convergent. Actually, we even do not need an initial force model, namely,
initial approximate values of the unknown parameters $\mathbf{p}$.

\section{Measurement-based condition adjustment with parameters}
In this section, we will briefly outline an alternative method to estimate the
gravitational field from measured orbits and satellite tracking measurements. The basic
idea now is to treat the integral equations as natural equality constraints on the
expectations of measurements and the unknown parameters, namely, the unknown harmonic
coefficients and the unknown initial condition values. These equality constraints will
automatically become a standard model of condition equations with unknown parameters,
which can then naturally be solved by using the condition (LS) adjustment with
parameters.

To start with, let us again assume the measured orbit $\mathbf{x}_o(\tau)$ ($t_0\leq \tau
\leq t$) and a number of satellite tracking measurements $y_i$. These tracking
measurements are assumed to be collected at different time epochs of
$(t_{y1},t_{y2},...\,,t_{yn})\in [t_0,\, t]$ and, each of $y_i$ is assumed, without loss
of generality, to be the function of the satellite position (and likely, also velocity)
at this particular epoch $t_{yi}$. If a measurement is involved with more than one LEO
satellite, then the corresponding measurement is the function of the positions and
velocities of all these satellites.

Under the above assumptions, we can rewrite the true position $\mathbf{x}(\tau)$ of a
satellite as the measured position $\mathbf{x}_o(\tau)$ plus its correction
$\bm{\xi}_{x\tau}$, namely, \beq \label{condEQPosition} \mathbf{x}(\tau) =
\mathbf{x}_o(\tau) + \bm{\xi}_{x\tau}. \eeq If the velocity of the satellite is not
directly measured, then we can treat the velocity as an unknown vector. Thus, we have the
equality constraint with unknowns as follows:
\begin{eqnarray} \label{GlobalV1EConstrained} \mathbf{v}(t_{yi}) & = & -
\int_{t_0}^{t_{yi}} \frac{GM}{r^3_{o\xi}}[\mathbf{x}_o(\tau)+\bm{\xi}_{x\tau}]d\tau +\sum
\limits^{N_{\scriptsize \textrm{max}}} \limits_{l=2} \sum \limits_{m=0} \limits^l C_{lm}
\int_{t_0}^{t_{yi}}  \mathbf{p}_{lm}^c(\alpha_{o\xi},
\zeta_{o\xi}, r_{o\xi}) d\tau  \nonumber \\
 &  & + \sum \limits^{N_{\scriptsize \textrm{max}}} \limits_{l=2} \sum \limits_{m=0}
  \limits^l S_{lm} \int_{t_0}^{t_{yi}}
\mathbf{p}_{lm}^s(\alpha_{o\xi}, \zeta_{o\xi}, r_{o\xi}) d\tau  + \mathbf{v}_0,
\end{eqnarray} where $[\alpha_{o\xi}, \zeta_{o\xi}, r_{o\xi}]$ are transformed from
$[\mathbf{x}_o(\tau) + \bm{\xi}_{x\tau}]$. In practice, $\mathbf{x}_o(\tau)$ ($t_0\leq
\tau \leq t$) is only given in a densely discrete format. For simplicity, we assume that
the orbit is sampled with an equal interval and denoted by $\mathbf{x}_o(t_i)$ ($0\leq i
\leq n_t$). Thus, all the integrals on the right hand side of
(\ref{GlobalV1EConstrained}) can only be computed numerically by using numerical
integration rules such as Newton-Cotes formulae or Gaussian integration rules. Since
$\bm{\xi}_{x\tau}$ are small at the level of measurement noise, if we neglect all the
terms of $C_{lm}\bm{\xi}_{x\tau}$ and $S_{lm}\bm{\xi}_{x\tau}$, the linearized version of
(\ref{GlobalV1EConstrained}) should then be equivalently written in the discretized form
as follows:
\begin{eqnarray} \label{GlobalV1EConstrainedFinal} \mathbf{v}(t_{yi}) & = & -
\int_{t_0}^{t_{yi}} \frac{GM}{r^3_o}\mathbf{x}_o(\tau)d\tau - \sum
\limits_{j=0}\limits^{m_{yi}} w_j\mathbf{A}^x(\mathbf{x}_o(t_j))\bm{\xi}_{xt_j} +\sum
\limits^{N_{\scriptsize \textrm{max}}} \limits_{l=2} \sum \limits_{m=0} \limits^l C_{lm}
\int_{t_0}^{t_{yi}}
\mathbf{p}_{lm}^c(\alpha_o(\tau), \zeta_o(\tau), r_o(\tau)) d\tau  \nonumber \\
 &  & + \sum \limits^{N_{\scriptsize \textrm{max}}} \limits_{l=2} \sum \limits_{m=0}
  \limits^l S_{lm} \int_{t_0}^{t_{yi}}
\mathbf{p}_{lm}^s(\alpha_o(\tau), \zeta_o(\tau), r_o(\tau)) d\tau  +
\mathbf{v}_0^0+\Delta \mathbf{v}_0,
\end{eqnarray} where $w_j$ are positive coefficients, which are given, depending solely on
the chosen rule of numerical integration. We still keep some integral notations in
(\ref{GlobalV1EConstrainedFinal}), mainly to emphasize that the integration rules for
$\bm{\xi}_{xt_j}$ can be different from those integrals without $\bm{\xi}_{xt_j}$. For
more details on numerical integration, the reader is referred to Phillips and Taylor
(1996) and Stoer and Burlirsch (2002).

If the velocity of the satellite is also measured, then (\ref{GlobalV1EConstrainedFinal})
should be replaced by
\begin{eqnarray} \label{GlobalV2EConstrainedFinal} \mathbf{v}_o(t_{yi}) + \bm{\xi}_{vt_{yi}} & = & -
\int_{t_0}^{t_{yi}} \frac{GM}{r^3_o}\mathbf{x}_o(\tau)d\tau - \sum
\limits_{j=0}\limits^{m_{yi}}
w_j\mathbf{A}^x(\mathbf{x}_o(t_j))\bm{\xi}_{xt_j} \nonumber \\
 &  & + \sum \limits^{N_{\scriptsize \textrm{max}}} \limits_{l=2} \sum \limits_{m=0}
 \limits^l C_{lm} \int_{t_0}^{t_{yi}} \mathbf{p}_{lm}^c(\alpha_o(\tau),
 \zeta_o(\tau), r_o(\tau)) d\tau  \nonumber \\
 &  & + \sum \limits^{N_{\scriptsize \textrm{max}}} \limits_{l=2} \sum \limits_{m=0}
  \limits^l S_{lm} \int_{t_0}^{t_{yi}}
\mathbf{p}_{lm}^s(\alpha_o(\tau), \zeta_o(\tau), r_o(\tau)) d\tau  +
\mathbf{v}_0^0+\Delta \mathbf{v}_0,
\end{eqnarray} where $\mathbf{v}_o(t_{yi})$ and $\bm{\xi}_{vt_{yi}}$ stand for the
measurements of the velocity and the corrections at the time epoch $t_{yi}$,
respectively.

For the measured orbital position $\mathbf{x}_o(t_i)$, we have the starting condition
equations:
\begin{eqnarray} \label{GlobalXEConstrained} \mathbf{x}_o(t_i) + \bm{\xi}_{xt_{i}}  & = & -
\int_{t_0}^{t_i}\int_{t_0}^{\eta}
\frac{GM}{r^3_{o\xi}}[\mathbf{x}_o(\tau)+\bm{\xi}_{x\tau}]
d\tau d\eta \nonumber \\
 &  & + \sum \limits^{N_{\scriptsize \textrm{max}}} \limits_{l=2} \sum
\limits_{m=0} \limits^l C_{lm} \int_{t_0}^{t_i}\int_{t_0}^{\eta}
\mathbf{p}_{lm}^c(\alpha_{o\xi},
\zeta_{o\xi}, r_{o\xi}) d\tau d\eta  \nonumber \\
 &  & + \sum \limits^{N_{\scriptsize \textrm{max}}} \limits_{l=2}
 \sum \limits_{m=0} \limits^l S_{lm} \int_{t_0}^{t_i}\int_{t_0}^{\eta}
\mathbf{p}_{lm}^s(\alpha_{o\xi}, \zeta_{o\xi}, r_{o\xi}) d\tau d\eta  +
\mathbf{v}_0(t_i-t_0) + \mathbf{x}_0.
\end{eqnarray} As in the case of velocity, if we neglect all the
terms of $C_{lm}\bm{\xi}_{x\tau}$ and $S_{lm}\bm{\xi}_{x\tau}$, then we can linearize the
equality condition equations (\ref{GlobalXEConstrained}) and obtain the final linearized
condition equations as follows:
\begin{eqnarray} \label{GlobalXEConstrainedFinal} \mathbf{x}_o(t_{i}) + \bm{\xi}_{xt_{i}}  & = & -
\int_{t_0}^{t_i}\int_{t_0}^{\eta} \frac{GM}{r^3_{o}(\tau)}\mathbf{x}_o(\tau)
d\tau d\eta \nonumber \\
 &  & - \sum \limits_{j=0}\limits^{i} w_j \sum \limits_{k=0}\limits^{j}
w_k\mathbf{A}^x(\mathbf{x}_o(t_k))\bm{\xi}_{xt_k} \nonumber \\
 &  & + \sum \limits^{N_{\scriptsize \textrm{max}}} \limits_{l=2} \sum
\limits_{m=0} \limits^l C_{lm} \int_{t_0}^{t_i}\int_{t_0}^{\eta}
\mathbf{p}_{lm}^c(\alpha_o(\tau), \zeta_o(\tau), r_o(\tau)) d\tau d\eta  \nonumber \\
 &  & + \sum \limits^{N_{\scriptsize \textrm{max}}} \limits_{l=2}
 \sum \limits_{m=0} \limits^l S_{lm} \int_{t_0}^{t_i}\int_{t_0}^{\eta}
\mathbf{p}_{lm}^s(\alpha_o(\tau), \zeta_o(\tau), r_o(\tau)) d\tau d\eta  \nonumber \\
 &  & + \mathbf{v}_0^0(t_i-t_0) + \Delta \mathbf{v}_0(t_i-t_0) + \mathbf{x}_0^0 + \Delta \mathbf{x}_0.
\end{eqnarray}

Satellite tracking measurements $\mathbf{y}$ of all types are functions of satellite
positions and velocities. For ground-based tracking systems, tracking measurements are
also functions of the positions of ground tracking stations, which are often assumed to
be given. Since ground stations are actually derived {\em a priori}, they may also be
treated as pseudo-measurements with random errors in satellite tracking systems. Thus, in
its most general form, a tracking measurement $y_i$ at the time epoch $t_{yi}$ must
theoretically satisfy the following physical and/or geometrical constraint, which can be
symbolically written as: \beq \label{SatTrackingOBS} E(y_i) = f(\mathbf{x}^s(t_{yi}),
\mathbf{v}^s(t_{yi}), \mathbf{x}^g), \eeq where $E(y_i)$ stands for the theoretical value
of the measurement $y_i$ (without biases), $\mathbf{x}^s(t_{yi})$ for the true position
of the satellite, $\mathbf{v}^s(t_{yi})$ for the true velocity of the satellite,
$\mathbf{x}^g$ for the true position of a ground tracking station, and $f(\cdot)$ is a
nonlinear functional, as defined in section~\ref{SectNumMeth}. By replacing the
theoretical/true values of the quantities in (\ref{SatTrackingOBS}) with the
corresponding measurements plus corrections, we can readily turn the theoretical
constraint (\ref{SatTrackingOBS}) into a condition equation. For example, let us assume
that except for $\mathbf{v}^s(t_{yi})$, all the other quantities in
(\ref{SatTrackingOBS}) are measured (and/or known {\em a priori} with random errors). As
a result, we have the nonlinear condition equation: \beq \label{SatTrackingCondOBS} y_i +
\xi_{yi} = f(\mathbf{x}^s_o(t_{yi})+\bm{\xi}^s_{xt_{yi}}, \mathbf{v}^s(t_{yi}),
\mathbf{x}^g_o+\bm{\xi}^g_x), \eeq where $\xi_{yi}$ stands for the correction to $y_i$,
$\bm{\xi}^s_{xt_{yi}}$ for the corrections to the satellite orbital coordinates
$\mathbf{x}^s_o(t_{yi})$, $\bm{\xi}^g_x$ for the corrections to the {\em a priori}
coordinates $\mathbf{x}^g_o$ of the ground tracking station. If the velocity of the
satellite is also measured, then we need to replace $\mathbf{v}^s(t_{yi})$ with
$[\mathbf{v}^s_o(t_{yi})+\bm{\xi}^s_{vt_{yi}}]$. Very often, one would go ahead to
linearize (\ref{SatTrackingCondOBS}), which is technically straightforward and will be
omitted here.

If a tracking measurement is involved with two satellites, then (\ref{SatTrackingOBS})
can be alternatively written as follows: \beq \label{SatSatTrackingOBS} E(y_i) =
f(\mathbf{x}^{s1}(t_{yi}), \mathbf{v}^{s2}(t_{yi}), \mathbf{x}^{s2}(t_{yi}),
\mathbf{v}^{s2}(t_{yi})), \eeq where the superscripts {\em s1} and {\em s2} stand for
satellites 1 and 2, respectively. By replacing the theoretical values with the
corresponding measurements plus their corrections, we can construct the corresponding
condition equation for (\ref{SatSatTrackingOBS}) as follows: \beq
\label{SatSatTrackingCondOBS} y_i + \xi_{yi} =
f(\mathbf{x}^{s1}_o(t_{yi})+\bm{\xi}^{s1}_{xt_{yi}}, \mathbf{v}^{s1}(t_{yi}),
\mathbf{x}^{s2}_o(t_{yi})+\bm{\xi}^{s2}_{xt_{yi}}, \mathbf{v}^{s2}(t_{yi})), \eeq if both
$\mathbf{v}^{s1}(t_{yi})$ and $\mathbf{v}^{s2}(t_{yi})$ are unknown and not measured. If
one or both of them are measured, we need to replace them with their corresponding
measurements plus their corrections in (\ref{SatSatTrackingCondOBS}).

Actually, the condition equations (\ref{GlobalV1EConstrainedFinal}),
(\ref{GlobalV2EConstrainedFinal}), (\ref{GlobalXEConstrainedFinal}),
(\ref{SatTrackingCondOBS}) and (\ref{SatSatTrackingCondOBS}) apply to all types of
satellite tracking measurements, including but not limited to orbital position
measurements, satellite velocity measurements, Doppler measurements, directional
measurements, ranges and range rates. By collecting the (linearized) condition equations
for all tracking measurements of any types together, and by collecting all the
corrections to measurements in the vector $\bm{\xi}$ and all the unknown parameters such
as the unknown harmonic coefficients and the corrections to initial satellite position
and velocity in the vector $\bm{\beta}$, we can symbolically write the final linearized
condition equations for all measurements as follows: \beq \label{ALLSatGTrackingOBS}
\mathbf{A}\bm{\xi} + \mathbf{B}\bm{\beta} + \mathbf{u} = \mathbf{0}, \eeq where
$\mathbf{A}$ and $\mathbf{B}$ are the known coefficient matrices, and $\mathbf{u}$ is the
misclosure vector of measurements. If we further assume that the satellite tracking
measurements $\mathbf{y}$ have the weighting matrix $\mathbf{W}$, then we can finally
estimate the Earth's gravitational model, together with other nuisance unknown
parameters, by solving the following minimization problem: \beq \label{MEarthCondLS}
\textrm{min:} \,\,\, \bm{\xi}^T \mathbf{W} \bm{\xi} \eeq subject to the equality
constraints (\ref{ALLSatGTrackingOBS}). If different types of satellite tracking
measurements are used to reconstruct the Earth's gravitational field, we may also have to
simultaneously estimate the different weighting factors of the measurements for this
typical kind of ill-posed inverse problems in Earth Sciences (see e.g., Xu et al. 2006;
Xu 2009b).

\section{Concluding remarks}
Differential equations with unknown parameters and the derived differential equations of
the partial derivatives with respect to the unknown parameters were originally published
by Gronwall (1919) and Ritt (1919) almost 100 years ago (see also Goddington and Levinson
1955; Howland and Vaillancourt 1961), which have been widely used for reconstruction of
the unknown parameters of differential equations from measurements, with applications in
many areas of science and engineering such as statistics, chemistry, physics, and
satellite gravimetry (see e.g., Lerch et al. 1974; Dickinson et al. 1976; Hwang et al.
1978; Long et al. 1989; Reigber 1989; Montenbruck and Gill 2000; Linga et al. 2006;
Ramsay et al. 2007; Wang and Enright 2013). The method was, likely independently,
re-discovered by Anderle (1965b) and Riley et al. (1967), now best known as the numerical
integration method in geodesy and aerospace engineering, and has since become a standard
technique in satellite gravimetry and been widely used routinely by major institutions
worldwide to produce global  gravitational models from satellite tracking measurements of
CHAMP and/or GRACE types. A precise gravitational model can serve as a precise global
static vertical datum surface (i.e. the geoid) in geodesy. Time-varying gravitational
models from CHAMP and/or GRACE tracking measurements have found widest possible
multidisciplinary applications in environmental monitoring, continental water variation,
seismology, the structure and dynamics of the core and mantle, and ocean dynamics (see
e.g., Nerem et al. 1995; NRC 1997; Wahr et al. 1998; Dickey 2000; Tapley et al. 2004a).
The most important element of the numerical integration method is to solve the partial
differential equations (\ref{PartialCS}) with the assumption of zero initial values (see
e.g., Gronwall 1919; Goddington and Levinson 1955; Howland and Vaillancourt 1961; Riley
et al. 1967; Lerch et al. 1974; Dickinson et al. 1976; Hwang et al. 1978; Ballani 1988;
Long et al. 1989; Reigber 1989; Montenbruck and Gill 2000; Linga et al. 2006; Ramsay et
al. 2007; Beutler et al. 2010; Wang and Enright 2013). Although the equations
(\ref{PartialCS}) are mathematically derived rigorously from the original Newton's
governing differential equations (\ref{EarthNewtonLAW}), the zero initial values cannot
be derived from (\ref{EarthNewtonLAW}) but are a claim without any mathematical/physical
support.

We have proved that the numerical integration method is groundless, mathematically and
physically. From the mathematical point of view, we have readily constructed counter
examples to invalidate the assumption of zero initial values for the partial derivatives.
Since an orbital position is nothing more than a mathematical point of the general
solution, and since any epoch can serve as an initial epoch, if initial values of the
partial derivatives in (\ref{PartialCS}) could be set to zero, then all the partial
derivatives could be logically set to zero as well. From the physical point of view, if
the initial values of the partial derivatives in (\ref{PartialCS}) could be set to zero,
this would imply that satellite tracking measurements would not contain any information
on the Earth's gravitational field; this certainly contradicts with the fact that
satellite tracking measurements indeed contain the physical information on the Earth's
gravitational field and can be used to determine it. The effect of incorrectly setting
the initial partial derivatives to zero on gravitational models produced by major
institutions worldwide for the geoscience community remains unclear and should be further
investigated in the future.

Given differential equations with unknown parameters and unknown initial conditions, and
assuming a nominal reference orbit, we have developed three different methods, namely,
linearization of the original differential equations (\ref{EarthNewtonLAW}), Euler and
modified Euler numerical integration methods with the unknown differential equation
parameters and unknown initial conditions, and the integral equation approach, to derive
local solutions to the differential equations (\ref{EarthNewtonLAW}), which, together
with satellite tracking measurements, can be used to estimate the unknown differential
equation parameters and unknown initial conditions, and as a result, to reconstruct
global gravitational models. Unlike the numerical integration method, these new solutions
require neither the differential equations of the partial derivatives nor the incorrect
assumption of zero initial values for the partial derivatives. The solutions are
represented in terms of the unknown corrections of parameters and initial conditions and
are said to be local, since they are valid in the neighbourhood of a nominal reference
orbit. The modelling errors will increase with time. In this case, if an orbital arc is
sufficiently lengthy, one will have to iteratively solve for the unknown gravitational
parameters. Deterministic global optimization methods can also be used to find the
optimal solution.

Orbits of LEO satellites can now be measured precisely and almost continuously, thanks to
the profound advance of space observation technology and GNSS receiver hardware. Modern
and next generation of space observation will become even more precise to an
unprecedented level. With precisely and almost continuously measured orbits of LEO
satellites, we have developed the measurement-based perturbation theory by turning the
nonlinear differential equations (\ref{EarthNewtonLAW}) into the nonlinear Volterra's
integral equations of the second kind and linearizing the nonlinear integral equations
with precisely measured orbits of satellites. As a result, we have constructed different
global uniformly convergent solutions. Theoretically speaking, the global uniformly
convergent solutions are able to fully use unprecedented  accuracy and continuity of
modern and next generation of space observation and are a mathematical guarantee to
extract smallest possible gravitational signals from satellite tracking measurements to
their technological limit of noise level. Thus, the global uniformly convergent solutions
can be used for high-precision high-resolution mapping of the Earth's gravitational field
from satellite tracking measurements. One more important advantage of measurement-based
perturbation theory is that no iteration will be needed to estimate the unknown
gravitational parameters from satellite tracking measurements, if the quasi-perturbation
solutions of position and velocity are used, because these solutions are global uniformly
convergent and linear with respect to the gravitational parameters. In this case, we do
not need any initial approximate values of the harmonic coefficients $\mathbf{p}$ either.
With precisely measured orbits of LEO satellites, we have also reformulated the
determination of the Earth's gravitational field from satellite tracking measurements as
a standard condition adjustment with unknown parameters.  \vspace{3mm} \\
{\bf Acknowledgement:} I thank Prof. Lars E.
Sj\"{o}berg very much for bringing the work of Bjerhammer (1967) to my attention and for
scanning and sending it to me. I also thank a reviewer for the very constructive comments
on the general relativistic (Lense-Thirring) effect.

\section*{References}
\bl
 \ite Anderle, R.J., 1965a.  Observations of resonance effects on satellite orbits arising from
 the thirteenth- and fourteenth-order tesseral gravitational coefficients, {\em J. geophys. Res.},
 {\bf 70}, 2453-2458. \\[-6mm]
 \ite Anderle, R.J., 1965b. Geodetic parameter set NWL-5E-6 based on Doppler satellite
 observations,  NWL Report No.1978, {\em U.S. Naval Weapons Laboratory}, Dahlgren,
 Virginia. \\[-6mm]
 \ite Ballani, L., 1988. Partielle Ableitungen und Variationsgleichungen zur Modellierung von
Satellitenbahnen und Parameterbestimmung. {\em Vermessungstechnik}, {\bf 36}, 192-194.
(Partial derivatives and variational equations for the modelling of satellite
orbits and parameter estimation) \\[-6mm]
 \ite Bender, P.L., Hall, J.L., Ye, J. \& Klipstein, W.M., 2003.
Satellite-satellite laser links for future gravity missions, {\em Space Sci.
Rev.}, {\bf 108}, 377-384. \\[-6mm]
 \ite Beutler, G., Jaggi, A., Mervart, L., \& Meyer, U., 2010. The celestial mechanics approach:
 theoretical foundations, {\em J. Geod.}, {\bf 84}, 605-624. \\[-6mm]
 \ite Biancale, R.,  Balmino, G., Lemoine, J., Marty, J., Moynot, B., Barlier, F., Exertier, P.,
 Laurain, O., Gegout, P., Schwintzer, P., Reigber, C., Bode, A., Konig, R., Massmann, F., Raimondo, J.,
 Schmidt, R. \& Zhu, S., 2000. A new global Earth's gravity field model from satellite
 orbit perturbations: GRIM5-S1, {\em Geophys. Res. Lett.}, {\bf 27}, 3611-3614. \\[-6mm]
 \ite Bjerhammer, A., 1967. {\em On the Energy Integral for Satellites}, Internal Report,
 Geodesy, The Royal Institute of Technology, Stockholm. \\[-6mm]
  \ite Bjerhammer, A., 1969. On the energy integral for satellites, {\em Tellus}, {\bf 21},
  1-9.  \\[-6mm]
 \ite Blitzer, L. \& Anderson, J.D., 1981. Theory of satellite orbit-orbit resonance, {\em
 Celest. Mech.}, {\bf 29}, 65-78. \\[-6mm] 
 \ite Bezd\v{e}k A., Sebera, J., Kloko\v{c}n\'{i}k, J., Kosteleck\'{y}, J., 2014. Gravity field
 models from kinematic orbits of CHAMP, GRACE and GOCE satellites, {\em Adv. Space Res.},
 {\bf 53}, 412-429. \\[-6mm]
 \ite Brouwer, D., 1959. Solution of the problem of artificial satellite theory without
 drag, {\em Astron. J.}, {\bf 64}, 378-396. \\[-6mm]
  \ite Brouwer, D. \& Clemence, G.M., 1961. {\em Methods of Celestial
Mechanics}, Academic Press, New York. \\[-6mm]
 \ite Buchar, E., 1958. Motion of the nodal line of the second Russian Earth satellite
 (1957$\beta$) and the flattening of the Earth, {\em Nature}, {\bf 182}, 198-199.
 \\[-6mm]  
 \ite Cary, J.R., 1981. Lie transform perturbation theory for Hamiltonian systems, {\em
 Physics Reports}, {\bf 79}, 129-159. \\[-6mm] 
   \ite Christophe, B., Boulanger, D., Foulon, B., Huynh,  P.-A., Lebat, V., Liorzou,  F.,
 Perrot, E. 2015. A new generation of ultra-sensitive electrostatic accelerometers for
 GRACE Follow-on and towards the next generation gravity missions, {\em Acta Astronaut.},
 {\bf 117}, 1-7. \\[-6mm]
 \ite Cook, A.H., 1961. Resonant orbits of artificial satellites and longitude terms in the Earth's
  external gravitational potential, {\em Geophys. J. Roy. astr. Soc.}, {\bf 4}, 53-72. \\[-6mm]
   \ite Cook, A.H., 1963. The contribution of observations of satellites to the
   determination of the Earth's gravitational potential, {\em Space Sci. Rev.},
{\bf 2}, 355-437. \\[-6mm]
  \ite Cook, A.H., 1967. Determination of the Earth's gravitational field from satellite
  orbits: methods and results, {\em Phil. Trans. Roy. Soc. London A: Math. Phys. Sci.}, {\bf
  262}, 119-132. \\[-6mm]
 \ite Dickey, J.O., 2000. Time variable gravity: an emerging frontier in interdisciplinary
 geodesy, in: {\em Gravity, Geoid, and Geodynamics 2000}, edited by M Sideris, Springer,
 New York, pp.1-5. \\[-6mm]
  \ite Dickinson, R.P., \&  Gelinas, R.J., 1976. Sensitivity analysis of ordinary
differential equation systems -- a direct method, {\em J. comput. Phys.}, {\bf 21},
123-143. \\[-6mm]
 \ite Ditmar, P., van der Sluijs, A.A., 2004. A technique for modeling the Earth's gravity field
 on the basis of satellite accelerations, {\em J. Geod.}, {\bf 78}, 12-33. \\[-6mm]
   \ite  Flechtner, F., Morton, P., Watkins, M., \& Webb, F. 2014. Status of the GRACE Follow-On
 Mission. In: Marti U. (eds) Gravity, Geoid and Height Systems.
 International Association of Geodesy Symposia, vol 141, Springer, Berlin, pp.117-121.
 \\[-6mm]
  \ite Gaposchkin, E.M., 1974. Earth's gravity field to the eighteenth degree and geocentric coordinates
  for 104 stations from satellite and terrestrial data, {\em J. geophys. Res.}, {\bf
  79}, 5377-5411. \\[-6mm]
  \ite Gaposchkin, E.M. \& Lambeck, K., 1970. {\em 1969 Smithsonian Standard Earth (II)}, SAO Special
  Report No.315, Smithsonian Institution Astrophysical Observatory, Cambridge. \\[-6mm]
  \ite Gaposchkin, E.M. \& Lambeck, K., 1971. Earth's gravity field to the sixteenth degree and
  station coordinates from satellite and terrestrial data, {\em J. geophys. Res.}, {\bf
  76}, 4855-4883. \\[-6mm]
 \ite Goddington, E.A. \& Levinson, N., 1955. {\em Theory of Ordinary Differential Equations},
McGraw-Hill, New York. \\[-6mm]
 \ite Grewal, M.S. \& Andrews, A.P., 1993. {\em Kalman Filtering},
 Prentice Hall, New Jersey. \\[-6mm]
 \ite  Gronwall, T.H., 1919. Note on the derivatives with respect to a parameter of the solutions
 of a system of differential equations, {\em Ann. Math.}, {\bf 20}, 292-296. \\[-6mm]
 \ite Grosser, M., 1964. The search for a planet beyond Neptune, {\em ISIS}, {\bf 55}, 163-183.
 \\[-6mm]
  \ite Guier, W.H., 1963. Determination of the non-zonal
harmonics of the geopotential from satellite Doppler data, {\em Nature},
{\bf 200}, 124-125. \\[-6mm]
 \ite Guier, W.H. \& Newton, R.R., 1965. The Earth's gravity field as deduced from the Doppler
 track- ing of five satellites, {\em J. geophys. Res.}, {\bf 70}, 4613-4626. \\[-6mm]
 \ite Gunter, B., Ries, J., Bettadpur, S. \& Tapley, B., 2006. A simulation study of the errors of
 omission and commission for GRACE RL01 gravity fields, {\em J. Geod.}, {\bf 80}, 341-351. \\[-6mm]
   \ite Hackbusch, W., 1995. {\em Integral Equations -- Theory and Numerical
 Treatment}, Birkh\"{a}user, Berlin. \\[-6mm]
 \ite Hagihara, Y., 1972. {\em Celestial Mechanics, Vol.II, Part 1: Perturbation Theory}, MIT
 Press, Cambridge. \\[-6mm]  
  \ite Heiskanen, W.A. \&  Moritz, H., 1967.
 {\em Physical Geodesy}, Freeman Publ Co, San Francisco. \\[-6mm]
 \ite Hotine, M. \& Morrison, F., 1969. First integrals of the equations of satellite motion,
 {\em Bull. Geod.}, {\bf 43}, 41-45. \\[-6mm]
 \ite  Howland, J.L. \& Vaillancourt, R., 1961. A generalized curve-fitting procedure, {\em SIAM J.
 appl. Math.}, {\bf 9}, 165-168. \\[-6mm]
 \ite Hubbell, H.G. \& Smith, R.W., 1992. Neptune in America: negotiating a discovery,
 {\em J. Hist. Astron.}, {\bf 23}, 261-291. \\[-6mm]
  \ite  Hwang J-T., Dougherty, E.P., Rabitz, S., \& Rabitz, H., 1978.  The Green's function
method of sensitivity analysis in chemical kinetics, {\em J. chem. Phys.}, {\bf 69},
5180-5191. \\[-6mm]
 \ite Ilk, K.H., Feuchtinger, M. \& Mayer-G\"{u}rr, T., 2005. Gravity field
 recovery and validation by analysis of short arcs of a satellite-to-satellite
 tracking experiment as CHAMP and GRACE, in: Sanso F. ed. {\em A Window on the
 Future of Geodesy}, pp.189-194, Springer, Berlin. \\[-6mm]
\ite Ilk, K.H.,  L\"{o}cher, A. \& Mayer-G\"{u}rr, T., 2008. Do we need new gravity field
recovery techniques for the new gravity field satellites? In: Xu P.L., Liu J.N. \&
Dermanis A. eds. {\em VI Hotine-Marussi Symp. Theor. Comput. Geodesy}, pp.3-8, Springer,
Berlin. \\[-6mm]
 \ite Iorio, L. 2012. Dynamical orbital effects of general relativity on the
 satellite-to-satellite range and range-rate in the GRACE mission:
 A sensitivity analysis, {\em Adv. Space Res.}, {\bf 50}, 334-345.\\[-6mm]
  \ite Iorio, L., Lichtenegger, H., Ruggiero, M.L. \& Cordu, C., 2011.
Phenomenology of the Lense-Thirring effect in the solar system, {\em Astrophys. Space
Sci.}, {\bf 331}, 351-395.\\[-6mm]
 \ite Iorio, L., Ruggiero, M.L. \& Cordu, C. 2013.  Novel considerations about the error
budget of the LAGEOS-based tests of frame-dragging with GRACE geopotential models, {\em
Acta Astronaut.}, {\bf 91}, 141-148.\\[-6mm]
 \ite Izsak, I.G., 1961. A determination of the ellipticity of the earths equator from
 the motion of 2 satellites, {\em Astron. J.}, {\bf 66}, 226-229. \\[-6mm]
 \ite Izsak, I.G., 1963. Tesseral harmonics in the geopotential, {\em Nature}, {\bf 199},
 137-139. \\[-6mm]
   \ite Jekeli, C., 1999. The determination of gravitational potential
 differences from satellite-to-satellite tracking, {\em Celest. Mech. Dynam.
 Astron.}, {\bf 75}, 85-101. \\[-6mm]
 \ite Jekeli, C. \& Garcia, R., 1997. GPS phase accelerations for moving-base gravimetry.
 {\em J. Geod.}, {\bf 71}, 630-639. \\[-6mm]
  \ite Kaula, W.M., 1961a. Analysis of gravitational and geometric aspects of
 geodetic utilization of satellites, {\em Geophys. J. Roy. astr. Soc.}, {\bf 5},
 104-133. \\[-6mm]
 \ite Kaula, W.M., 1961b. A geoid and world geodetic system based on a
combination of gravimetric, astrogeodetic and satellite data, {\em J. geophys.
Res.}, {\bf B66}, 1799-1811. \\[-6mm]
  \ite Kaula, W.M., 1963. Determination of the Earth's gravitational field, {\em Rev.
  Geophys.}, {\bf 1}, 507-551. \\[-6mm]
  \ite Kaula, W.M., 1966. {\em Theory of Satellite Geodesy}. Blaisdell Publishing
Company, London. \\[-6mm]
 \ite Kim, J., 2000. {\em Simulation Study of a Low-low Satellite-to-Satellite
Tracking Missions}, PhD Dissertation, The University of Texas at Austin. \\[-6mm]
 \ite King-Hele, D.G. \& Merson, R.H., 1959. A new value for the Earth's flattening, derived from
 measurements of satellite orbits, {\em Nature}, {\bf 183},
 881-882. \\[-6mm] 
 \ite King-Hele, D.G. \& Walker, D.M.C., 1982. Geopotential harmonics of order 29, 30 and 31
 from analysis of resonant orbits, {\em Planet. Space Sci.}, {\bf 30}, 411-425. \\[-6mm]
 \ite Kloko\v{c}n\'{i}k, J. \& Posp\'{i}\v{s}ilov\'{a}, L., 1981. Intercomparison of Earth models
 by means of lumped coefficients, {\em Planet. Space Sci.}, {\bf 29}, 653-671. \\[-6mm]
 \ite Kloko\v{c}n\'{i}k, J., Gooding, R.H., Wagner, C.A., Kosteleck\'{y}, J. \& Bezd\v{e}k, A.,
 2013. The use of resonant orbits in satellite geodesy: A review, {\em Surv. Geophys.}, {\bf
 34}, 43-72.  \\[-6mm]
 \ite Kondo, J., 1991. {\em Integral Equations}, Clarendon Press, Oxford. \\[-6mm]
 \ite Koop, R., 1993. {\em Global Gravity Field Modelling Using Satellite Gravity
Gradiometry}, Netherlands Geodetic Commission, Publ. Geod. New Series No.38, Delft. \\[-6mm]
 \ite Kozai, Y., 1959. The motion of a close Earth satellite, {\em Astron. J.},
{\bf 64}, 367-377. \\[-6mm]
\ite Kozai, Y., 1961. Tesseral harmonics of the gravitational potential of the Earth
 as derived from satellite motions, {\em Astron. J.}, {\bf 66}, 355-358. \\[-6mm]
 \ite Kozai, Y., 1962. Second-order solution of artificial satellite theory
 without air drag, {\em Astron. J.}, {\bf 67}, 446-461. \\[-6mm]
  \ite Kozai, Y., 1966. The Earth gravitational potential derived from satellite
 motion, {\em Space Sci. Rev.}, {\bf 5}, 818-879. \\[-6mm]
 \ite Lambeck, K. \& Coleman, R., 1983. The Earth's shape and gravity field: a report of
 progress from 1958 to 1982, {\em Geophys. J. Roy. astr. Soc.}, {\bf 74}, 25-54. \\[-6mm]
 \ite Lambeck, K. \& Coleman, R., 1986. Reply to comments by Lerch et al. on 'The Earth's
 shape and gravity field: a report of progress from 1958 to 1982',
 {\em Geophys. J. Roy. astr. Soc.}, {\bf 86}, 665-668. \\[-6mm]
 \ite Lequeux, J., 2013. {\em Le Verrier -- Magnificent and Detestable Astronomer},
 Astrophysics and Space Science Library 397, Springer, New York. \\[-6mm]
 \ite Lerch, F.J., Wagner, C.A., Richardson, J.A. \& Brownd, J.E., 1974. {\em Goddard Earth Models (5 and
 6)}, Technical Report NASA-TM-X-70868, Goddard Space Flight Center, Maryland. \\[-6mm]
 \ite  Linga, P., Al-Saifi, N. \& Englezos, P., 2006. Comparison of the Luus-Jaakola optimization and
 gauss-newton methods for parameter estimation in ordinary differential equation models, {\em
  Ind. Eng. Chem. Res.}, {\bf 45}, 4716-4725. \\[-6mm]
  \ite Liang, H. \& Wu, H., 2008. Parameter estimation for differential equation models using a framework
  of measurement error in regression models, {\em J. Amer. statist. Ass.}, {\bf 103},
  1570-1583. \\[-6mm]
 \ite Long, A.C., Cappellari, J.O., Velez, C.E. \& Fuchs, A.J., 1989. {\em Goddard Trajectory
 Determination System (GTDS) Mathematical Theory}, Technical Report FDD/552-89/0001 and CSC/TR-89/6001,
 Goddard Space Flight Center, Maryland. \\[-6mm]
 \ite Lowell, P., 1915. {\em Memoir on a Trans-Neptunian Planet}, Memoirs of the Lowell Observatory,
 Vol.1, No.1, Thos. P. Ncholas \& Son, Mass. \\[-6mm]
 \ite Marsh, J.G., Lerch, F.J., Putney, B.H., Christodoulidis, D.C., Smith, D.E.,
Felsentreger, T.L., Sanches, B.V., Klosko, S.M., Pavlis, E.C., Martin, T.V., Robbins,
J.W., Williamson, R.G., Colombo, O.L., Rowlands, D.D., Eddy, W., Chandler, N.L., Rachlin,
K.E., Patel, G.B., Bhati, S. \& Chinn, D.S., 1988. A new gravitational model for the
Earth from satellite tracking data: GEM-T1, {\em J. geophys. Res.}, {\bf 93},
 6169-6215. \\[-6mm]
\ite Marsh, J.G., Lerch, F.J., Putney, B.H., Felsentreger, T.L., Sanches, B., Klosko,
S.M., Patel, G.B., Robbins, J.W., Williamson, R.G., Engelis, T.L., Eddy, W., Chandler,
N.L., Chinn, D.S., Kapoor, S., Rachlin, K.E., Braatz, L.E. \& Pavlis, E.C., 1990. The
GEM-T2 gravitational model, {\em J. geophys. Res.}, {\bf 95}, 22043-22071. \\[-6mm]
   \ite Mayer-G\"{u}rr, T., Ilk, K.H., Eicker, A. \& Feuchtinger, M., 2005.
 ITG-CHAMP01: a CHAMP gravity field model from short kinematic arcs over a
 one-year observation period, {\em J. Geod.}, {\bf 78}, 462-480. \\[-6mm]
 \ite Merson, R.H. \& King-Hele, D.G., 1958. Use of artificial satellites to explore the Earth's
 gravitational field: Results from SPUTNIK 2 (1957$\beta$), {\em Nature}, {\bf 182},
 640-641. \\[-6mm] 
 \ite Montenbruck, O. \& Gill, E., 2000. {\em Satellite Orbits}, Berlin: Springer. \\[-6mm]
 \ite O'Keefe, J.A., Eckels, A. \& Squires, R.K., 1959. Vanguard measurements give pearl-shaped
 component of Earth's figure, {\em Science}, {\bf 129}, 565-566. \\[-6mm]
 \ite National Research Council (NRC), 1997. {\em Satellite Gravity and the Geosphere: Contributions
 to the Study of the Solid Earth and Its Fluid Envelopes}, National Academy Press,
 Washington DC. \\[-6mm]
 \ite Nayfeh, A.H., 2004. {\em Perturbation methods}, Wiley, New York. \\[-6mm]
 \ite Nerem, R.S., Jekeli, C. \& Kaula, W.M., 1995. Gravity field determination and characteristics:
 Retrospective and prospective, {\em J. geophys. Res.}, {\bf 100}, 15,053-15,074.
 \\[-6mm]
 \ite Pierce, R., Leitch, J., Stephens, M., Bender, P. \& Nerem, R., 2008. Intersatellite range monitoring
 using optical interferometry, {\em Appl. Opt.}, {\bf 47}, 5007-5018. \\[-6mm]
 \ite Phillips, G.M. \& Taylor, P.J., 1996. {\em Theory and Applications of Numerical Analysis}  (2nd
 Ed), Elsevier, Amsterdam. \\[-6mm]
\ite Prussing, J.E. \& Conway, B.A., 1993. {\em Orbital Mechanics},
Oxford University Press, Oxford. \\[-6mm]
 \ite  Ramsay, J.O., Hooker, G., Campbell, D. \& Cao, J., 2007.
Parameter estimation for differential equations: a generalized smoothing approach (with
discussions), {\em J. Roy. Statist. Soc.}, {\bf B69}, 741-796. \\[-6mm]
 \ite Reubelt, T., Austen, G. \& Grafarend, E., 2003. Harmonic analysis of the
Earth's gravitational field by means of semi-continuous ephemerides of a low Earth
orbiting GPS-tracked satellite. Case study: CHAMP, {\em J. Geod.}, {\bf 77}, 257-278. \\[-6mm]
 \ite Reigber, C., 1989. Gravity field recovery from satellite tracking data, in: Sanso F and Rummel R
 (eds.) {\em Theory of satellite geodesy and gravity field determination},
 Lecture notes in Earth Sciences, vol.25, Springer, Berlin, pp.197-234 \\[-6mm]
 \ite Reigber, C., Balmino, G., Schwintzer, P., Biancale, R., Bode, A., Lemoine,
J.-M., K\"{o}nig, R., Loyer, S., Neumayer, K.-H., Marty, J.-C., Barthelmes, F.,
 Perosanz, F. \& Zhu, S.Y., 2003. Global gravity field recovery using solely
 GPS tracking and accelerometer data from CHAMP, {\em Space Sci. Rev.}, {\bf
 108}, 55-66. \\[-6mm] 
  \ite Renzetti, G. 2012. Are higher degree even zonals really harmful for the LARES/LAGEOS
 frame-dragging experiment? {\em Can. J. Phys.}, {\bf 90}, 883-888.\\[-6mm]
 \ite Renzetti, G. 2013. History of the attempts to measure orbital frame-dragging with
 artificial satellites, {\em Cent. Eur. J. Phys.}, {\bf 11}, 531-544.\\[-6mm]
\ite Riley, J.D., Bennett, M.M. \& McCormick, E., 1967. Numerical integration
of variational equations, {\em Math. Comput.}, {\bf 21}, 12-17. \\[-6mm]
 \ite Ritt, J.F., 1919. On the differentiability of the solution of a differential equation
 with respect to a parameter, {\em Ann. Math.}, {\bf 20}, 289-291. \\[-6mm]
  \ite Rowlands, D.D., Ray, R.D., Chinn, D.S. \& Lemoine, F.G., 2002. Short-arc
 analysis of intersatellite tracking data in a gravity mapping mission, {\em J.
 Geod.}, {\bf 76}, 307-316. \\[-6mm]
 \ite Rummel, R., 1986. Satellite gradiometry, In: {\em Mathematical and Numerical
Techniques in Physical Geodesy}, edited by H S\"{u}nkel, Springer, Berlin, pp.317-363 \\[-6mm]
 \ite Rummel, R., Horwath, M., Yi, W.Y., Albertella, A., Bosch, W., \& Haagmans, R.,
 2011a. GOCE, satellite gravimetry and antarctic mass transports, {\em Surv. Geophys.}, {\bf 32},
 643-657. \\[-6mm]
 \ite Rummel, R., Yi, W.Y. \& Stummer, C., 2011b. GOCE gravitational gradiometry,
 {\em J. Geod.}, {\bf 85}, 777-790.  \\[-6mm]
  \ite Seeber, G., 2003. {\em Satellite Geodesy}, 2nd edn., Walter de Gruyter,
 Berlin. \\[-6mm]
 \ite Schwintzer, P., Reigber, C., Bode, A., Kang, Z., Zhu, S., Massmann, F., Raimondo, J., Biancale, R.,
 Balmino, G., Lemoine, J., Moynot, B., Marty, J., Barlier, F. \&  Boudon, Y., 1997. Long-wavelength global
 gravity field models: GRIM4-S4, GRIM4-C4, {\em J. Geod.}, {\bf 71}, 189-208. \\[-6mm]
  \ite Sheard, B.S., Heinzel, G., Danzmann, K., Shaddock, D.A., Klipstein, W.M.
  \& Folkner, W.M., 2012. Intersatellite  laser ranging instrument for the GRACE
  follow-on mission, {\em J. Geod.}, {\bf 86},   1083-1095. \\[-6mm]
  \ite Stengel, R.F., 1986. {\em Optimal Control and Estimation}, Wiley, New York. \\[-6mm]
  \ite Stoer, J. \& Burlirsch, R., 2002. {\em Introduction to Numerical Analysis}, 3rd edn, Springer,
  Berlin. \\[-6mm]
   \ite \v{S}vehla, D. \& Rothacher, M., 2005. Kinematic positioning of LEO and
 GPS satellites and IGS stations on the ground, {\em Adv. Space Res.}, {\bf
 36}, 376-381. \\[-6mm]
   \ite Taff, L.G., 1985. {\em Celestial Mechanics: A Computational Guide for
  the Practitioners}, Wiley-Interscience, New York. \\[-6mm]
  \ite Tapley, B.D., 1989. Fundamentals of orbit determination, in: Sanso F and Rummel R
 (eds.) {\em Theory of satellite geodesy and gravity field determination},
 Lecture notes in Earth Sciences, vol.25, Springer, Berlin, pp.235-260 \\[-6mm]
 \ite Tapley, B.D., Bettadpur, S., Watkins, M. \& Reigber, C., 2004a. The
 gravity recovery and climate experiment: mission overview and early results,
 {\em Geophys. Res. Lett.}, {\bf 31}, L09607. \\[-6mm] 
 \ite Tapley, B.D., Schutz, B.E. \& Born, G.H., 2004b. {\em Statistical Orbit Determination},
 Elsevier, Amsterdam. \\[-6mm] 
 \ite Teodorescu, P., St\u{a}nescu, N. \& Pandrea, N., 2013. {\em Numerical Analysis with
 Applications in Mechanics and Engineering}, Wiley, New Jersey. \\[-6mm]
  \ite Turyshev, S.G., Sazhin, M.V. \& Toth, V.T., 2014. General relativistic laser interferometric
  observables of the GRACE-Follow-On mission, {\em Phys. Rev. D}, {\bf 89}, art.105029. \\[-6mm]
 \ite  Wang, B. \& Enright, W., 2013. Parameter estimation for odes using across-entropy
 approach, {\em SIAM J. sci. Comput.}, {\bf 35}, A2718-A2737. \\[-6mm]
 \ite Wahr, J.M., Molenaar, M. \& Bryan, F., 1998. Time-variability of the Earth's gravity
 field: Hydrological and oceanic effects and their possible detection using GRACE,
 {\em J. geophys. Res.}, {\bf 103}, 30205-30230. \\[-6mm]
   \ite Wolff, M., 1969. Direct measurements of the Earth's gravitational
 potential using a satellite pair, {\em J. geophys. Res.}, {\bf 74}, 5295-5300.
 \\[-6mm]
 \ite Xu, P.L., 2008. Position and velocity perturbations for the determination
of geopotential from space geodetic measurements. {\em Celest. Mech. dynam. Astr.},
 {\bf 100}, 231-249. \\[-6mm]
\ite Xu, P.L., 2009a. Zero initial partial derivatives of satellite orbits with respect
to force parameters violate the physics of motion of celestial bodies, {\em Science China
Series D: Earth Sciences}, {\bf 52}, 562-566. \\[-6mm]
 \ite Xu, P.L., 2009b. Iterative generalized cross-validation for fusing
 heteroscedastic data of inverse ill-posed problems, {\em Geophys. J. Int.},
 {\bf 179}, 182-200.
 \ite Xu, P.L.,  2012. Mathematical challenges arising from
earth-space observation: mixed integer linear models, measurement-based perturbation
theory and data assimilation for ill-posed problems, Invited talk, Joint Mathematical
Meeting of American Mathematical Society, Boston, Jan 4-7. \\[-6mm]
  \ite Xu, P.L., 2015a. Zero initial partial derivatives of satellite orbits with respect
 to force parameters nullify the mathematical basis of the numerical integration
 method for the determination of standard gravity models from space geodetic
 measurements, European Geosciences Union, Vienna, Apr 12-17. \\[-6mm]
 \ite Xu, P.L., 2015b. Mathematical foundation for the next generation of global gravity models
 from satellite gravity missions of CHAMP/GRACE types, presented at the 26th IUGG General
 Assembly, Prague, June 21 - July 2. \\[-6mm]
  \ite Xu, P.L., Shen, Y.Z., Fukuda, Y. \& Liu, Y.M., 2006. Variance component
 estimation in inverse ill-posed linear models,  {\em J. Geod.}, {\bf 80}, 69-81. \\[-6mm]
 \ite Xu, P.L., Shi, C., Fang, R.X., Liu, J.N., Niu, X.J. Zhang, Q. \& Yanagidani, T., 2013.
 High-rate precise point positioning (PPP) to
measure seismic wave motions: An experimental comparison of GPS PPP with inertial
measurement units, {\em J. Geod.}, {\bf 87}, 361-372, DOI 10.1007/s00190-012-0606-z \\[-6mm]
 \ite Yionoulis, S.M., 1965. A study of the resonance effects due to the Earth's potential
 function, {\em J. geophys. Res.}, {\bf 70}, 5991-5996.

 \el

\section*{Appendix: Derivation of $\Delta\mathbf{z}(t_{yi})$ for the modified Euler
method} For convenience, we rewrite the recursive formula of the modified Euler method as
follows: \beq \label{AppMEulerStart} \mathbf{z}(t_j) = \mathbf{z}(t_{j-1}) + \frac{h}{2}
[ \mathbf{g}(t_{j-1}, \mathbf{z}(t_{j-1}),\mathbf{p}) + \mathbf{g}\{t_j,
\mathbf{z}(t_{j-1})+h\mathbf{g}(t_{j-1}, \mathbf{z}(t_{j-1}),\mathbf{p}),\mathbf{p}\}],
\eeq for $j=1,2,...,m$, with the nominal reference orbit $\mathbf{z}^0(t,
\mathbf{z}^0(t),\mathbf{p}^0)$.

We will now derive the representation of $\Delta\mathbf{z}(t_{yi})$ in terms of the
corrections $\Delta\mathbf{z}_0$ and $\Delta\mathbf{p}$.

To start with, we set $j=1$ in (\ref{AppMEulerStart}) and have \beq
\label{AppMEulerStartXT1} \mathbf{z}(t_1) = \mathbf{z}(t_0) + \frac{h}{2} [
\mathbf{g}(t_0, \mathbf{z}(t_0),\mathbf{p}) + \mathbf{g}\{t_1,
\mathbf{z}(t_0)+h\mathbf{g}(t_0, \mathbf{z}(t_0),\mathbf{p}),\mathbf{p}\}]. \eeq
Linearizing both $\mathbf{g}(t_0, \mathbf{z}(t_0),\mathbf{p})$ and $\mathbf{g}\{t_1,
\mathbf{z}(t_0)+h\mathbf{g}(t_0, \mathbf{z}(t_0),\mathbf{p}),\mathbf{p}\}$, and
neglecting the terms of $h\Delta\mathbf{z}_0$ and $h\Delta\mathbf{p}$ (because of the
coefficient $h/2$ before the brackets in (\ref{AppMEulerStartXT1})), we have \alpheqn
\beq \label{gFunT0} \mathbf{g}(t_0, \mathbf{z}(t_0),\mathbf{p}) = \mathbf{g}(t_0,
\mathbf{z}^0(t_0),\mathbf{p}^0) + \mathbf{G}_{gz0}\Delta\mathbf{z}_0 +
\mathbf{G}_{gp0}\Delta\mathbf{p}, \eeq and \begin{eqnarray} \label{gFunT1}
 &  & \mathbf{g}\{t_1, \mathbf{z}(t_0)+h\mathbf{g}(t_0,
\mathbf{z}(t_0),\mathbf{p}),\mathbf{p}\} \nonumber \\
& = & \mathbf{g}(t_1,
\mathbf{z}^0(t_1),\mathbf{p}^0) + \mathbf{G}_{gz1} \{ \mathbf{z}(t_0)+h\mathbf{g}(t_0,
\mathbf{z}(t_0),\mathbf{p}) - \mathbf{z}^0(t_1) \} +
\mathbf{G}_{gp1}\Delta\mathbf{p} \nonumber \\
& = & \mathbf{g}(t_1, \mathbf{z}^0(t_1),\mathbf{p}^0) + \mathbf{G}_{gz1} \{
\mathbf{z}^0(t_0)+\Delta\mathbf{z}_0 + h\mathbf{g}(t_0, \mathbf{z}^0(t_0),\mathbf{p}^0) -
\mathbf{z}^0(t_1) \} + \mathbf{G}_{gp1}\Delta\mathbf{p} \nonumber \\
& = & \mathbf{g}(t_1, \mathbf{z}^0(t_1),\mathbf{p}^0) + \mathbf{G}_{gz1}
\delta\mathbf{z}^0_{01} + \mathbf{G}_{gz1}\Delta\mathbf{z}_0 +
\mathbf{G}_{gp1}\Delta\mathbf{p}.
\end{eqnarray}\reseteqn\setcounter{EQ1App}{\value{equation}}

Inserting (\ref{gFunT0}) and (\ref{gFunT1}) into (\ref{AppMEulerStartXT1}) yields
\begin{eqnarray} \label{AppMEulerStartXT1S1} \mathbf{z}(t_1) & = & \mathbf{z}^0(t_1) +
\Delta\mathbf{z}(t_1) \nonumber \\
& = & \mathbf{z}^0_0 + \Delta\mathbf{z}_0 + \frac{h}{2} [ \mathbf{g}(t_0,
\mathbf{z}^0(t_0),\mathbf{p}^0) + \mathbf{G}_{gz0}\Delta\mathbf{z}_0 +
\mathbf{G}_{gp0}\Delta\mathbf{p}] \nonumber \\
&  & + \frac{h}{2} [\mathbf{g}(t_1, \mathbf{z}^0(t_1),\mathbf{p}^0) + \mathbf{G}_{gz1}
\delta\mathbf{z}^0_{01} + \mathbf{G}_{gz1}\Delta\mathbf{z}_0 +
\mathbf{G}_{gp1}\Delta\mathbf{p}] \nonumber \\
& = & \mathbf{z}^0_0 +\frac{h}{2} [\mathbf{g}(t_0, \mathbf{z}^0(t_0),\mathbf{p}^0) +
\mathbf{g}(t_1, \mathbf{z}^0(t_1),\mathbf{p}^0) ] + \frac{h}{2}\mathbf{G}_{gz1}
\delta\mathbf{z}^0_{01} \nonumber \\
 &   & + [ \mathbf{I}_6 + \frac{h}{2}(\mathbf{G}_{gz0}+\mathbf{G}_{gz1})]
 \Delta\mathbf{z}_0 + \frac{h}{2}(\mathbf{G}_{gp0}+\mathbf{G}_{gp1})
 \Delta\mathbf{p}, \end{eqnarray} which can also be rewritten in terms of
 $\Delta\mathbf{z}(t_1)$ as follows:
\begin{eqnarray} \label{AppMEulerStartXT1Final} \Delta\mathbf{z}(t_1) & = &
\mathbf{z}^0_0 + \frac{h}{2}[\mathbf{g}(t_0, \mathbf{z}^0(t_0),\mathbf{p}^0) +
\mathbf{g}(t_1, \mathbf{z}^0(t_1),\mathbf{p}^0) ] + \frac{h}{2}\mathbf{G}_{gz1}
\delta\mathbf{z}^0_{01} - \mathbf{z}^0(t_1) \nonumber \\
 &  & + [ \mathbf{I}_6 + \frac{h}{2}(\mathbf{G}_{gz0}+\mathbf{G}_{gz1})]
 \Delta\mathbf{z}_0 + \frac{h}{2}(\mathbf{G}_{gp0}+\mathbf{G}_{gp1})
 \Delta\mathbf{p} \nonumber \\
 & = & \delta\mathbf{z}^{0M}_{01} + \frac{h}{2}\mathbf{G}_{gz1}
\delta\mathbf{z}^0_{01} \nonumber \\
 &  & + [ \mathbf{I}_6 + \frac{h}{2}(\mathbf{G}_{gz0}+\mathbf{G}_{gz1})]
 \Delta\mathbf{z}_0 + \frac{h}{2}(\mathbf{G}_{gp0}+\mathbf{G}_{gp1})
\Delta\mathbf{p}, \end{eqnarray} where $$ \delta\mathbf{z}^{0M}_{01} = \mathbf{z}^0_0 +
\frac{h}{2}[\mathbf{g}(t_0, \mathbf{z}^0(t_0),\mathbf{p}^0) + \mathbf{g}(t_1,
\mathbf{z}^0(t_1),\mathbf{p}^0) ] -\mathbf{z}^0(t_1). $$

For $j=2$, we have \beq \label{AppMEulerStartXT2} \mathbf{z}(t_2) = \mathbf{z}(t_1) +
\frac{h}{2} [ \mathbf{g}(t_1, \mathbf{z}(t_1),\mathbf{p}) + \mathbf{g}\{t_2,
\mathbf{z}(t_1)+h\mathbf{g}(t_1, \mathbf{z}(t_1),\mathbf{p}),\mathbf{p}\}]. \eeq In a
similar manner, we linearize both $\mathbf{g}(t_1, \mathbf{z}(t_1),\mathbf{p})$ and
$\mathbf{g}\{t_2, \mathbf{z}(t_1)+h\mathbf{g}(t_1,
\mathbf{z}(t_1),\mathbf{p}),\mathbf{p}\}$ in the formula (\ref{AppMEulerStartXT2}),
neglect the terms of $h\Delta\mathbf{z}_0$ and $h\Delta\mathbf{p}$ (again due to the
reason of the coefficient $h/2$) and obtain
\begin{eqnarray} \label{AppMEulerStartXT2S1} \mathbf{z}(t_2) & = & \mathbf{z}^0(t_2) +
\Delta\mathbf{z}(t_2) \nonumber \\
& = & \mathbf{z}^0(t_1) + \Delta\mathbf{z}(t_1) + \frac{h}{2} [ \mathbf{g}(t_1,
\mathbf{z}^0(t_1),\mathbf{p}^0) + \mathbf{G}_{gz1}\Delta\mathbf{z}(t_1) +
\mathbf{G}_{gp1}\Delta\mathbf{p}] \nonumber \\
&  & + \frac{h}{2} \{\mathbf{g}(t_2, \mathbf{z}^0(t_2),\mathbf{p}^0) + \mathbf{G}_{gz2}[
 \mathbf{z}(t_1)+h\mathbf{g}(t_1,
\mathbf{z}(t_1),\mathbf{p}) - \mathbf{z}^0(t_2) ] + \mathbf{G}_{gp2}\Delta\mathbf{p} \}
\nonumber \\
& = & \mathbf{z}^0(t_1) + \Delta\mathbf{z}(t_1) + \frac{h}{2} \{ \mathbf{g}(t_1,
\mathbf{z}^0(t_1),\mathbf{p}^0) + \mathbf{G}_{gz1}[\delta\mathbf{z}^{0M}_{01}
+\Delta\mathbf{z}_0] + \mathbf{G}_{gp1}\Delta\mathbf{p}\} \nonumber \\
&  & + \frac{h}{2} \{\mathbf{g}(t_2, \mathbf{z}^0(t_2),\mathbf{p}^0) + \mathbf{G}_{gz2}[
\delta\mathbf{z}^0_{12} + \delta\mathbf{z}^{0M}_{01} +\Delta\mathbf{z}_0] +
\mathbf{G}_{gp2}\Delta\mathbf{p} \}.  \end{eqnarray}

Substituting $\Delta\mathbf{z}(t_1)$ of (\ref{AppMEulerStartXT1Final}) into
(\ref{AppMEulerStartXT2S1}) and after some re-arrangement, we have
\begin{eqnarray} \label{AppMEulerStartXT2Final}
\Delta\mathbf{z}(t_2) & = & \mathbf{z}^0(t_1) - \mathbf{z}^0(t_2) +
\delta\mathbf{z}^{0M}_{01} + \frac{h}{2}\mathbf{G}_{gz1}
\delta\mathbf{z}^0_{01} \nonumber \\
 &  & + [ \mathbf{I}_6 + \frac{h}{2}(\mathbf{G}_{gz0}+\mathbf{G}_{gz1})]
 \Delta\mathbf{z}_0 + \frac{h}{2}(\mathbf{G}_{gp0}+\mathbf{G}_{gp1})
\Delta\mathbf{p} \nonumber \\
 &  & + \frac{h}{2} \{ \mathbf{g}(t_1,
\mathbf{z}^0(t_1),\mathbf{p}^0) + \mathbf{G}_{gz1}\delta\mathbf{z}^{0M}_{01} +
\mathbf{G}_{gz1}\Delta\mathbf{z}_0 +
\mathbf{G}_{gp1}\Delta\mathbf{p}\} \nonumber \\
&  & + \frac{h}{2} \{\mathbf{g}(t_2, \mathbf{z}^0(t_2),\mathbf{p}^0) + \mathbf{G}_{gz2}
 [ \delta\mathbf{z}^0_{12} + \delta\mathbf{z}^{0M}_{01} +\Delta\mathbf{z}_0 ] +
\mathbf{G}_{gp2}\Delta\mathbf{p} \}
\nonumber \\
 & = & \delta\mathbf{z}^{0M}_{01} + \delta\mathbf{z}^{0M}_{12} + \frac{h}{2}\mathbf{G}_{gz1}
\delta\mathbf{z}^0_{01} + \frac{h}{2}\mathbf{G}_{gz2} \delta\mathbf{z}^0_{12} +
\frac{h}{2} [ \mathbf{G}_{gz1}  +
\mathbf{G}_{gz2}] \delta\mathbf{z}^{0M}_{01} \nonumber \\
 &  & + [ \mathbf{I}_6 + \frac{h}{2}(\mathbf{G}_{gz0}+\mathbf{G}_{gz1})]
 \Delta\mathbf{z}_0 + \frac{h}{2}(\mathbf{G}_{gp0}+\mathbf{G}_{gp1})
\Delta\mathbf{p} \nonumber \\
 &  & + \frac{h}{2}(\mathbf{G}_{gz1}+\mathbf{G}_{gz2})]
 \Delta\mathbf{z}_0 + \frac{h}{2}(\mathbf{G}_{gp1}+\mathbf{G}_{gp2})
\Delta\mathbf{p} \nonumber \\
 & = & \delta\mathbf{z}^{0M}_{02} + \frac{h}{2}\sum\limits_{l=1}\limits^{1}[
\mathbf{G}_{gzl} + \mathbf{G}_{gz(l+1)}]\delta\mathbf{z}^{0M}_{0l}  +
\frac{h}{2}\sum\limits_{l=0}\limits^{1}\mathbf{G}_{gz(l+1)}
\delta\mathbf{z}^0_{l(l+1)} \nonumber \\
&  & + \left[\mathbf{I}_6 + \frac{h}{2}\sum\limits_{l=0}\limits^{1}\{
\mathbf{G}_{gzl}+\mathbf{G}_{gz(l+1)}\}\right]\Delta\mathbf{z}_0 \nonumber \\
&  & + \frac{h}{2}\sum\limits_{l=0}\limits^{1} \{
\mathbf{G}_{gpl}+\mathbf{G}_{gp(l+1)}\}\Delta\mathbf{p},
 \end{eqnarray} where $$ \delta\mathbf{z}^{0M}_{02} = \mathbf{z}^0_0 +
\frac{h}{2}\sum\limits_{l=0}\limits^{1}[\mathbf{g}(t_l, \mathbf{z}^0(t_l),\mathbf{p}^0) +
\mathbf{g}(t_{l+1}, \mathbf{z}^0(t_{l+1}),\mathbf{p}^0) ] -\mathbf{z}^0(t_2). $$

For $j=3$, we simply list the representation of $\Delta\mathbf{z}(t_j)$ as follows:
\begin{eqnarray} \label{AppMEulerStartXT3Final}
\Delta\mathbf{z}(t_3) & = & \delta\mathbf{z}^{0M}_{03} +
\frac{h}{2}\sum\limits_{l=1}\limits^2[
\mathbf{G}_{gzl} + \mathbf{G}_{gz(l+1)}]\delta\mathbf{z}^{0M}_{0l} \nonumber \\
&  & + \frac{h}{2}\sum\limits_{l=0}\limits^2\mathbf{G}_{gz(l+1)}
\delta\mathbf{z}^0_{l(l+1)} \nonumber \\
&  & + \left[\mathbf{I}_6 + \frac{h}{2}\sum\limits_{l=0}\limits^2\{
\mathbf{G}_{gzl}+\mathbf{G}_{gz(l+1)}\}\right]\Delta\mathbf{z}_0 \nonumber \\
&  & + \frac{h}{2}\sum\limits_{l=0}\limits^2 \{
\mathbf{G}_{gpl}+\mathbf{G}_{gp(l+1)}\}\Delta\mathbf{p}, \end{eqnarray} where $$
\delta\mathbf{z}^{0M}_{03} = \mathbf{z}^0_0 +
\frac{h}{2}\sum\limits_{l=0}\limits^{2}[\mathbf{g}(t_l, \mathbf{z}^0(t_l),\mathbf{p}^0) +
\mathbf{g}(t_{l+1}, \mathbf{z}^0(t_{l+1}),\mathbf{p}^0) ] -\mathbf{z}^0(t_3). $$

Repeating the same procedure and by induction to summarize, we can finally obtain the
representation of $\Delta\mathbf{z}(t_{yi})$ in terms of $\Delta\mathbf{z}_0$ and
$\Delta\mathbf{p}$ as follows:
\begin{eqnarray} \label{ModifiedEulerFinalYiApp} \Delta\mathbf{z}(t_{yi}) & = &
\delta\mathbf{z}^{0M}_{0t_{yi}} + \frac{h}{2}\sum\limits_{j=1}\limits^{m_{yi}-1}[
\mathbf{G}_{gzj} + \mathbf{G}_{gz(j+1)}]\delta\mathbf{z}^{0M}_{0j} \nonumber \\
&  & + \frac{h}{2}\sum\limits_{j=0}\limits^{m_{yi}-1}\mathbf{G}_{gz(j+1)}
\delta\mathbf{z}^0_{j(j+1)}  + \left[\mathbf{I}_6 +
\frac{h}{2}\sum\limits_{j=0}\limits^{m_{yi}-1}\{
\mathbf{G}_{gzj}+\mathbf{G}_{gz(j+1)}\}\right]\Delta\mathbf{z}_0 \nonumber \\
&  & + \frac{h}{2}\sum\limits_{j=0}\limits^{m_{yi}-1} \{
\mathbf{G}_{gpj}+\mathbf{G}_{gp(j+1)}\}\Delta\mathbf{p}, \end{eqnarray} where $$
\delta\mathbf{z}^{0M}_{0k} = \mathbf{z}_0^0 + \frac{h}{2}\sum\limits_{j=0}\limits^{k-1} [
\mathbf{g}(t_j, \mathbf{z}^0(t_j),\mathbf{p}^0) + \mathbf{g}\{t_{j+1},
\mathbf{z}^0(t_{j+1}),\mathbf{p}^0\}] - \mathbf{z}^0(t_k), $$ and
$$ \delta\mathbf{z}^0_{j(j+1)} = \mathbf{z}^0(t_j)+ h\mathbf{g}(t_j,
\mathbf{z}^0(t_j),\mathbf{p}^0) - \mathbf{z}^0(t_{j+1}). $$

\end{document}